\documentclass[useAMS,usenatbib]{mn2e}
\usepackage{graphicx}
\usepackage{times}
\usepackage{natbib}
\usepackage{subfigure}
\usepackage{url}
\usepackage{float,lscape}
\usepackage{amsmath}
\usepackage[dvipsnames]{xcolor}
\usepackage{amssymb}
\usepackage{multirow}
\usepackage{pdfpages}
\usepackage{titlesec}
\usepackage{tikz}
\usepackage[titletoc,page]{appendix}
 
\newcommand*{\textquotedouble}[1]{\textquotedblleft #1\textquotedblright}
\def\hii{\mbox{H\,{\sc ii}}}
\def\cii{\mbox{C\,{\sc ii}}}

  \title[Kinematics of G351.69--1.15 and G351.63--1.25]{Gas Kinematics in the \hii~regions G351.69--1.15 and G351.63--1.25}

\author[Veena et al.]{V. S. Veena$^{1}$\thanks{E-mail: veenavs.13@iist.ac.in}, S. Vig$^1$, A. Tej$^1$, N. G. Kantharia$^2$, S. K. Ghosh$^2$\\
$^1$Indian Institute of Space Science and Technology, Thiruvananthapuram, 695 547, India \\
$^2$National Centre for Radio Astrophysics (NCRA-TIFR), Pune, 411 007, India}

\begin{document}

\date{}

\pagerange{\pageref{firstpage}--\pageref{lastpage}} \pubyear{}

\maketitle

\label{firstpage}
  
\begin{abstract}
We probe the structure and kinematics of two neighbouring \hii~regions identified as cometary and bipolar, using radio recombination lines (RRL). The H172$\alpha$ RRLs from these \hii~regions: G351.69--1.15 and G351.63--1.25, are mapped using GMRT, India. We also detect carbon RRLs C172$\alpha$ towards both these regions. The hydrogen RRLs display the effects of pressure and dynamical broadening in the line profiles, with the dynamical broadening ($\sim15$~km/s) playing a major role in the observed profile of G351.69--1.15. We investigate the kinematics of molecular gas species towards this \hii~region from the MALT90 pilot survey. The molecular gas is mostly distributed towards the north and north-west of the cometary head. The molecular line profiles indicate signatures of turbulence and outflow in this region. The ionized gas at the cometary tail is blue shifted by $\sim$8~km/s with respect to the ambient molecular cloud, consistent with the earlier proposed champagne flow scenario. The relative velocity of $\sim$5~km/s between the northern and southern lobes of the bipolar \hii~region G351.63--1.25 is consistent with the premise that the bipolar morphology is a result of the expanding ionized lobes within a flat molecular cloud. 
 
\end{abstract}
   
\begin{keywords}   
\hii~regions -- stars : formation -- ISM : kinematics and dynamics -- ISM : lines and bands -- radio lines : ISM  -- ISM: individual: G351.69--1.15, G351.63--1.25 
\end{keywords}

\section{INTRODUCTION}

\hii~regions are photoionized nebulae formed around massive young stars. As these young stars are embedded in their parent molecular clouds, \hii~regions are visible mostly in the infrared and longer wavelengths as they can penetrate the thick layers of gas and dust in the surrounding cocoons. The hypercompact and ultracompact \hii~regions with their small sizes ($<$0.1~pc) and high electron densities ($n_e>$10$^6$cm$^{-4}$) are located in the interior of the molecular clouds and are believed to trace the early phases of \hii~region evolution. High resolution radio interferometric observations of ultracompact \hii~regions reveal that they appear in a wide range of morphological classifications such as spherical, core-halo, shell, irregular and cometary \citep{1989ApJS...69..831W}.  An additional morphology designated as bipolar  has been also included recently in the classification scheme \citep{{1994AAS...185.8417D},{2015A&A...582A...1D}}. According to \citet{1989ApJS...69..831W}, 43\% of the young \hii~regions are spherical or unresolved sources, 16\% are core-halo, 4\% are shell-like, 17\% are irregular or multiply peaked,  and 20\% are cometary in morphology. These findings are further supported by the studies of \citet{1994ApJS...91..659K}. The diverse morphologies of \hii~regions depend on several factors such as age, ionized and molecular gas dynamics, density structure of the ambient interstellar medium (ISM) and relative motion of the \hii~region with respect to the ambient medium \citep{2002ARA&A..40...27C}. However, the details of such interaction mechanisms are not clearly understood. In addition, studies have shown that the morphology of a \hii~region also depends on the spatial scale at which it is observed and one class of \hii~region morphology could lead to another at an alternate resolution \citep{{1989ApJS...69..831W},{1992AJ....103..234F},{2004ApJ...605..285S}}. Although the classification scheme is primarily based on ultracompact \hii~regions, these morphologies have also been identified on much larger scales (order of few parsecs). Few models that explain the cometary morphology predict large scale structures extending to few parsecs \citep[e.g.][]{{1988ApJ...333..801R},{1994ApJ...429..268G},{2014MNRAS.438.1335R}}. The common appearance of a specific morphology hints that there are ordered physical processes associated with massive star formation \citep{2007prpl.conf..181H}. 

\par Our knowledge of the physical environments of \hii~regions can be improved by a thorough understanding of their morphologies.
 According to \citet{1989ApJS...69..831W}, a spherical or unresolved source could potentially reveal structures at higher resolution. In addition, there is a possibility that the spherical morphology is the result of a cometary object viewed along the axis of symmetry. Core-halo morphology is characterized by the presence of a bright peak enveloped within a low surface brightness halo. Shell-like regions appear as bright rings of emission where the morphology could be a consequence of optical depth effects. 
Irregular \hii~regions show multiple peaks along with an extended envelope. Cometary \hii~regions are characterized by their resemblance to a comet with a sharp, parabolic ionization front at the head, and the surface brightness decreasing towards the tail. The bipolar morphology is characterized by a \hii~region bright in its central part with two ionizing diffuse lobes on either side \citep{2015A&A...582A...1D}. In the present work, we investigate two \hii~regions: G351.69--1.15 and G351.63--1.25. While G351.69--1.15 has a cometary morphology,  G351.63--1.25 resembles a bipolar \hii~region.

\par The cometary appearance of \hii~regions are explained using several models. Among these, the most prominent ones are the \textquotedouble{champagne flow} and the \textquotedouble{bow shock} models. The \textquotedouble{champagne flow} model predicts that the cometary morphology is the result of density gradients in the molecular cloud and assumes the ionizing star to be stationary \citep{{1978A&A....70..769I},{1979A&A....71...59T}}. In this model, the \hii~regions expand out towards regions of lower density. The effect of strong stellar winds have also been incorporated in the recent models \citep{{1994ApJ...432..648G},{2006ApJS..165..283A}}.
Regarding the \textquotedouble{bow shock} model, \citet{1985ApJ...288L..17R} first suggested that the cometary  appearance of \hii~regions is due to the  relative motion of the ionizing star with respect to the ambient medium. This model was further modified by \citet{1990ApJ...353..570V} by incorporating the effects of a stellar wind, and they proposed that the cometary morphology is created due to the motion of wind blowing massive stars moving supersonically through the molecular cloud.  More recently, \citet{2003ApJ...596..344C} predicted hybrid models in the DR21 region that is a  combination of stellar wind and stellar motion (bow shock) along a density gradient (champagne flow). This is further confirmed by sensitive, high resolution studies of the same region by \citet{2014A&A...563A..39I}.

%%%%%%%%%%%%%%%%%%%% Fig1 %%%%%%%%%%%%%%%%%%%%%%%%%%%%%%%%%%%%%%%%%%%%%%%%%

\begin{figure}
\hspace*{-0.4cm}
\centering
\includegraphics[scale=0.3]{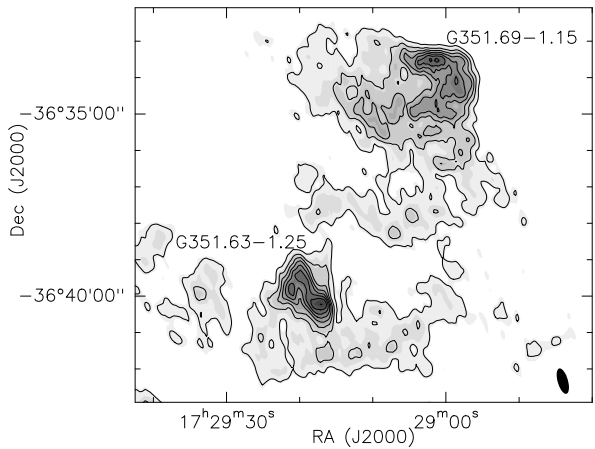}\caption{Radio continuum map of G351.69--1.15  and G351.63-1.25 at 325~MHz. The contour levels are from 25~mJy/beam to 250~mJy/beam in steps of 30~mJy/beam. The beam size is 21.7$\arcsec\times\rm8.5\arcsec$ and is shown as a filled ellipse in the bottom right corner of the image.}
\label{325}
\end{figure}

%%%%%%%%%%%%%%%%%%%%%%%%%%%%%%%%%%%%%%%%%%%%%%%%%%%%%%%%%%%%%%%%%%%%%

\par Among the proposed explanations of the cometary morphology, the two prominent models discussed predict different velocity structures within the \hii~region. Consequently, the velocity field of the ionized gas with respect to the ambient medium can be analyzed to test the hypothesis of each model. High resolution observations of radio recombination lines (RRLs) being excellent tracers of velocity distribution in the ionized gas can be used for this purpose \citep[e.g.,][]{{1979A&A....75...34P},{2001AJ....121.2681L},{2008ApJ...678L.109K},{2008ApJ...681..350S}}. RRLs arise from atomic transitions between large principal quantum number levels (typically n$\geq$40), where the small difference between energy levels entails emission of photons that belong to the radio regime. RRLs serve as the best tracers of ionized plasma in the \hii~regions due to their well-understood physics and high immunity to extinction \citep{2015aska.confE.126T}. The detailed analysis of RRLs enables the estimation of the physical properties of ionized gas such as  kinematic distance, electron temperature, electron density and metallicity \citep{{1970A&A.....5...53H},{1980A&A....91..279S},{1994ApJS...91..713M},{2005EAS....15..271K}}. Moreover, the kinematics and dynamics of the ionized gas can also be examined in finer detail to shed light on the interaction mechanisms.

\par The morphology of a bipolar \hii~region resembles an hourglass when projected in the sky \citep{2004ApJ...604L.105F}. This kind of morphology is observed in some massive star forming regions, for example, K-350A \citep{1994ApJ...428..670D} and NGC 7538 IRS1 \citep{1995ApJ...438..776G}. The most common characteristics of the identified bipolar \hii~regions are the presence of enhanced molecular or dust densities in a plane perpendicular to the axis of ionized flow and very high velocity gradients \citep{1998MNRAS.298...33R}. Only few of these objects have been identified so far and there are limited studies on the origin of bipolar morphology. Two dimensional hydrodynamic simulations of \citet{1979ApJ...233...85B} have shown that the expansion of a \hii~region within a sheet-like molecular cloud could result in a bipolar morphology. If the cloud is surrounded by a low density medium, the growing \hii~region can expand out towards the opposites faces of the flat molecular cloud, forming a double cone structure.

%%%%%%%%%%%%%%%%%%%% Fig2 %%%%%%%%%%%%%%%%%%%%%%%%%%%%%%%%%%%%%%%%%%%%%%%%%

\begin{figure*}
\hspace*{-0.4cm}
\centering
\includegraphics[scale=0.28,angle=90]{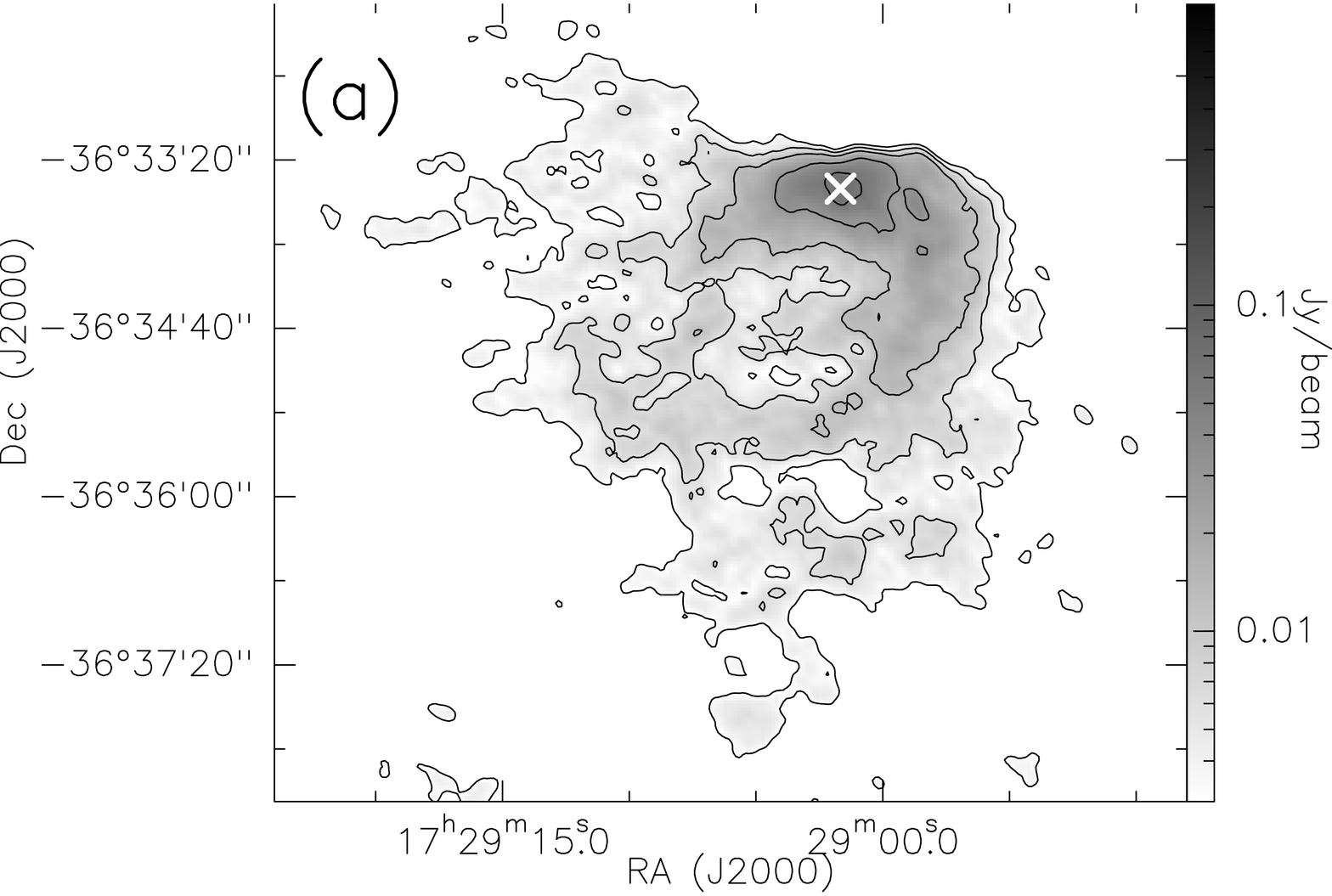} \quad \includegraphics[scale=0.28, angle=90]{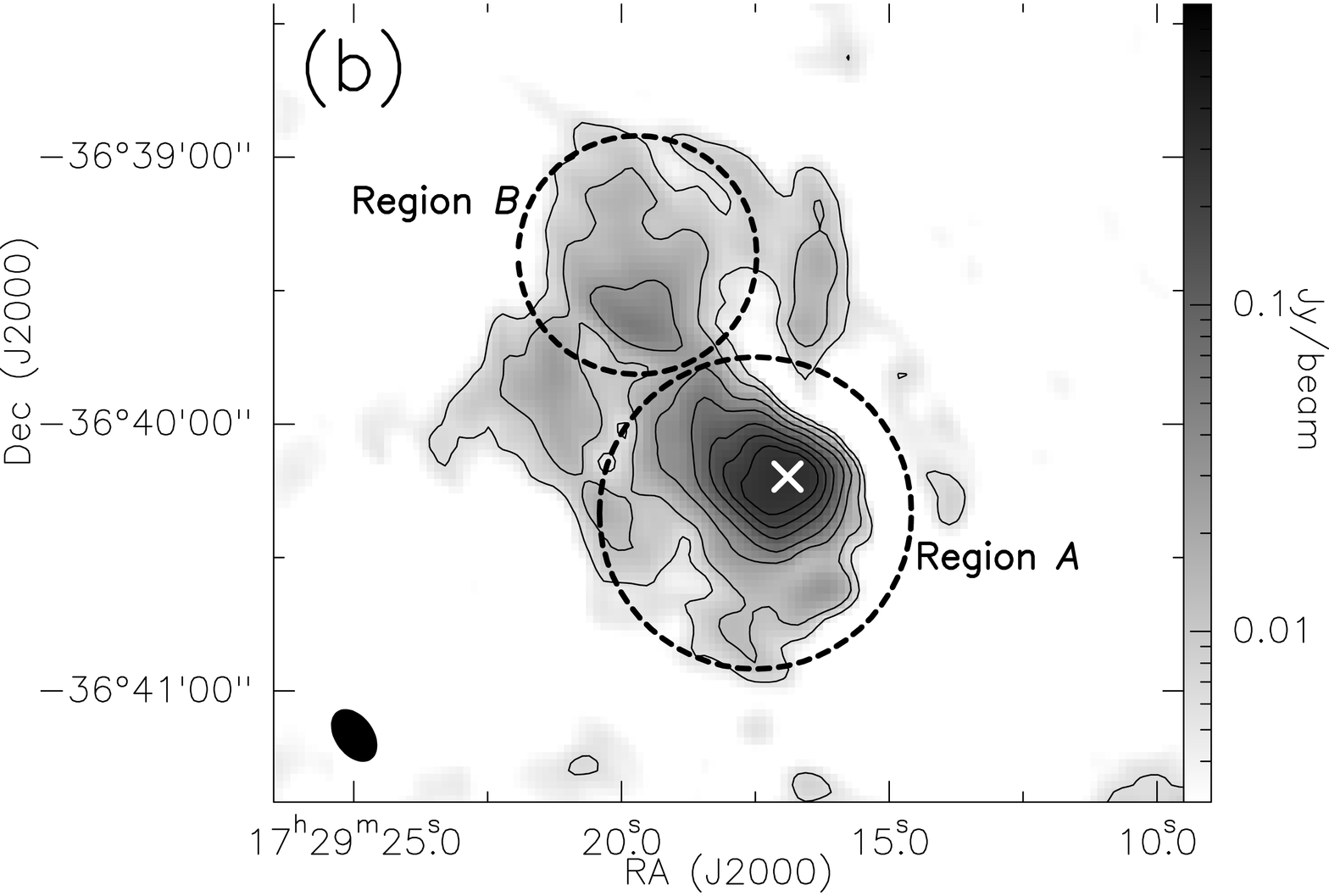} \caption{ (a) Radio continuum map of G351.69--1.15 at 1280~MHz. The contour levels are 3, 6, 12, 24, 48, 96 and 192~mJy/beam. (b) Radio continuum map of G351.63--1.25 at 1280~MHz. The contour levels are at 6, 12, 24, 48, 96,192 and 272~mJy/beam . The beam size is 6.7$\arcsec\times\rm4.5\arcsec$ and is shown as a filled ellipse in the bottom left corner of both panels. Locations of the peak emission are marked with crosses.}
\label{continuum}
\end{figure*}

%%%%%%%%%%%%%%%%%%%%%%%%%%%%%%%%%%%%%%%%%%%%%%%%%%%%%%%%%%%%%%%%%%%%%
\par The two \hii~regions investigated in this work, G351.69--1.15 and G351.63--1.25 are separated by $\sim$7$\arcmin$. The structure, morphology and star formation activity towards the cometary shaped G351.69--1.15 (also known as IRAS~17256--3631) that is located at a distance of 2~kpc, has been examined at multiwavelengths by us earlier in \citet[][hereafter Paper I]{2016MNRAS.456.2425V}. In the near-infrared deep images, we detected a star cluster that is partially embedded. The mid-infrared warm dust emission is diffuse and extended,  and possesses a sharp, arc-like edge towards the north. Several filamentary structures are also discerned in this region. The radio spectral index maps point towards the presence of thermal and non-thermal emission in the region. The zero age main sequence (ZAMS) spectral type of the ionizing star estimated from the radio emission is O7--O7.5. We also used simple analytic calculations in conjunction with the morphology of dust and gas emission to show that the cometary morphology is better explained with the champagne flow model. G351.63--1.25, on the other hand, is a bipolar \hii~region located at a distance of 2.4~kpc \citep[][hereafter Paper II]{2014MNRAS.440.3078V}. G351.63--1.25 harbors an infrared cluster embedded in fan-shaped nebulous emission. Several filamentary structures are visible in the warm dust emission. Earlier studies also reported the presence of a cold dust clump at 1.2~mm in this region \citep{2004A&A...426...97F}. The morphology of ionized gas emission is similar to that of warm dust emission with radio emission comprising six high density regions encompassed in diffuse emission spanning a region 1.5$\times$1.0~pc$^2$. The ZAMS spectral type of the bright radio core is estimated as O7.5. The bipolar nature is explained as arising due to density gradients within a flat molecular cloud. 

\par In the current paper, we present the 172$\alpha$ RRL observations of these two \hii~regions, carried out with the Giant Metrewave Radio Telescope (GMRT). We use the RRLs to investigate the large scale motion of ionized gas. Complementary molecular line observations from MALT90 survey towards G351.69--1.15 has also been used to analyze the motion of the ambient molecular medium. Towards G351.63--1.25,  far-infrared dust emission observed by the ATLASGAL has been used to infer about the distribution of molecular cloud. These sensitive observations provide a vital clue towards improving our understanding of the morphology of the \hii~regions. The organization of the paper is as follows. The details of observations and data reduction are given in Section 2. We discuss the results of continuum emission as well as RRL emission in Sections 3 and 4, respectively. The results of the molecular line analysis towards G351.69--1.15 is presented and discussed in Section 5. The proposed kinematic models of G351.69--1.15 and G351.63--1.25 are discussed in Section 6. Finally in Section 7, we present our conclusions.

\section{OBSERVATIONS AND DATA REDUCTION}
\subsection{Radio Recombination Line Observations}
The 172$\alpha$ radio recombination lines towards G351.69--1.15 and G351.63--1.25 are probed using the L-band (1280~MHz) spectral mode of GMRT \citep{1991CuSc...60...95S}. The GMRT comprises of 30 antennas each of diameter 45~m, arranged in a Y-shaped configuration. Twelve antennas are randomly placed in a central square of area 1~km$^2$. The remaining 18 antennas are stretched out along three arms of length $\sim14$~km each. The minimum and maximum baselines are 105~m and $\sim25$~km respectively. The angular extent of the largest structure observable with GMRT is $\sim$7$\arcmin$ at 1280~MHz.

\par The observed field was centered at G351.69--1.15 ($\rm{\alpha_{J2000}}$: 17$^h$29$^m$01.1$^s$, $\rm{\delta_{J2000}}$: -36$\degr$33$\arcmin$38.0$\arcsec$). The second \hii~region G351.63--1.25 is located within the field-of-view of the primary beam ($\sim$20$\arcmin$). The rest frequency of H172$\alpha$ line is 1281.175~MHz and C172$\alpha$ is 1281.815~MHz. We selected the central frequency as 1281.69~MHz and a total bandwidth of 2~MHz, divided into 512 channels. The observable frequency was determined keeping in view 
the LSR velocity of $\sim-11$~km/s of these regions \citep{2006ApJS..165..338Q} as well as the motions of Earth and Sun. These settings correspond to a spectral resolution of 4.06~kHz (or velocity resolution of 0.95~km/s). 3C286 and 3C48 were used as the primary flux calibrators while 1626-298 and 1830-360 were used as phase calibrators. The latter was also used for bandpass calibration.  As our targets lie in the southern sky and are available for only $\sim$7 hours per day above the elevation limit of GMRT, the targets were observed for 5 days (28 February, 01-03 March and 21 July, 2015). The total on-source integration time was 23.2~hrs. The calibration and flagging of each day's data was carried out separately before co-adding the UV data.

\par The data reduction was carried out using the NRAO Astronomical Image Processing System (AIPS).  Each data set was carefully checked for radio frequency interference and poor quality (due to non-working antennas, bad baselines, etc.) using the tasks {\tt UVPLT} and {\tt VPLOT}. Subsequent editing was carried out using the tasks {\tt TVFLG}, {\tt UVFLG} and {\tt SPFLG}. The final calibrated data set was cleaned and deconvolved using the task {\tt IMAGR} to produce a continuum map. Next, we subtracted the continuum (created from line-free channels) using the task {\tt UVLIN}. Of the 5 data sets taken on different days, one was found to be very noisy. We  excluded this dataset from further analysis (leading to a total on-source integration time of 18.8~hrs). The remaining four data sets were combined together using the task {\tt DBCON} and a line map was generated using the task {\tt IMAGR}.  In order to improve the signal-to-noise ratio (SNR), we produced  a lower resolution map by selecting {\tt UVTAPER} of 10~k$\lambda$ in {\tt IMAGR}. The primary beam correction was applied and the image was rescaled by a factor of 1.3 to correct for the contribution of Galactic plane emission to the system temperature (see Paper~I for details). The final resolution of the spectral cube is  27.4$\arcsec\times$18.2$\arcsec$ while that of the continuum map is 6.7$\arcsec\times$4.5$\arcsec$.    

%%%%%%%%%%%%%%%%%%%% Fig3 %%%%%%%%%%%%%%%%%%%%%%%%%%%%%%%%%%%%%%%%%%%%%%%%%

\begin{figure*}
\hspace*{-0.55cm}
\centering
\includegraphics[scale=0.34]{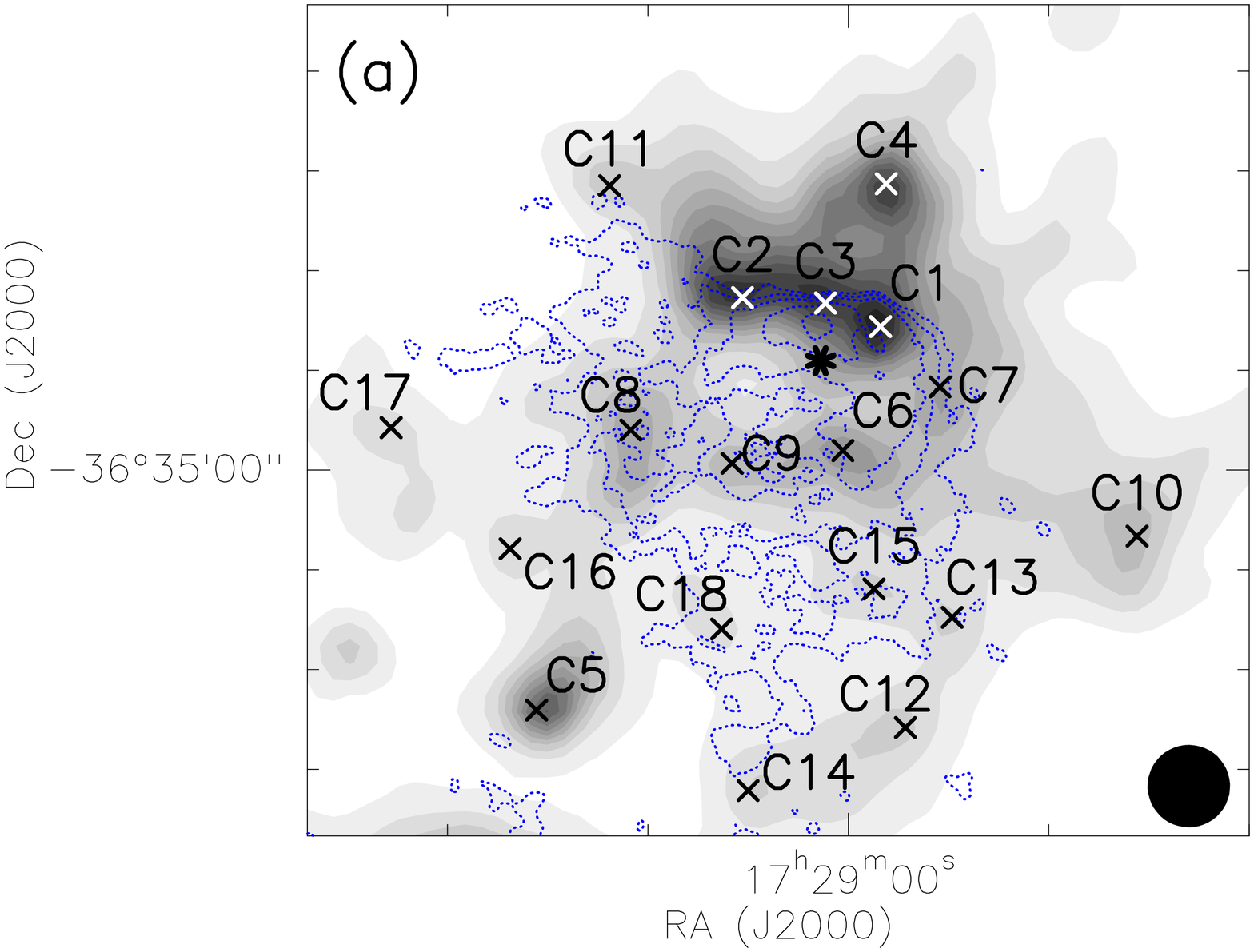}  \hspace*{-0.4cm} \quad  \includegraphics[scale=0.22]{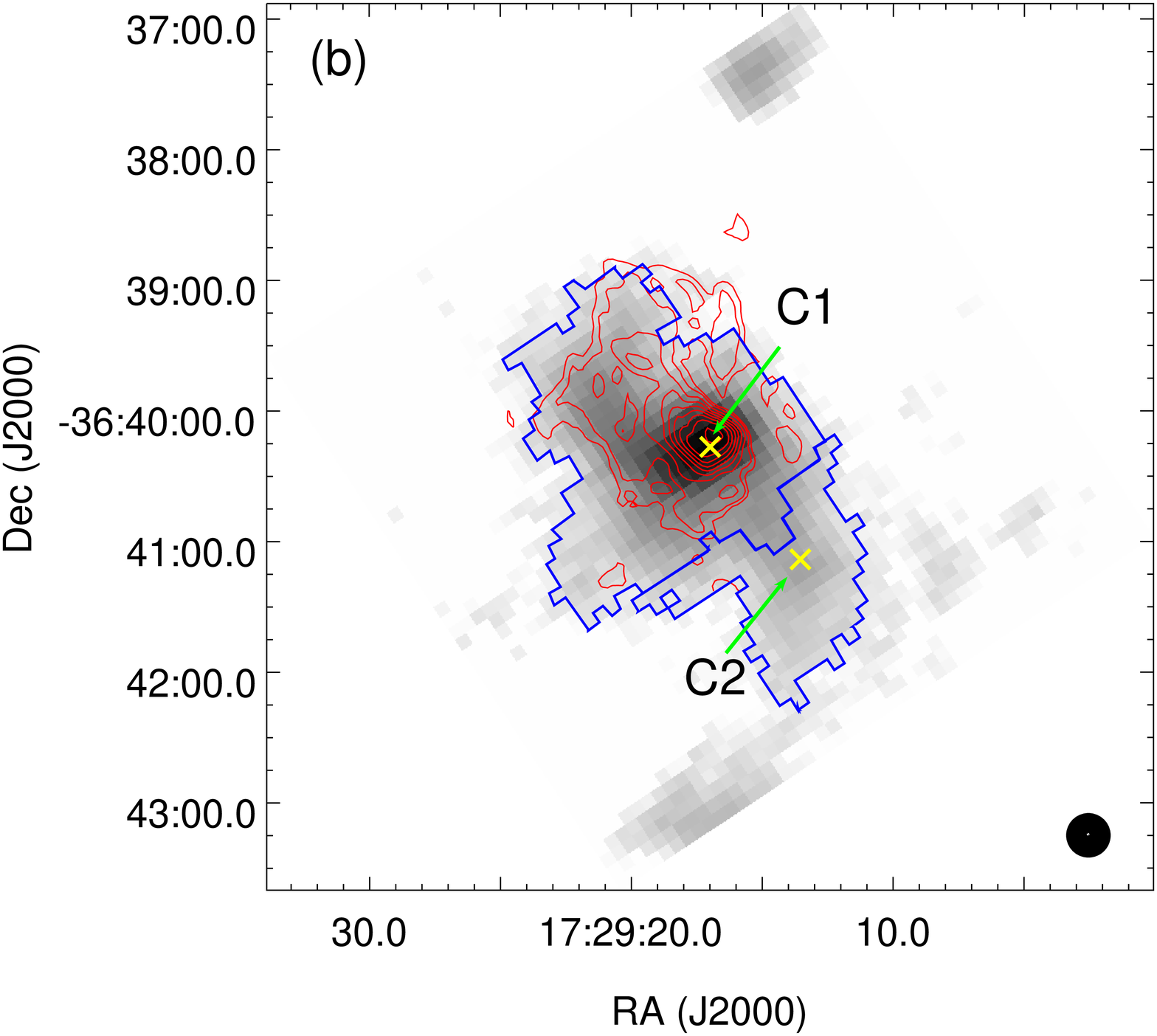} \caption{ (a) Cold dust emission from G351.69--1.15 at 350~$\mu$m overlaid with 1280~MHz radio contours. The peak positions of 18 clumps identified in Paper I are marked and labeled. The asterisk represents IRS-1 which is believed to be the ionizing source (Paper I). (b) Cold dust emission at 870~$\mu$m from G351.63--1.25 overlaid with 1280~MHz radio contours. The peak positions of the identified clumps are marked with cross points. The clump apertures from $\it{Clumpfind}$ algorithm are shown in blue. The contour levels are same as that of Fig.~\ref{continuum}. The respective beam sizes are given in bottom right of each panel.}
\label{350}
\end{figure*}

%%%%%%%%%%%%%%%%%%%%  %%%%%%%%%%%%%%%%%%%%%%%%%%%%%%%%%%%%%%%%%%%%%%%%%

\subsection{MALT90 Archival Data}
In addition to the radio observations, we use data from the Millimetre Astronomy Legacy Team 90~GHz Pilot Survey \citep[MALT90;][]{{2011ApJS..197...25F},{2013PASA...30...57J}} to discern the morphology and kinematics of the molecular gas. The \hii~region G351.69--1.15 is covered by the MALT90 Pilot Survey, while G351.63--1.25 is not. The MALT90 observations were carried out using the  8~GHz wide Mopra Spectrometer (MOPS). The data reduction was conducted by the MALT90 team using an automated reduction pipeline. The spectral and angular resolutions achieved are 0.11~km/s and 72$\arcsec$, respectively. The data products that include spectral cubes of frequencies around transitions of 16 molecular species were taken from the MALT90 website (http://malt90.bu.edu/pilot.html).  These molecular species offer an excellent combination of optically thick and thin tracers. The data analysis was carried out the using the GILDAS software (http://www.iram.fr/IRAMFR/GILDAS). 

\subsection{ATLASGAL Survey}
In order to obtain the distribution of gas around G351.63--1.25 whose MALT90 data is not available, we  used the submillimeter 870~$\mu$m data from ATLASGAL survey that samples the emission from cold dust in this region. Our motive is to examine the cold dust clumps in this region. The Apex Telescope Large Area Survey of the GALaxy (ATLASGAL) using the Large Apex Bolometer Camera (LABOCA) is a systematic survey of the Galaxy carried out at 870~$\mu$m to observe the continuum emission from interstellar dust at submillimeter wavelengths. The resolution and pixel size are 18.2$\arcsec$ and 6$\arcsec$ respectively \citep{2009A&A...504..415S}. The images available from the archive are tiles of size 5$\arcmin\times$5$\arcmin$.

\section{CONTINUUM EMISSION}

In this section, we discuss the continuum emission arising from ionized gas at radio wavelengths and from dust at far-infrared wavelengths. The latter traces the expanse of the associated molecular cloud. The relative position of the two \hii~regions is illustrated in the low frequency ionized gas emission at 325~MHz in Fig.~\ref{325}, taken from Papers I and II. The large scale diffuse emission encompasses the \hii~regions. Considering that the kinematic distances of these regions are similar, it is possible that both of them are associated with the same giant molecular cloud.

\subsection{G351.69--1.15}
\par The radio continuum emission at 1280~MHz towards G351.69--1.15 is presented in Fig.~\ref{continuum}(a). The image displays a sharp, arc-like edge with a nebulous tail characteristic of cometary \hii~regions. The cometary axis is aligned in the NW-SE direction. Large scale diffuse emission that spans an area of 3$\times$3~pc$^2$ is perceived towards the southern region. We estimated the ZAMS spectral type of the ionizing source to be O7--O7.5 based on the radio flux density as described in Paper I. The distribution of the ionized gas with respect to cold dust emission is shown in Fig.~\ref{350}(a).  We identified 18 cold dust clumps in this region and using an extensive multiwavelength analysis established that they represent early stages in the star formation process (see Paper I). These clumps are marked in Fig.~\ref{350}(a).

%%%%%%%%%%%%%%%%%%%% Fig4 %%%%%%%%%%%%%%%%%%%%%%%%%%%%%%%%%%%%%%%%%%%%%%%%%

\begin{figure*}
\hspace*{-0.4cm}

\centering
\includegraphics[scale=0.4]{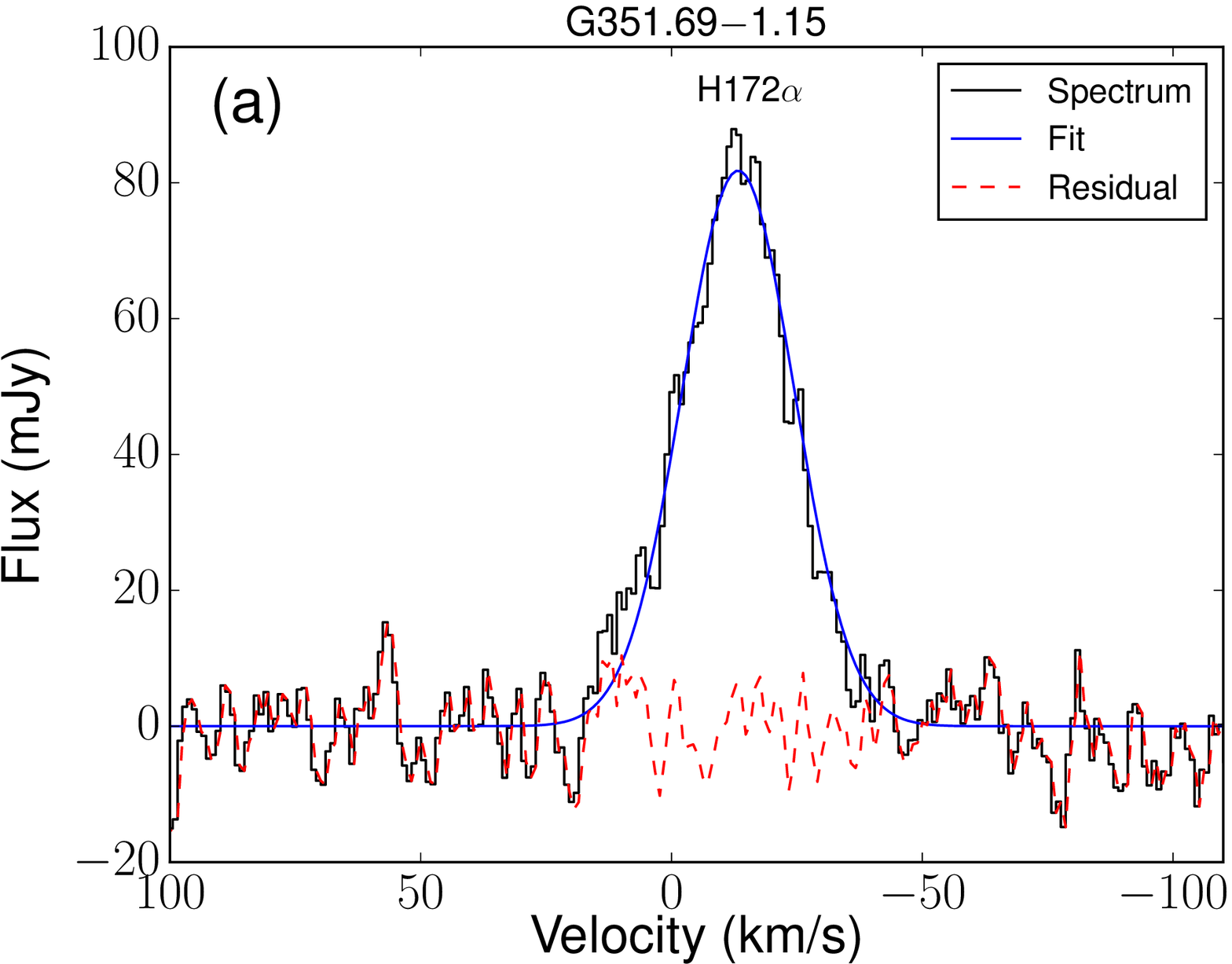} \quad \includegraphics[trim={0 6.5cm 0 6cm},clip, scale=0.4]{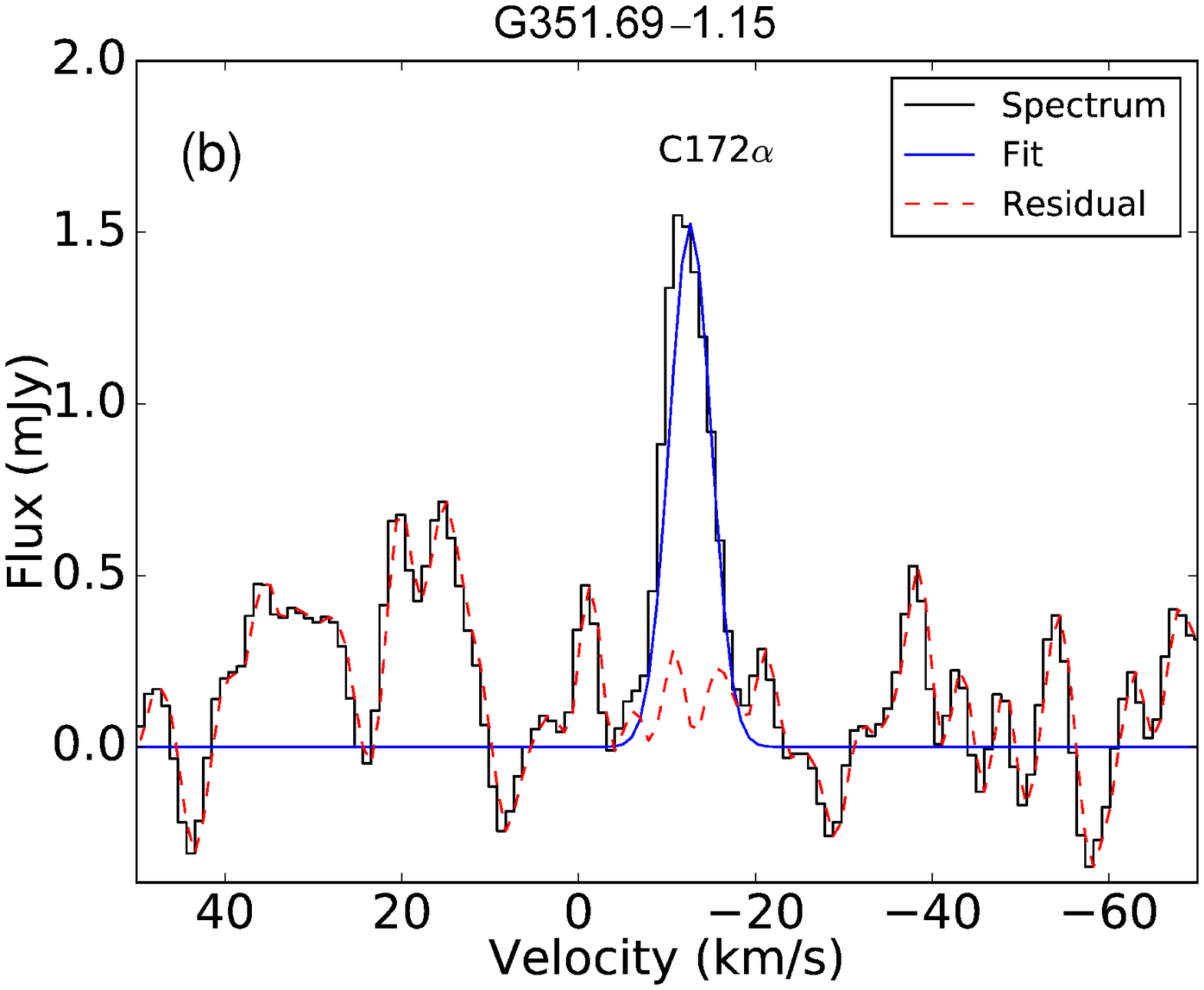} 
\caption{Spectra of hydrogen and carbon RRLs towards G351.69--1.15. The spectra are boxcar smoothed by 3 channels that corresponds to a velocity smoothing of 2.9~km/s.}
\label{17256spec}
\end{figure*}

%%%%%%%%%%%%%%%%%%%% %%%%%%%%%%%%%%%%%%%%%%%%%%%%%%%%%%%%%%%%%%%%%%%%%

We have estimated the average electron density of this region from the radio continuum emission using the following expression \citep{2016A&A...588A.143S}.

\begin{equation}
\begin{split}
\left[\frac{n_e}{\textrm{cm}^{-3}}\right]=2.576\times10^6\,\left[\frac{F_\nu}{\textrm{Jy}}\right]^{0.5}\left[\frac{T_e}{\textrm{K}}\right]^{0.175}\left[\frac{\nu}{\textrm{GHz}}\right]^{0.05}\\ \quad \times \left[\frac{\theta_{source}}{\textrm{arcsec}}\right]^{-1.5}\left[\frac{D}{\textrm{pc}}\right]^{-0.5}
\end{split}
\label{ne}
\end{equation}

\noindent where $n_e$ is the electron density, $F_\nu$ is the flux density at frequency $\nu$, $T_e$ is the electron temperature,  $\theta_{source}$ is the source diameter and $D$ is the distance to the source. We have adopted the value of electron temperature as 7560~K based on the studies of \citet{2006ApJ...653.1226Q}. The flux density $F_\nu$ is estimated within the 3$\sigma$ level of radio emission. This corresponds to a region of area 24 arcmin$^2$ that gives $\theta_{source}\sim 5.5\arcmin$ assuming spherical symmetry. By application of the above expression, we obtain average electron density across the entire region $\bar{n}_e\sim152$~cm$^{-3}$. 

%%%%%%%%%%%%%%%%%%%% Fig 5 %%%%%%%%%%%%%%%%%%%%%%%%%%%%%%%%%%%%%%%%%%%%%%%%%

\begin{figure}
\centering
\includegraphics[scale=0.3,angle=90]{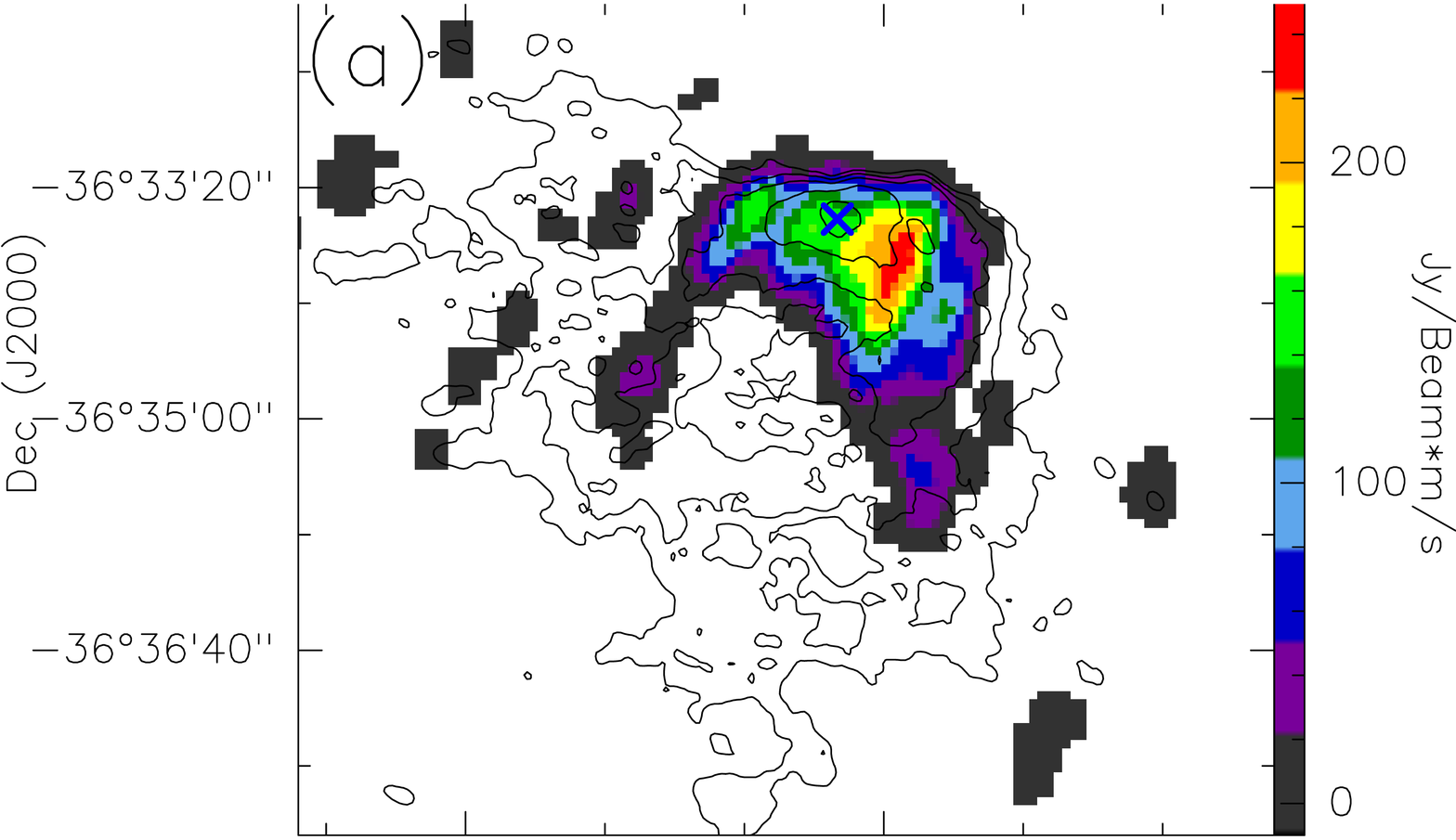}  

\vspace{0.05mm}

\quad  \includegraphics[scale=0.3,angle=90]{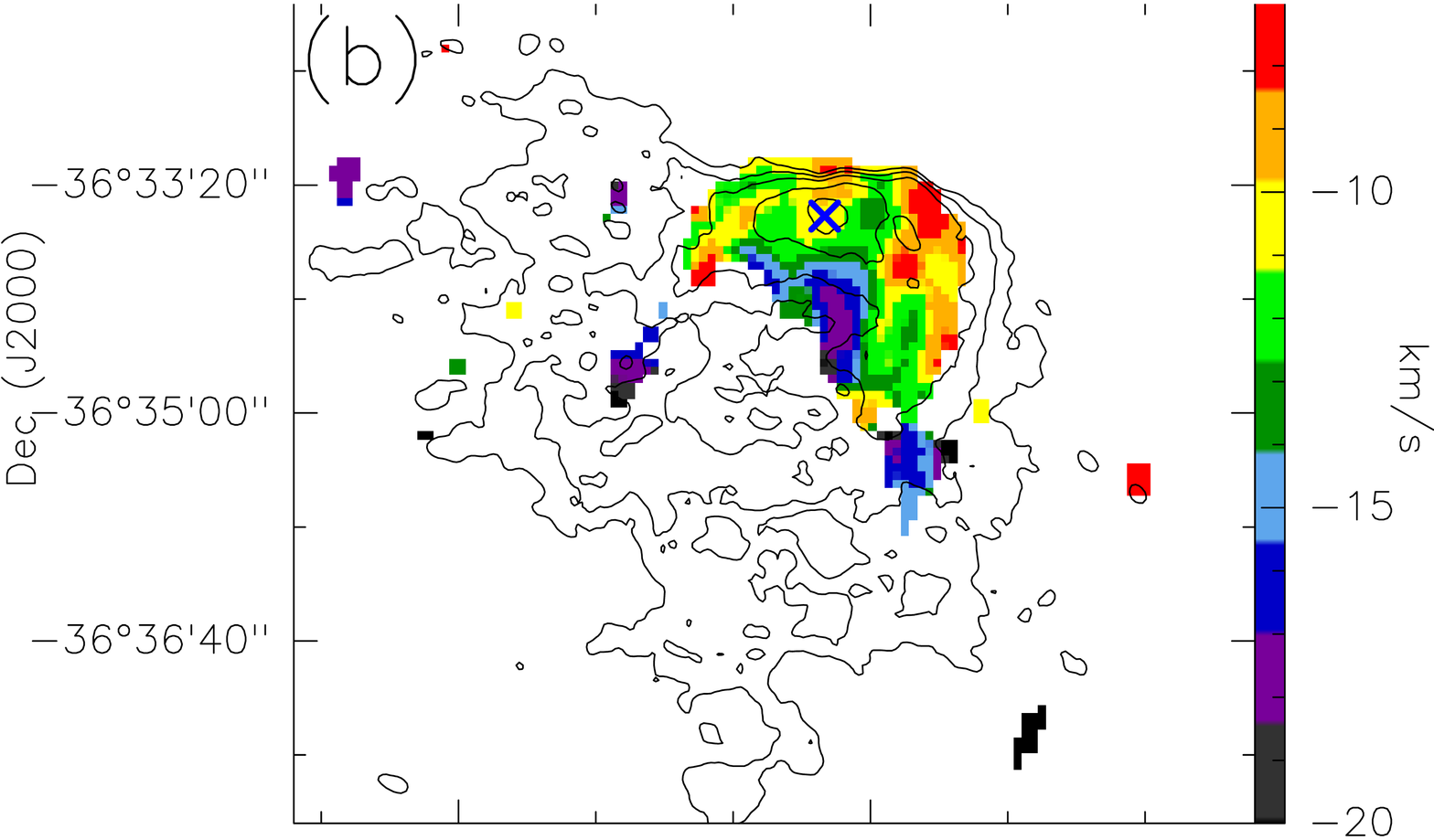} 

\vspace{0.05mm}

\quad \includegraphics[scale=0.3,angle=90]{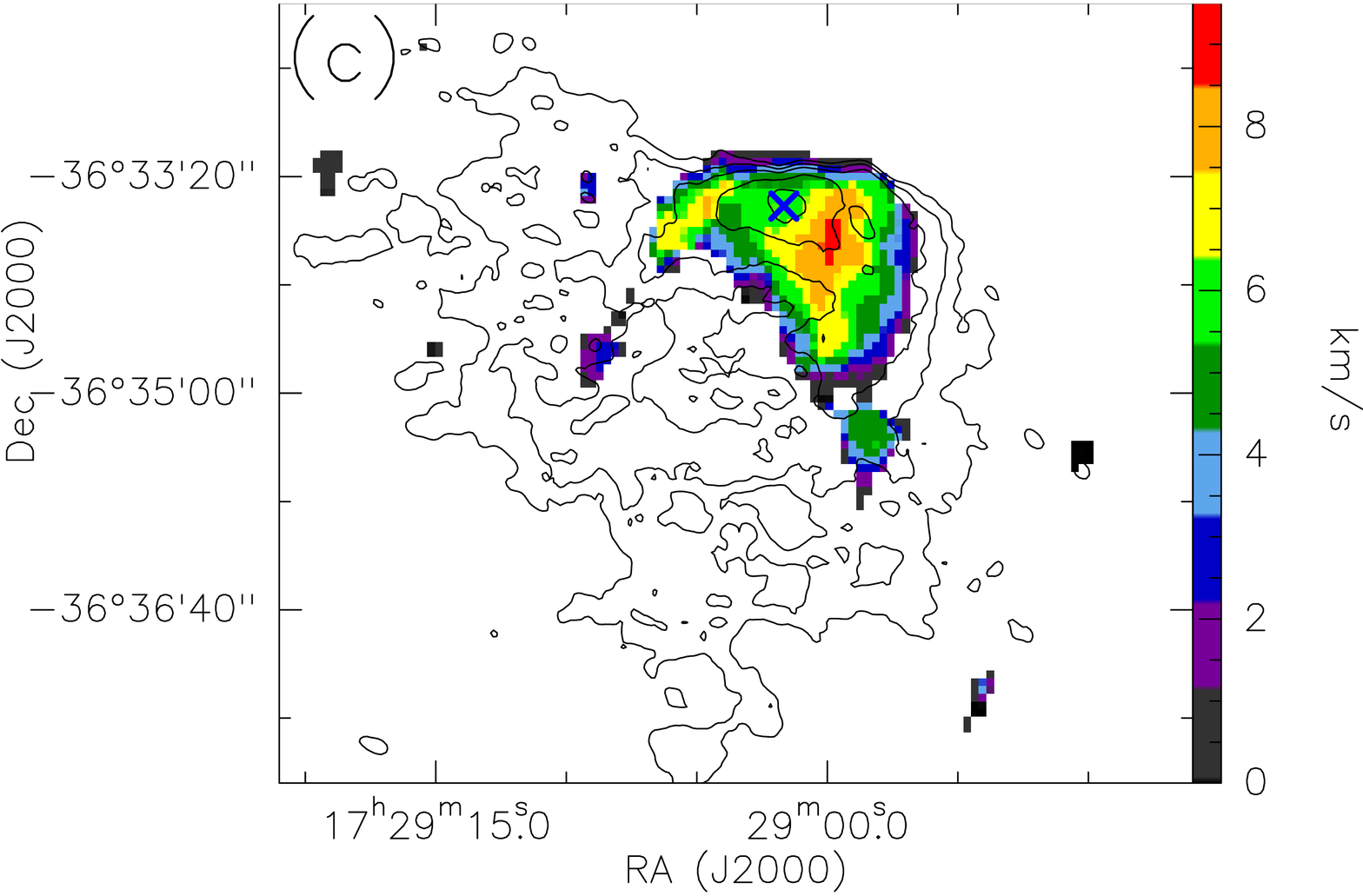} 
\caption{ (a) Integrated intensity (zeroth moment), (b) Central velocity (first moment), and (c) Velocity dispersion (second moment) of the hydrogen RRL towards G351.69--1.15. Contours represent the 1280~MHz continuum emission. The contour levels are the same as those shown in Fig.~\ref{continuum}(a). The radio continuum peak is marked with a cross in all the three panels.}
\label{17256mom}
\end{figure}

%%%%%%%%%%%%%%%%%%%%%%%%%%%%%%%%%%%%%%%%%%%%%%%%%%%%%%%%%%%%%%%%%%%%%
\subsection{G351.63--1.25}

\subsubsection{Radio continuum emission}

\par The 1280~MHz radio continuum map of this region is presented in Fig.~\ref{continuum}(b). The \hii~region is aligned in the NE-SW direction. We note the elongated region of ionized gas suggesting a bipolar morphology, consistent with that described in Paper II. The two brightest radio regions are designated as \textit{A} and \textit{B} in the present work, shown in Fig.~\ref{continuum}(b). The ZAMS spectral type assuming a single ionizing source is  O7.5 (Paper II). Using Eqn. (\ref{ne}), we estimated the average electron density from the radio continuum emission towards regions $A$ and $B$ distinctly. For an electron temperature of 6490~K \citep{2006ApJ...653.1226Q}, $\theta_{source}$ of 1.1$\arcmin$ and 1.6$\arcmin$, for $A$ and $B$ respectively, we estimate $\bar{n}_e(A) \sim$ 816 and $\bar{n}_e(B) \sim$ 326~cm$^{-3}$. 

\subsubsection{Dust continuum emission}

\par In order to investigate the cold dust emission from this region, we examined the sensitive 870~$\mu$m ATLASGAL map presented in Fig.~\ref{350}(b). From the image, we observe that the diffuse emission tracing the large scale molecular cloud extends up to $2.8\times2.0$~pc$^2$ in the NE-SW direction. We have used the 870~$\mu$m cold dust map to identify the dust clumps for two reasons: (i) the 870~$\mu$m map has the best possible resolution among dust emission maps available for this region: Herschel maps at 350 \& 500~$\mu$m, 870~$\mu$m ATLASGAL and 1.2~mm map from SEST-SIMBA \citep{2004A&A...426...97F}, and (ii) the emission is optically thin at this wavelength. The \textit{Herschel} maps of this region are shown in Appendix A. We identified two clumps in the cold dust map. This is confirmed by the application of the 2D-$\it{Clumpfind}$ algorithm \citep{1994ApJ...428..693W} to this image. We considered a threshold of 0.8~Jy/beam corresponding to 5$\sigma$ flux level and a step size of 5$\sigma$. The peak of the dust emission lies close to the radio peak. 
%The small scale structure of the core within the cloud is oriented NW-SE. 
The clump masses are estimated using the expression \citep{2006A&A...447..221B}

\begin{equation}
\textrm{M}_{clump} = \frac{g\,S_\nu\,d^2}{\kappa_\nu\,B_\nu(T_d)}
\label{cmass}
\end{equation}

\noindent where $S_\nu$ is the flux density at frequency $\nu$, $d$ is the distance to the source (2.4~kpc adopted from Paper II), $\kappa_\nu$ is the dust mass opacity coefficient, $g$ is the gas-to-dust ratio $\sim100$ and $B_\nu(T_d)$  is the Planck function for a blackbody at dust temperature $T_d$. We have considered a dust temperature of 42~K from the studies of \citet{2004A&A...426...97F}. The assumed dust opacity is calculated using the expression \citep{2010A&A...518L..92W}.

\begin{equation}
\kappa_\nu = 0.1(\nu/1000~\textrm{GHz})^\beta
\end{equation}

\noindent In the above expression, $\beta$ is the dust emissivity index, taken as 2 \citep{{2002MNRAS.329..257W},{2010A&A...518L..92W}}. Using this expression, we find $\kappa_\nu$ as 1.2~cm$^2$g$^{-1}$. Incorporating this in Eqn.~(\ref{cmass}), the clump masses are estimated as 2480 and 380~M$_\odot$ for clumps C1 and C2 respectively. This is nearly two times larger than the previous mass estimate of 1400~M$_\odot$ by \citet{2004A&A...426...97F} using 1.2~mm observations. This could be attributed to the larger clump aperture considered as a result of better sensitivity of the 870~$\mu$m map. In addition, we note that \citet{2004A&A...426...97F} have used $\kappa_\nu\sim1$~cm$^2$/g at 1.2~mm.

\section{RRL EMISSION}

In this section, we probe the RRL emission from the two \hii~regions and carry out an investigation of the distribution and observed velocity field of the ionized gas. We also assess the effect of line broadening mechanisms on the velocity profile of H172$\alpha$ RRL.

\subsection{Moment Maps}

In this subsection, we discuss the morphology and properties of the line emission towards each \hii~region.
\let\cleardoublepage\clearpage
\subsubsection{G351.69--1.15}

\par The radio spectrum towards the \hii~region G351.69--1.15 integrated over the entire emission region is presented in Fig.~\ref{17256spec}. We detect both hydrogen as well as carbon RRLs. We fitted Gaussian profiles to the spectra to  estimate the velocity and width of the RRLs. The LSR velocity of the hydrogen line is found to be $-13.2$~km/s and the line width is 26.3 km/s, also listed in Table.~\ref{tb1}. The spatial distribution and velocity structure of RRLs can be studied by generating the moment maps. We have created the moment maps using the task {\tt MOMNT} in AIPS software. The zeroth, first and second moments correspond to the integrated intensity, central velocity and velocity dispersion $\sigma_v$, respectively. To generate these moment maps, the line emission has been integrated within the velocities: $+17$ to $-44$~km/s. The zeroth moment map of the Hydrogen RRL towards G351.69--1.15 is shown in Fig.~\ref{17256mom}(a). We note that the line emission peaks westward of the radio continuum peak. This is attributed to optical depth effects near the continuum peak. From the first moment map shown in Fig.~\ref{17256mom}(b), we see that there appears a velocity gradient in this region, $-8$ to $-20$~km/s. This is analyzed and discussed later in Section 4.3. The second moment map displaying the dispersion in velocity is shown in Fig.~\ref{17256mom}(c). The zeroth and second moment maps are quite similar in appearance where the maximum dispersion is observed towards peak emission of the integrated intensity map. The line emission is not seen in the outer envelope of this \hii~region owing to poor SNR.

\par Apart from the hydrogen RRL, we also detected emission from the carbon RRL. Carbon RRLs are believed to  originate from  photodissociation regions \citep[PDRs;][]{1978ARA&A..16..445B}. The LSR velocity of the carbon line in this region is found to be $-12.6$~km/s while the width of the line is 5.4~km/s. The carbon line has a poorer SNR. Hence, we solely present the spectrum of C172$\alpha$ line from this region, shown in Fig~\ref{17256spec}(b).

\citet{2006ApJS..165..338Q} detected H92$\alpha$ and C92$\alpha$ RRL emission towards this \hii~region. The LSR velocity and line width are $-12.0$~km/s and 23.6~km/s for H92$\alpha$ RRL and $-11.9$~km/s and 4.7~km/s for the C92$\alpha$ RRL. These values are consistent with values obtained by us (see Table~\ref{tb1}). Assuming local thermodynamic equilibrium and negligible pressure broadening effects for H172$\alpha$ RRL, we have calculated the electron temperature from the line to continuum ratio using Eqn.~(1) of \citet{2006ApJ...653.1226Q}. We obtained $T_e$ as 4038~K. This is lower than the $T_e$ value of 7960~K obtained using H92$\alpha$ RRL by \citet{2006ApJ...653.1226Q}. Our estimate can be considered as a lower limit as the emission is assumed to be optically thick towards the radio continuum peak.
%%%%%%%%%%%%%%%%%%%% Fig6 %%%%%%%%%%%%%%%%%%%%%%%%%%%%%%%%%%%%%%%%%%%%%%%%%

\begin{figure*}
\hspace*{-0.9cm}
\centering
\includegraphics[scale=0.32]{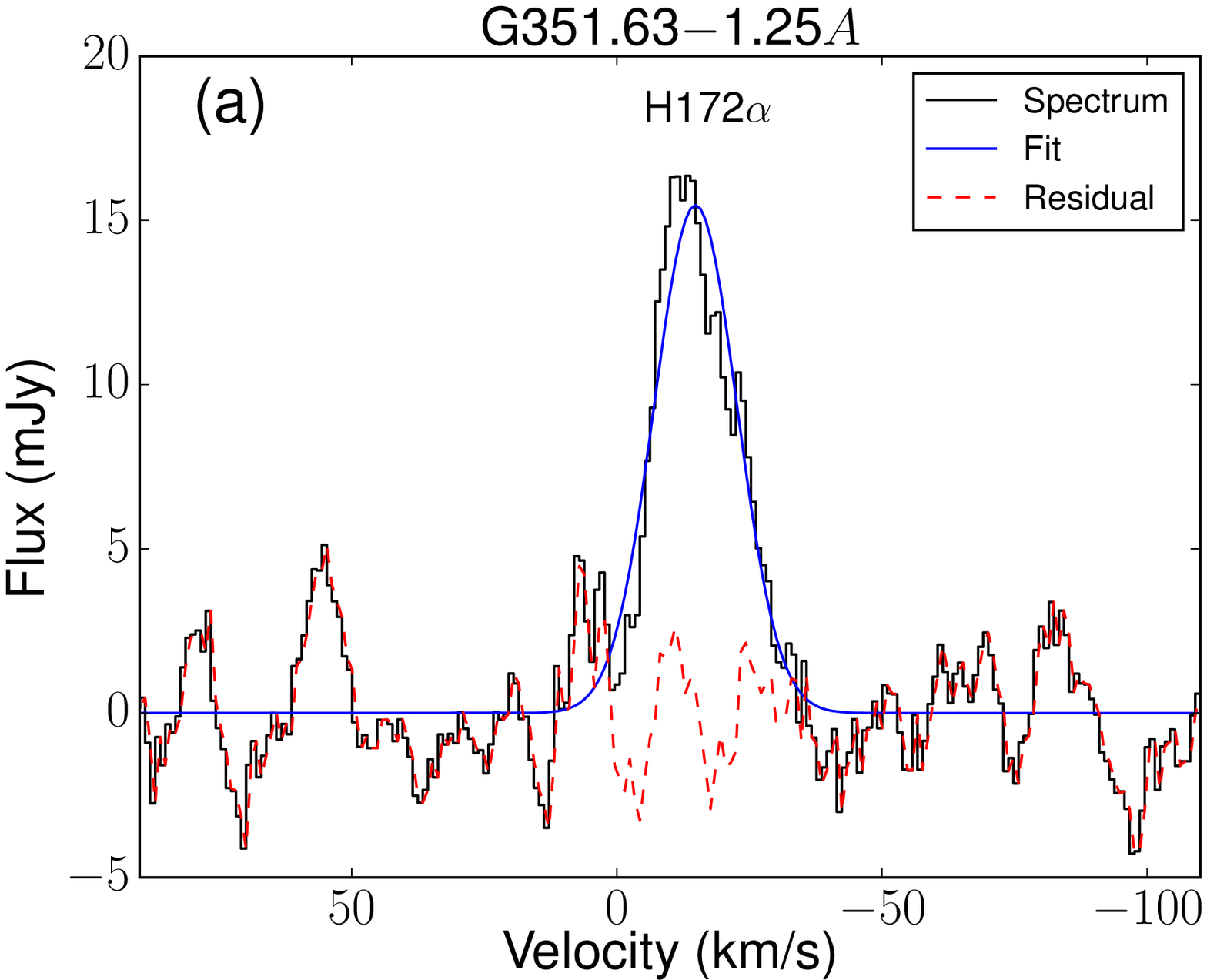} \hspace*{-0.9cm}\quad \includegraphics[scale=0.32]{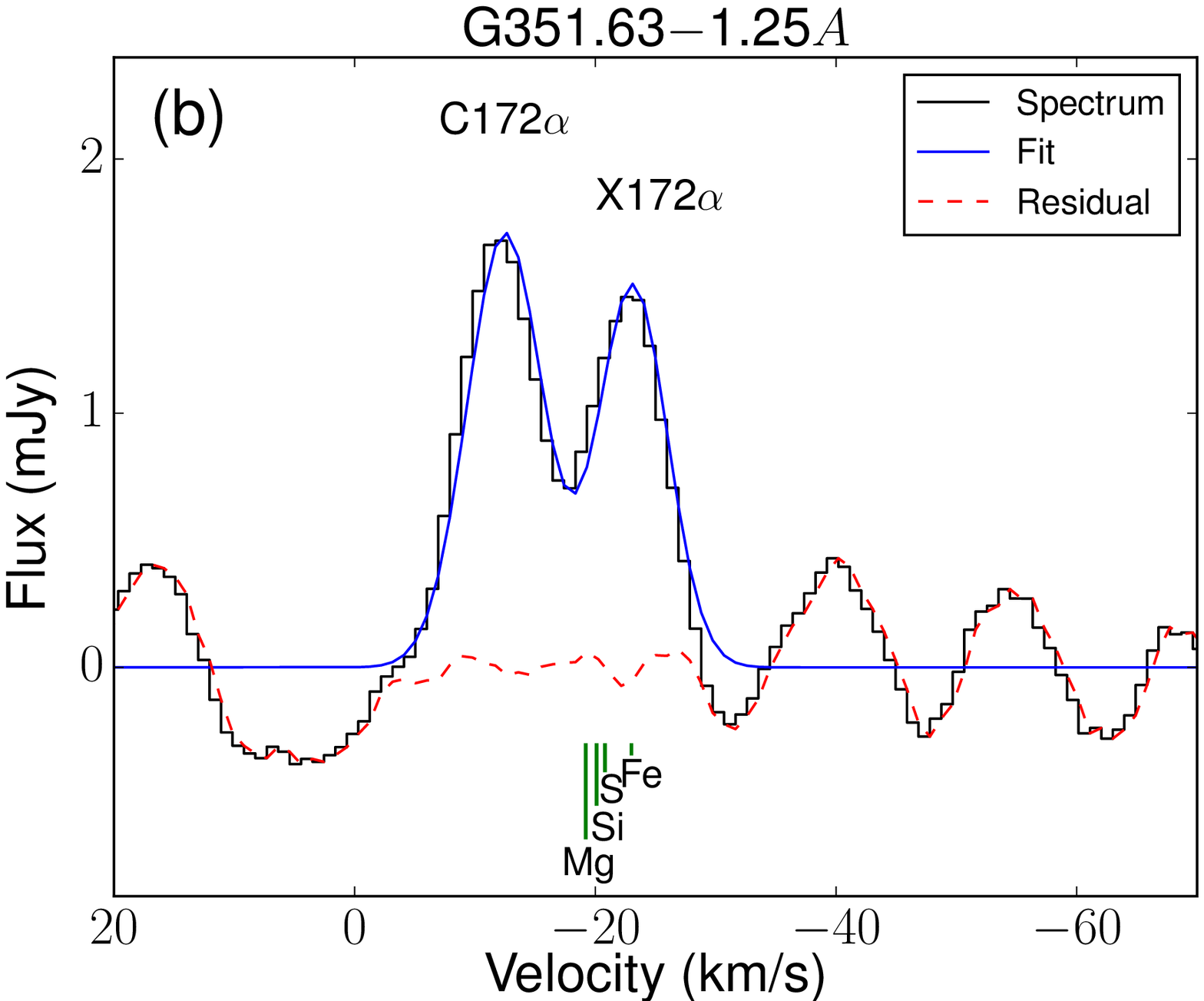} \hspace*{-0.9cm}\quad \includegraphics[scale=0.32]{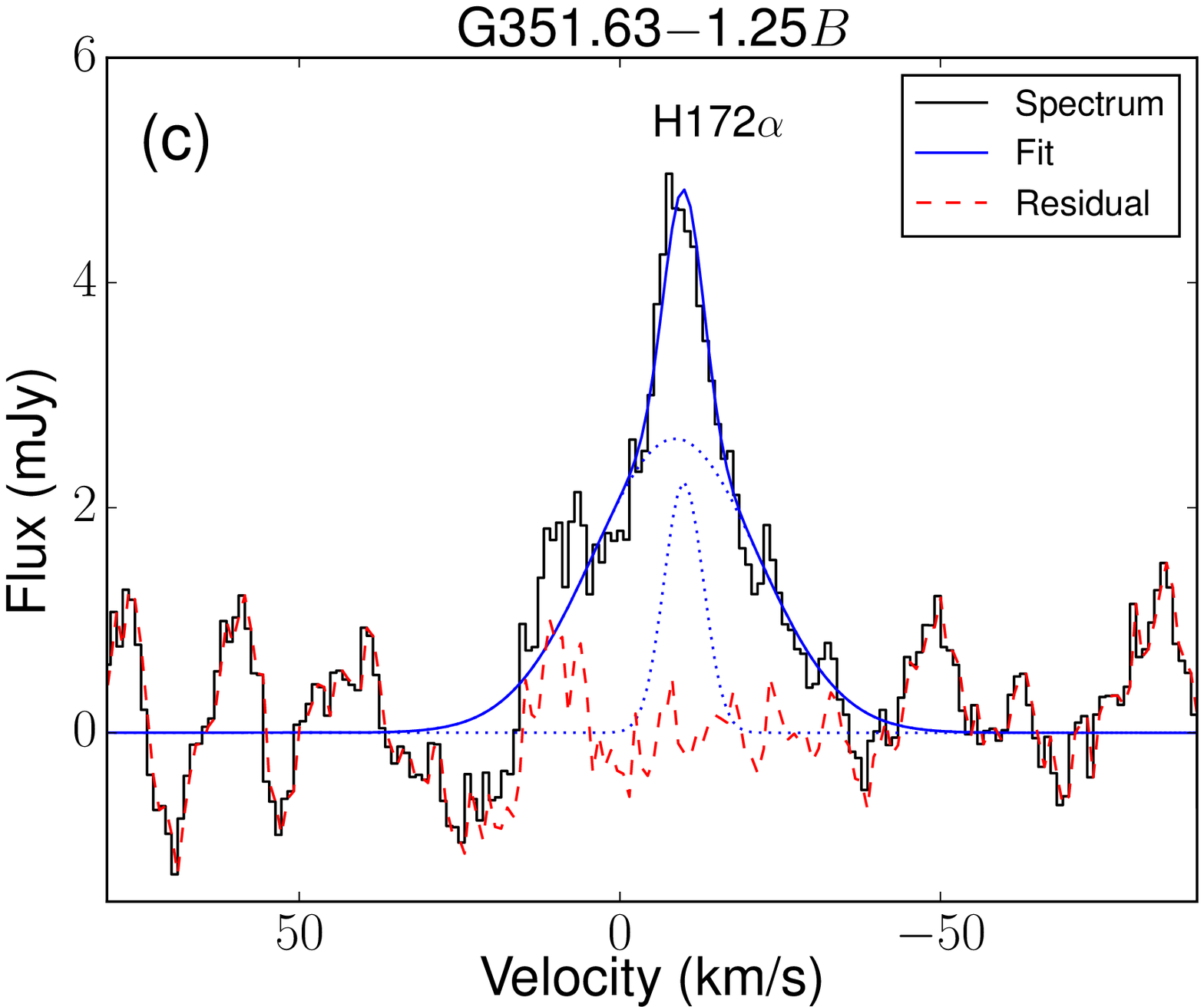}
\caption{Spectra of hydrogen and carbon RRLs toward G351.63--1.25\textit{A} and G351.63--1.25\textit{B}. The spectra are boxcar smoothed to 7 channels that corresponds to a velocity resolution of 6.3~km/s. The expected velocities of Mg, Si, S and Fe RRLs are also marked in panel (b). The estimation of velocities are under the assumption that these lines possess same LSR velocities as that of the carbon RRL.}
\label{17258spec}
\end{figure*}

%%%%%%%%%%%%%%%%%%%% %%%%%%%%%%%%%%%%%%%%%%%%%%%%%%%%%%%%%%%%%%%%%%%%%
\subsubsection{G351.63--1.25}
The RRL emission towards G351.63--1.25 is extracted for regions \textit{A} and \textit{B} separately. Our intention was to scrutinize for  differences in the spectra towards the two regions. Towards G351.63--1.25\textit{A}, we detected both hydrogen and carbon RRLs. The spectra towards the entire G351.63--1.25\textit{A} region are shown in Fig.~\ref{17258spec}(a) and (b). The LSR velocity of the hydrogen line is $-14.8$~km/s whereas it is $-12.4$~km/s for the carbon line (see Table.~\ref{tb1}). The line widths $\Delta$V (full width at half maximum: FWHM) are 18.5 and 6.4~km/s, respectively. There is a difference in the velocities of the two lines, of $\sim2.4$~km/s. We speculate on this difference in Section 4.4.2. 

Apart from the carbon line, we also detect another spectral emission feature blue shifted by $10.8$~km/s with respect to the carbon line, visible in the (b) panel of Fig.~\ref{17258spec}. The width of this line is 6.2~km/s, which is nearly the same as that of the carbon line. This line could be a RRL feature from an element heavier than carbon or alternately, it could be another doppler shifted RRL. A similar feature is detected in a few other star forming regions and is mostly believed to be a RRL feature originating from the sulfur atom \citep{{1972ApJ...173L.131C}, {1977A&A....57..341P}, {2005ApJ...626..253R}}. RRL of other heavier elements such as Si, Mg and Fe have also been detected in some \hii~regions \citep[e.g.,][]{1984ApJ...278..604S}. Assuming an LSR velocity of $-12.4$~km/s, similar to that of the carbon RRL, we estimated the approximate positions of these lines in the spectrum and marked them in Fig.~\ref{17258spec}(b). 
 The line strength of a single, heavy element RRL is expected to be $\sim1/3$ that of the carbon RRL \citep{1980ASSL...80..111P}. But in our case, this line has nearly the same intensity as that of carbon line. We, therefore, speculate that the origin of the second line is an admixture of multiple elements such as Mg, Si, S and Fe.

%%%%%%%%%%%%%%%%%%%% Fig 7 %%%%%%%%%%%%%%%%%%%%%%%%%%%%%%%%%%%%%%%%%%%%%%%%%

\begin{figure}

\centering
\includegraphics[scale=0.28,angle=90]{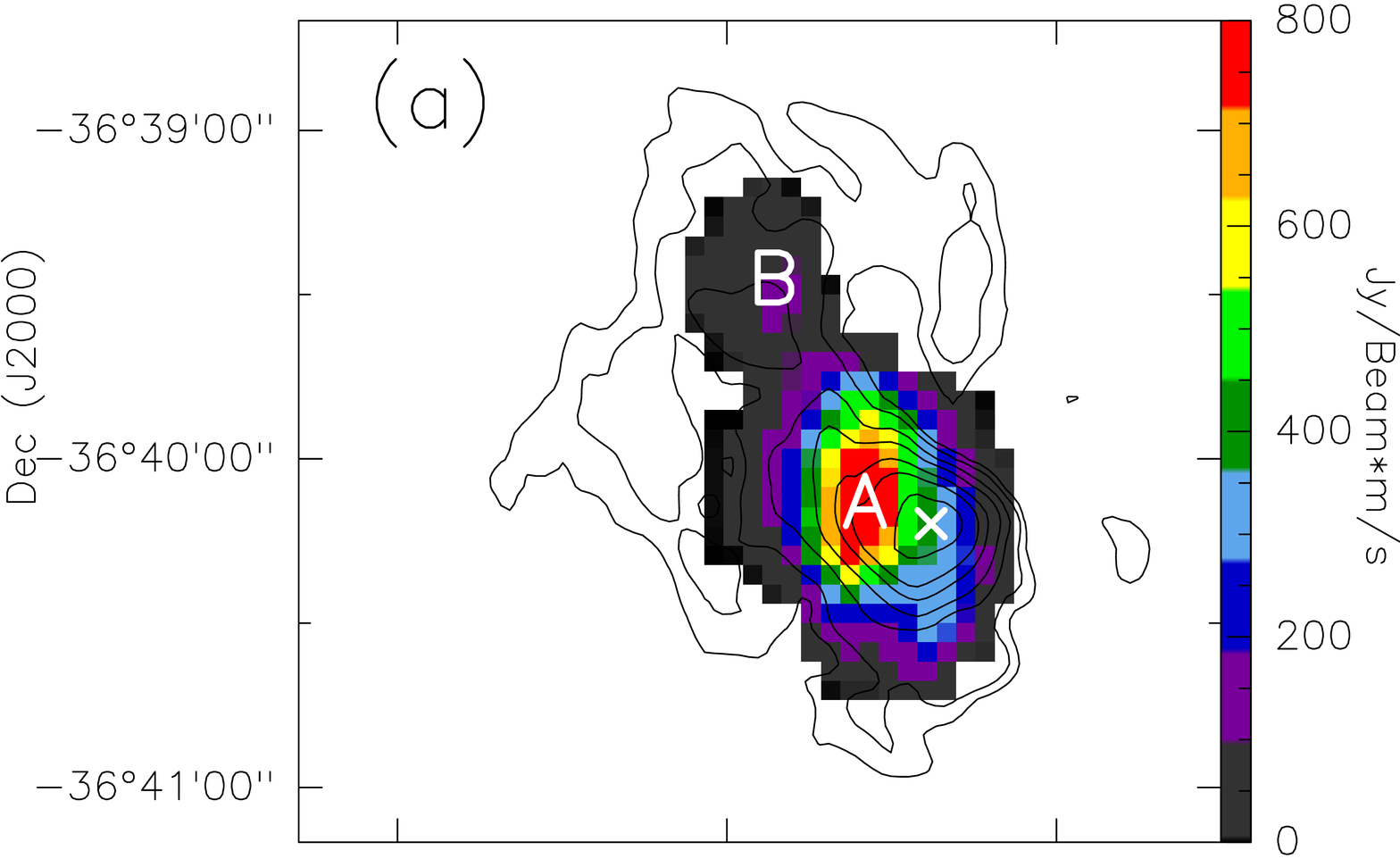} \quad \includegraphics[scale=0.28,angle=90]{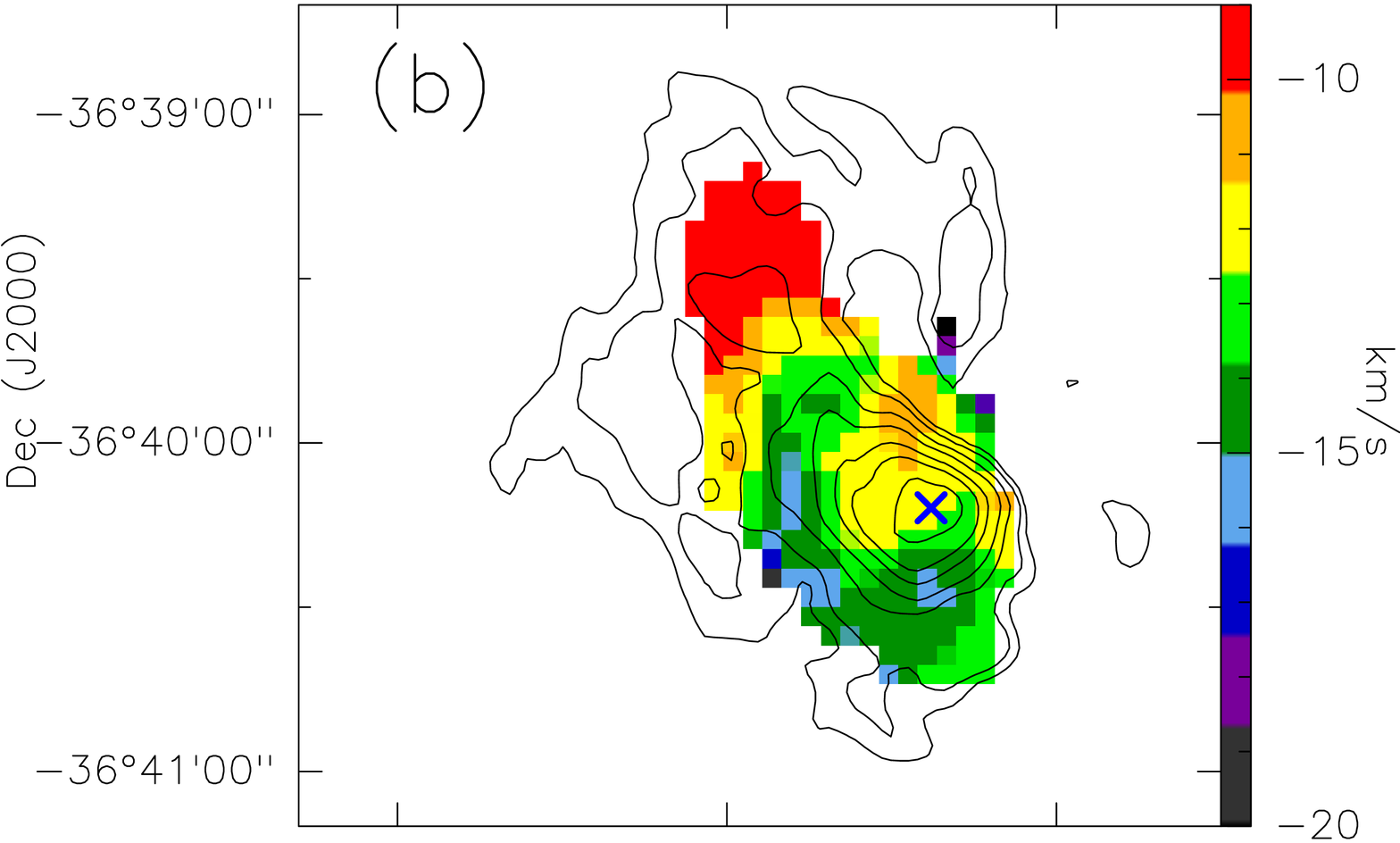} \quad \includegraphics[scale=0.28,angle=90]{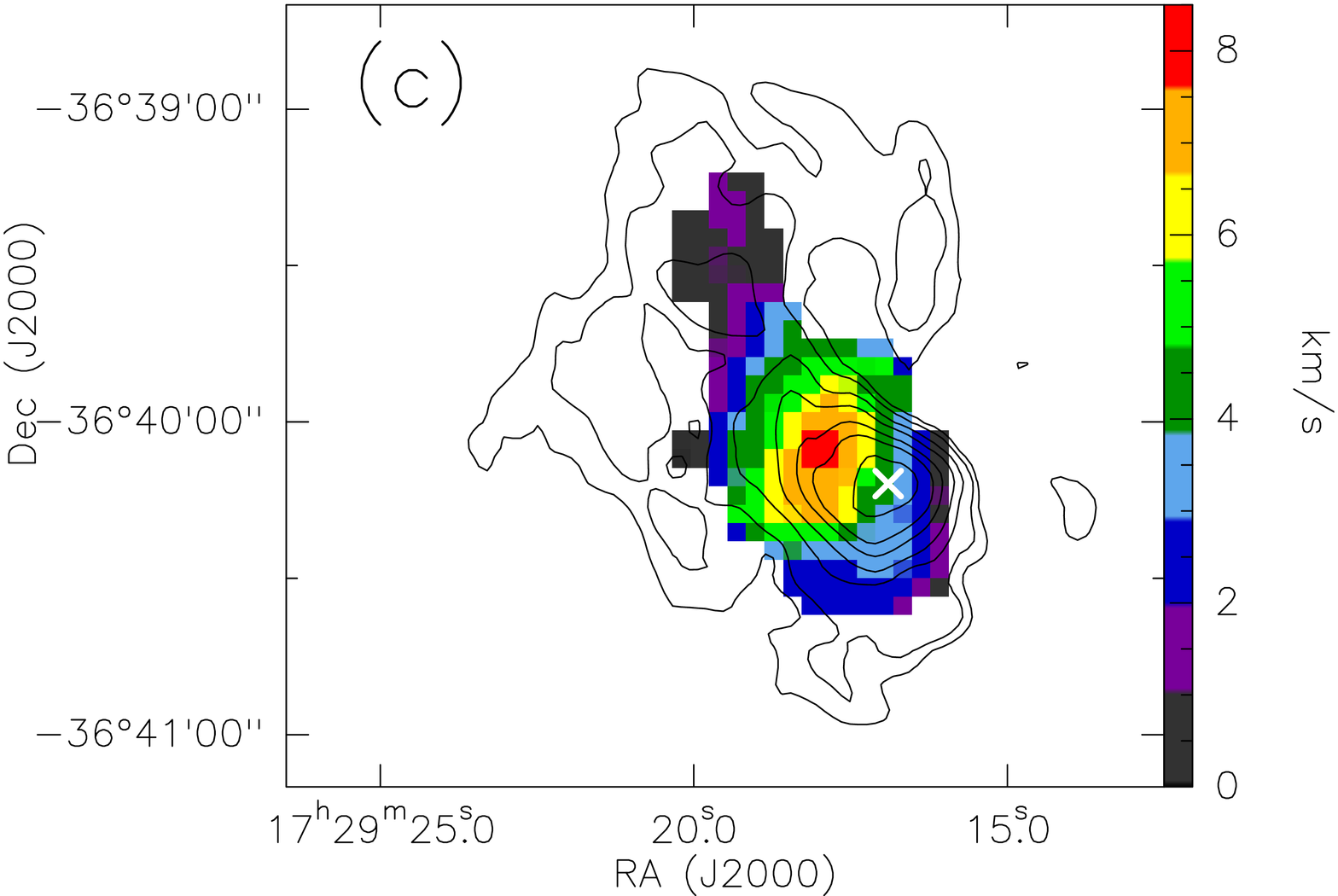} 
\caption{(a) Integrated intensity (zeroth moment), (b) Central velocity (first moment), and (c) Velocity dispersion (second moment) of the hydrogen RRL towards G351.63--1.25. The contours represent 1280~MHz continuum emission. The contour levels are the same as those shown in Fig.~\ref{continuum}(b). The two regions \textit{A} and \textit{B} are marked in panel (a). The radio continuum peak is marked with crosses in all the panels.}
\label{17258mom}
\end{figure}

%%%%%%%%%%%%%%%%%%%% %%%%%%%%%%%%%%%%%%%%%%%%%%%%%%%%%%%%%%%%%%%%%%%%%
\par The spectrum of G351.63--1.25\textit{B}, displayed in Fig.~\ref{17258spec}(c), shows the presence of hydrogen RRL from this region. The carbon RRL is not detected here. The non-detection of carbon RRL towards region \textit{B} could either be due to poor SNR or lack of significant emission from the PDR associated with this region. A comparison of the hydrogen RRLs from regions \textit{A} and \textit{B} shows differences in the spectra, both in terms of widths and central velocities. This hints at the possibility that \textit{A} and \textit{B} are two distinct regions with different properties. The hydrogen RRL towards region \textit{B} is broader than that from region \textit{A} and a single component Gaussian fit to the former is inadequate. We therefore, fit two Gaussian functions to the observed profile: one broad ($\Delta$V=30.1~km/s) and one narrow ($\Delta$V=7.2~km/s). The broad component is believed to arise from the \hii~region, whereas the narrow component component could be from the much cooler partially ionized medium (PIM) in the vicinity of the \hii~region \citep{{1977A&A....57..341P},{1998ApJ...504..375K}}. The central velocity of these lines (broad: $-8.7$ and narrow:$-9.9$~km/s) are relatively redshifted with respect to \textit{A}. The LSR velocity and line width of H92$\alpha$ from \citet{2006ApJS..165..338Q} using single dish measurements are $-13.9$~km/s and 28.4~km/s respectively. The LSR velocity and line width obtained by them for the C92$\alpha$ RRL are $-10.0$~km/s 9.7~km/s, respectively. The carbon RRL line width obtained by them is larger than that of C172$\alpha$ by 3.3~km/s (Table~\ref{tb1}). This could be attributed to the sampling of emission from a larger region (beam~$\sim3.2'$). We also estimated the electron temperature of G351.63--1.25$A$ region using the formalism mentioned in Section 4.1.1. Our estimate of 5906~K is lower compared to estimate of \citet{2006ApJ...653.1226Q} which is 6490~K. This could be due to optical depth effects similar to what is observed towards G351.69--1.15.

\par The moment maps towards the region associated with G351.63--1.25 are shown in Fig.~\ref{17258mom}. The morphology is similar to that of continuum morphology  but the spatial extent of line emission is again lower compared to the continuum emission (due to low SNR). Similar to the case of G351.69--1.15, the peak of line emission towards G351.63--1.25 is shifted with respect to the continuum peak. This suggests that the emission is optically thick at the continuum peak. The velocity distribution is non-uniform in this region, $-7.0$ to $-16.3$~km/s. We also notice that the velocity towards \textit{B} is red-shifted with respect to \textit{A}. This is consistent with the velocities obtained from the Gaussian fit to RRL integrated over the entire region and tabulated in Table~\ref{tb1}. Within G351.63--1.25\textit{A}, we observe a gradual increase in velocity as we move from central peak towards the south-east. This could be due to expansion of the \hii~region. It is difficult to interpret and correlate the velocity structures of regions \textit{A} and \textit{B} owing to the complex nature of G351.63--1.25 but an attempt has been made to explain the observed velocity fields in terms of projection effects  later in Section 6.1.2. \\

%%%%%%%%%%%%%%%%%%%%%%%%%%%%%%%Table 1%%%%%%%%%%%%%%%%%%%%%%%%%%%%%%%%%%%%%%%%%%%%%%%%%%%%%%%%%%%%%%%%%%%%%%%%%%
\begin{table*}
\footnotesize
\setlength{\tabcolsep}{2.5pt}
\caption{Line parameters estimated for the integrated emission across the entire region. The errors represent 1-$\sigma$ error obtained from the fit.}
\begin{center}
\hspace*{-0.5cm}
\begin{tabular}{l l c c c c c} \hline \hline
 & & & &\\
Source            && V (km/s) &   $\Delta$V (km/s) &S$^\star_I$ (mJy)&S$^{\star\star}_C$ (Jy) &$T^*_{e,max}$ (K)\\
\hline
\multirow{2}{*}{G351.69--1.15} &Hydrogen& $-13.2\pm0.2$& 26.3$\pm$0.6 &85.5 &3.6&15246\\
&Carbon& $-12.6\pm0.3$& 5.4$\pm$0.7&1.65&0.2&7639\\
\hline
\multirow{3}{*}{G351.63--1.25 \textit{A}}&Hydrogen & $-14.8\pm0.9$ &18.5 $\pm$2.1&16.4&0.8 &7544\\
&Carbon& $-12.4\pm0.6$& 6.4$\pm$1.5&1.96&0.3&10728 \\
&X172$\alpha^\dagger$& $-14.6\pm0.7$& 6.2$\pm$1.6 &1.64&0.3&26833\\
\hline
\multirow{3}{*}{G351.63--1.25 \textit{B}}&Hydrogen & $-8.7\pm0.9^a$ &30.1 $\pm$2.8 &\multirow{2}{*}{4.9}&\multirow{2}{*}{0.4}&19967\\
&& $-9.9\pm0.5^b$& 7.2$\pm$1.6&&&1145\\
\hline
\multicolumn{6}{l}{%
\begin{minipage}{7.5cm}
\scriptsize{$^\star$ : RRL flux density \\ $^{\star\star}$ : Radio continuum flux density \\ $^*$ : Upper limit assuming line width is primarily due to thermal broadening \\ $^\dagger$ : LSR velocity assuming X to be sulphur\\ $^a$ : Broad component\\ $^b$ : Narrow component}  
\end{minipage}%
}\\
\end{tabular}
\label{tb1}
\end{center}
              
\end{table*}
%%%%%%%%%%%%%%%%%%%%%%%%%%%%%%%%%%%%%%%%%%%%%%%%%%%%%%%%%%%%%%%%%%%%%%%%%%%%%%%%%%%%%%%%%%%%%%%%%%%%%%%%%%%%%%%%% 

\subsection{Line broadening in H172$\alpha$}
In this section, we attempt to estimate the effect of line broadening mechanisms on the observed H172$\alpha$ RRL towards both the \hii~regions. Apart from natural broadening, three important mechanisms are responsible for the broadening of radio recombination lines  \citep{2008ApJ...672..423K}: (i) Thermal/microturbulence broadening arising from the thermal motion of the gas particles and packets of gas that are much smaller than the beam, (ii) Dynamical broadening due to the large scale gas motions and (iii) Pressure broadening originating from the high electron densities in the region. We investigate the impact of each of these effects on the observed RRL and obtain an estimate of the electron density values.

The expression for thermal broadening $\Delta\nu_{t}$ depends on the line frequency $\nu_0$ and temperature $T_e$ as follows

\begin{equation}
\Delta\nu_{t}= \sqrt{\frac{8\,k\,T_e\,ln2}{m\,c^2}} \nu_0
\label{thermal}
\end{equation}

\noindent Here $\it{m}$ is the mass of the particle, $\it{k}$ is the Boltzmann constant and $\it{c}$ is the speed of light. Under the assumption that the line width is predominantly due to thermal broadening and the effects of dynamical and pressure broadening are negligible, we estimate the upper limits to the electron temperature, $T_{e,max}$ corresponding to the H172$\alpha$, C172$\alpha$ and X172$\alpha$ RRLs for both the \hii~regions. Substituting $\Delta\nu_{t}$=$\Delta$V in Eqn.~(\ref{thermal}), the upper limits are calculated. These values are given in Column 6 of Table~\ref{tb1}. We notice that the upper limit values $T_{e,max}$ obtained are high except for the hydrogen RRLs towards G351.63--1.25$A$ and the narrow component towards G351.63--1.25$B$. For the latter, an upper limit of 1145~K implies that this line originates in the cooler region of the PIM. The typical electron temperatures of the \hii~regions are $\leq$10$^4$~K. This suggests that the higher values of upper limits obtained for the 172$\alpha$ RRLs except for H172$\alpha$ RRL in G351.6--1.3$A$ and narrow component in G351.6--1.3$B$ are direct evidences for the effects of dynamical and pressure broadening.

\par To estimate the effects of dynamical and pressure broadening, we consider the following equations. The thermal width of the RRL is a Gaussian profile. This combines with the dynamical broadening $\Delta\nu_d$  to give a Gaussian profile $\Delta\nu_\textrm{G}$ for the RRL line. The dynamical broadening is the result of large scale gas motions in the \hii~region.

\begin{equation}
\Delta\nu_\textrm{G}=\sqrt{\Delta\nu_t^2+\Delta\nu_d^2}
\label{dynamical}
\end{equation}

\par For RRLs  with large principal quantum numbers ($\it{N}$) arising from high density gas, pressure broadening due to electron impacts are believed to be a major source of line broadening \citep{1967ApJ...148..547G}. The line profile arising from the pressure broadening $\Delta\nu_\textrm{P}$ is a Lorentzian and it has higher intensity in the line wings compared to the Gaussian profile. The Lorentzian profile combines with the Gaussian to give a resultant Voigt profile $\Delta\nu_\textrm{V}$ of the following form \citep{2008ApJ...672..423K}. 

\begin{equation}
\Delta\nu_\textrm{V}=0.53\Delta\nu_\textrm{P}+\sqrt{\Delta\nu_\textrm{G}^2+(0.47\Delta\nu_\textrm{P})^2} 
\label{pressure}
\end{equation}

\par The electron density can be estimated from the ratio of pressure and thermal broadening. For $\alpha$ ($\Delta \mathrm{N}=1$) transitions, the expression for electron density is defined as follows \citep{1995ApJ...444..765K}

\begin{equation}
\frac{\Delta\nu_\textrm{P}}{\Delta\nu_t}=1.2\left(\frac{n_e}{10^5}\right)\left(\frac{\rm N}{92}\right)^7
\label{eden}
\end{equation}

\noindent where $n_e$ is the electron density and $\mathrm{N}$ is the principal quantum number. 

\subsubsection{G351.69--1.15}

A comparison of our estimates of the line parameters for H172$\alpha$ RRL with that of H92$\alpha$ RRL by \citet{2006ApJS..165..338Q} shows that the width ($\Delta$V) of H172$\alpha$ line is larger by $\sim$3~km/s. This could be attributed to the pressure broadening effects if we assume both these lines originate within the same volume of gas and hence suffer the same amount of dynamical broadening. However, such a comparison is not fully appropriate, as \citet{2006ApJS..165..338Q} sample a larger region owing to their larger beam size using single dish, 140 foot Green Bank Telescope observations. In addition, there could be differing opacity effects at the distinct frequencies considered.

\par To analyze the various line broadening mechanisms towards G351.69--1.15, we first consider the thermal broadening $\Delta\nu_{t}$. From the observations of H92$\alpha$ RRLs towards G351.69--1.15 by \citet{2006ApJ...653.1226Q}, the estimated electron temperature of this region is 7560~K. Using this temperature value and $\it{m}$ as the mass of hydrogen atom, i.e. 1.67$\times$10$^{-27}$~kg, we determine $\Delta\nu_{t}$ from Eqn.~(\ref{thermal}) as $\sim$18.6~km/s. To get an estimate of the microturbulence present in the region, we have followed the method prescribed by \citet{2008ApJ...681..350S}. The average molecular line FWHM in this region is $\sim$2.5~km/s \citep[from CO and CS line observations of][]{2005A&A...432..921F}. This is similar to the  widths of HCO$^+$ and H$^{13}$CO$^+$ lines discussed later. Considering that the line width due to thermal broadening in molecular lines is $<$0.5~km/s, the estimated FWHM can be considered as that arising mostly from the turbulence, unless there are large scale effects. Thus, the combined width due to thermal and turbulence broadening (added in quadrature) is 18.8~km/s. This implies that the effect of microturbulence is negligible in this region. We next estimate the effect of pressure broadening along one line-of-sight (i.e single pixel). We select the pixel with peak emission as it corresponds to the maximum SNR. The observed line width towards this line-of-sight as estimated from the second moment map is $\sim21.3$~km/s (FWHM = 2.35$\sigma_v$). Considering the microturbulence of 2.5~km/s and other gas motions to be negligible towards the single pixel, we substitute $\Delta\nu_\textrm{V}\sim21.3$~km/s and $\Delta\nu_\textrm{G}=18.8$~km/s in Eqn.~(\ref{pressure}) and estimate the pressure broadening term  $\Delta\nu_\textrm{P}\sim4.3$~km/s. We use this value of pressure broadening to estimate the dynamical broadening across the entire \hii~region. From the integrated profile towards the full region, we get $\Delta\nu_\textrm{V}$ = 26.3~km/s. Our estimate of pressure broadening from the pixel with the peak RRL emission can be considered as sort of large value as 
there are density gradients present within the ionized nebula. Using this upper limit and $\Delta\nu_\textrm{G}=18.8$~km/s, the dynamical broadening due to large scale gas motion is found to be $\Delta\nu_d\sim14.7$~km/s from Eqns.~(\ref{dynamical}) and (\ref{pressure}). This is a conservative estimate and the actual value is expected to be higher.

\par A comparison of the estimates obtained for the broadening effects shows that the magnitude of pressure broadening is nearly one-third that of dynamical broadening while the magnitude of dynamical broadening is comparable to that of thermal broadening. This is not surprising as we observe a velocity gradient in this region that confirms the presence of high speed gas motions. The reverse has also been observed, where the effect of pressure broadening is larger than that of dynamical broadening \citep[e.g., G01.4--0.4;][]{2008ApJ...672..423K}. However, these mostly correspond to hypercompact \hii~regions that are characterized by their small sizes and high electron densities (10$^5$-10$^6$~cm$^{-3}$) and pressure broadening due to particle collisions is the dominant mechanism in the broadening of RRL. 

\par We next endeavor to estimate the electron density using the values obtained earlier for $\Delta\nu_\textrm{P}$ and $\Delta\nu_t$. Substituting N=172 in Eqn.~(\ref{eden}), we obtain $\it{n_e}$ as $\sim$241~cm$^{-3}$ for $\Delta\nu_\textrm{P}$ of 4.3~km/s. We would like to emphasize that the pressure broadening and hence the electron density values are expected to be larger as the emission is believed to be optically thick towards the continuum peak. Therefore, the line emission detected is predominantly from the outer surface of the \hii~region. The densities towards the continuum peak, therefore, could be higher than that of average value obtained. 

\par All our calculations are based on a Gaussian fit to the observed profile. Cases have been reported where there is a direct observable detection of Voigt profile of the RRLs \citep[e.g.,][]{{1973Ap&SS..20..187S},{1998ApJ...506..758K},{2007ApJ...667..248F},{2010PASP..122..354V}}. In few sources, apart from pressure broadening, the Lorentzian profile has a contribution from radiation broadening due to the Galactic background as well as bright background sources \citep{1998ApJ...506..758K}. In these cases, the profile deviates from the Gaussian considerably and is detected observationally. In our case, as the magnitude of pressure broadening is considerably lower than the other broadening mechanisms, it is unlikely to produce an observable effect on our line profile.  
%%%%%%%%%%%%%%%%%%%%%%%%%%%%%%%%%%% Table 2 %%%%%%%%%%%%%%%%%%%%%%%%%%%%%%%%%%%%
\begin{table*}
\footnotesize
\caption{Properties of the individual \hii~regions from radio continuum and H172$\alpha$ RRL}
\begin{center}
\setlength{\tabcolsep}{3pt}
\hspace*{-0.5cm}
\begin{tabular}{l c c c c c c} \hline \hline
 &  \\
Source      &$\Delta\nu_{t}$ (km/s)&$\Delta\nu_{d}$ (km/s)&$\Delta\nu_\textrm{P}$ (km/s)&$n_e$ (cm$^{-3}$)&$\bar{n}_e$ (cm$^{-3}$)&$\Delta\nu_\textrm{P,min}$ (km/s)\\
\hline\\
G351.69--1.15 &18.6  &14.6&4.8&273&152&2.7\\
G351.63--1.25$A$&17.2 &3.3&1.8&110&816&13.4\\
G351.63--1.25$B$&... &...&...&...&326&5.4 \\
\hline
\multicolumn{7}{l}{%
\begin{minipage}{7.5cm}
\scriptsize{\scriptsize{$\Delta\nu_{t}$ : Thermal broadening \\ $\Delta\nu_{d}$ : Dynamical broadening \\ $\Delta\nu_\textrm{P}$ : Pressure broadening \\ $n_e$ : Local electron density estimated from H172$\alpha$ RRL \\ $\bar{n}_e$ : Average electron density from radio continuum emission \\ $\Delta\nu_\textrm{P,min}$ : Minimum pressure broadening estimated from average electron density}}  
\end{minipage}%
}\\ 
\end{tabular}
\label{RRLline}
\end{center}
\end{table*}

%%%%%%%%%%%%%%%%%%%%%%%%%%%%%%%%%%%  %%%%%%%%%%%%%%%%%%%%%%%%%%%%%%%%%%%%

\subsubsection{G351.63--1.25}

A comparison of line width of H172$\alpha$ with H92$\alpha$ is difficult in this region as the line width of the latter has contributions from both the regions $A$ and $B$. Hence, we cannot draw any conclusions about the line broadening mechanisms apriori. To estimate the effects of various broadening mechanisms and to determine the average electron density towards G351.63--1.25, 
we proceed with the analysis similar to that for G351.69--1.15 (Sect. 4.2.1). We carry out the analysis separately for the two distinct regions, \textit{A} and \textit{B}. We first estimate the magnitude of thermal broadening using a temperature of 6490~K estimated by \citet{2006ApJ...653.1226Q}.  These authors used the  the line-to-continuum ratio from the single dish H92$\alpha$ observations  (Half Power Beam Width $\sim3.2\arcmin$) to determine the temperature. With this resolution, they could not distinguish the two regions $A$ and $B$ separated by $0.8\arcmin$ implying that they used emission from the entire \hii~region to estimate the electron temperature. We proceed by using the same electron temperature for both these regions. However, the results are treated with caution as the local electron temperatures for the individual regions may vary from the value considered.
Using Eqn.~(\ref{thermal}), we estimate the value of thermal broadening as $\Delta\nu_t\sim17.2$~km/s. 

\par We first consider G351.63--1.25\textit{A}. From the second moment map, the velocity dispersion corresponding to the peak line emission is $\Delta\nu_\textrm{V}\sim18.2$~km/s. Considering that the effect of microturbulence in this region to be similar to G351.69--1.17 ($\sim$2.5~km/s), the combined line width due to thermal and microturbulence broadening is 17.4~km/s.  Using this, we estimate the upper limit on pressure broadening as $\Delta\nu_\textrm{P}\sim1.6$~km/s. Now if we consider the integrated profile from the entire \textit{A} region,  $\Delta\nu_\textrm{V}\sim18.5$~km/s. Using this, the estimated lower limit of dynamical broadening is $\Delta\nu_d\sim3.2$~km/s. The scenario here is different from that of G351.69--1.15. The thermal broadening is the dominant broadening mechanism in this region.  The pressure broadening is negligible while the effect of dynamical broadening is also lower. 
Also, we could not find very large velocity gradients in the first moment map towards \textit{A}. 
From the pressure and thermal broadening estimates, the lower limit to the electron density is 91~cm$^{-3}$. 

For the northern region G351.63--1.25\textit{B}, we do not attempt to interpret the line broadening effects owing to the complex profile of hydrogen RRL. In Section 4.1.2,  we speculated that the narrow component originates in the PIM region while the broad component is from the \hii~region. Comparing the magnitude of thermal broadening $\Delta\nu_t\sim17.2$~km/s with the width of narrow RRL that is 7.2~km/s (see Table~\ref{tb1}), it is evident that the narrow component cannot be associated with the \hii~region. We can estimate the electron temperature of this PIM region assuming that the broadening is entirely due to thermal effects. Hence, a thermal width of 7.2~km/s corresponds to an electron temperature of $\sim1145$~K. This is comparable to the electron temperature values observed towards PDRs and PIMs associated with other \hii~regions \citep[][]{{1996A&A...315L.281T},{1998ApJ...501..710G}}. It is difficult to decipher the broadening mechanisms for the broad component as it is compounded with the narrow component in emission. In addition, the emission from \textit{B} is relatively weak and the noise effects start dominating in the higher order moment maps. \\

\par Table~\ref{RRLline} lists the various parameters obtained for the hydrogen RRLs in both the \hii~regions. Columns 2, 3 and 4 gives the magnitudes of thermal, pressure and dynamical broadening. Columns 5 and 6 lists the local electron density estimated from the pressure broadening and average electron density from the continuum measurements. We have also used the average column density $\bar{n}_e$ estimated from the radio continuum flux to calculate the minimum pressure broadening suffered by hydrogen RRLs. The estimated lower limits using Eq.~(\ref{eden}) are given in Column 7 of Table~\ref{RRLline}. For G351.63--1.25$A$, the average electron density from the continuum is much larger (factor of 7) than that estimated using hydrogen RRL. This is likely due to optical depth effects as we may be sampling emission only from the outer layer of the \hii~region where the electron densities are expected to be lower.

\subsection{Velocity gradient in G351.69--1.15}

\begin{figure}
\hspace*{-0.5cm}
\centering
\includegraphics[scale=0.33,angle=90]{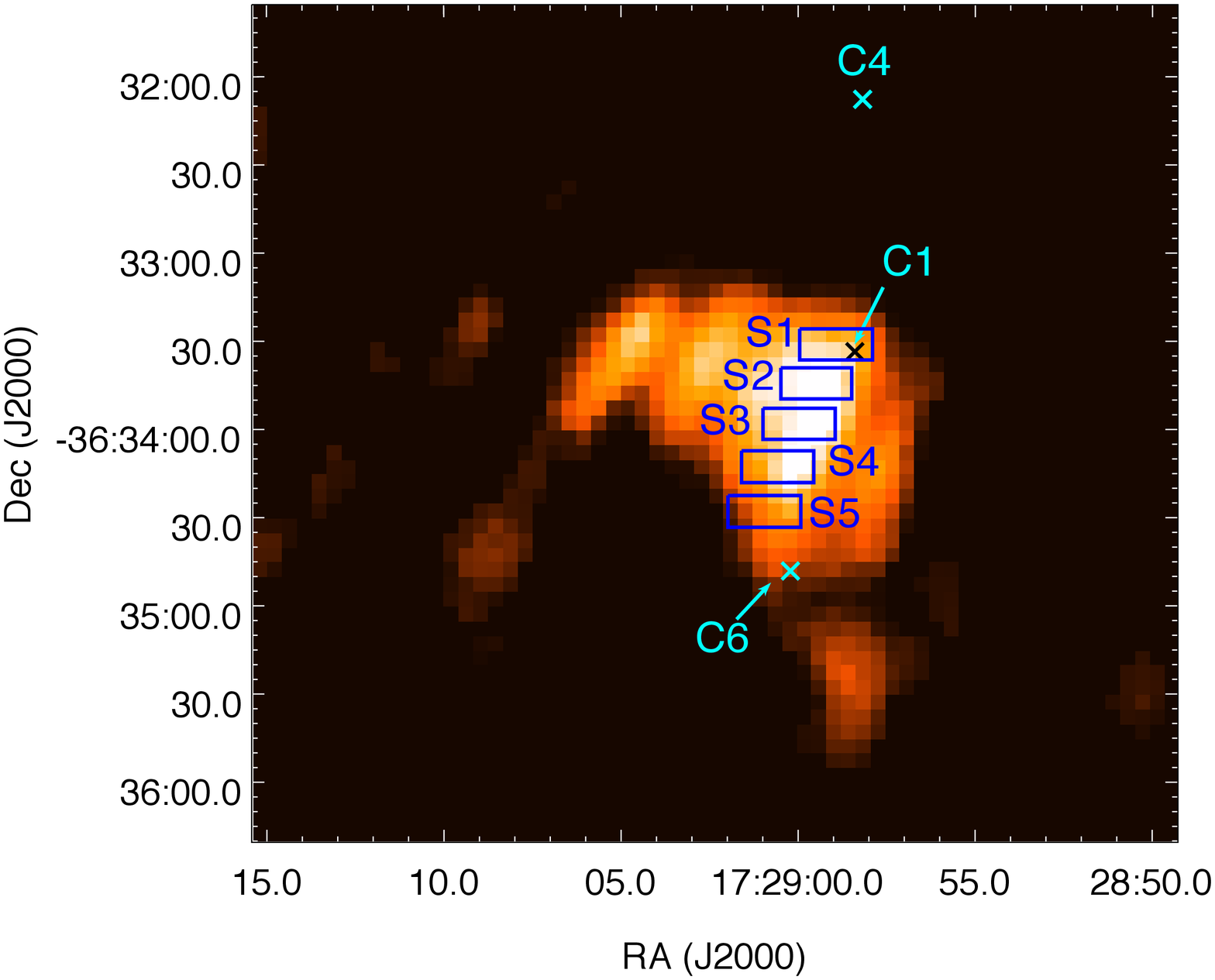}
\caption{Integrated intensity map of hydrogen RRL towards G351.69--1.15. The boxes show areas over which emission is integrated to study the velocity gradient across the source. The peak positions of 3 clumps C1, C4 and C6 whose spectrum are presented in a later section are also marked and labeled.} 
\label{slitpos}
\end{figure}

\begin{figure}
\hspace*{-0.4cm}
\centering
\includegraphics[scale=0.5]{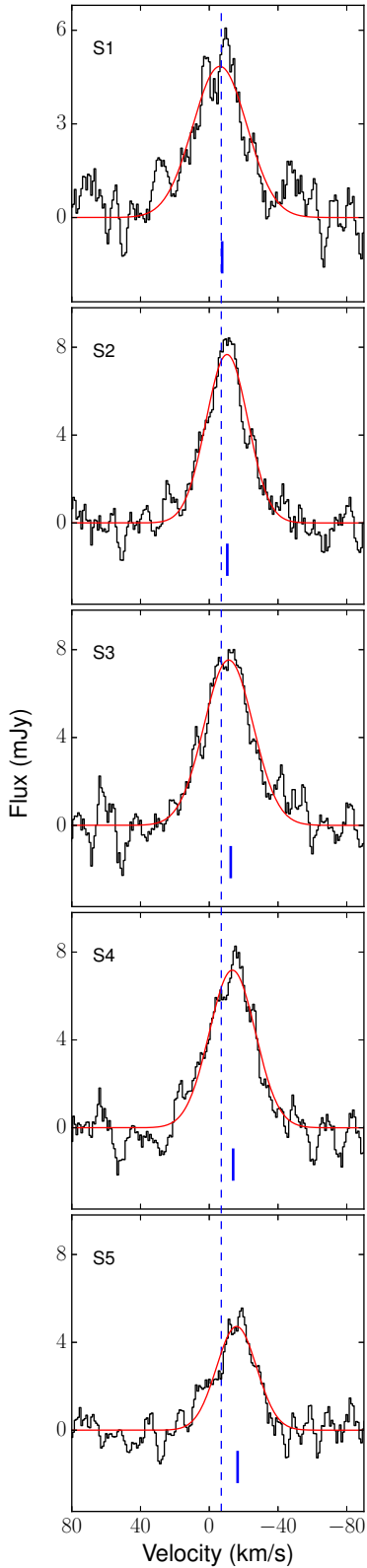}
\caption{Velocity profile of the H172$\alpha$ line towards G351.69--1.15 at different aperture positions. The dashed vertical line denotes the central velocity towards location S1 ($-8.6$~km/s). The small line in the individual panels mark the location of Gaussian peak of individual spectra.}
\label{slit}
\end{figure}

From the velocity map of G351.69--1.15, we discern a velocity gradient in the NW-SE direction that is the order of $\sim12$~km/s. In order to correlate the velocity with position along the cometary axis, we first consider a region with good SNR. We consider five rectangular apertures: S1, S2, S3, S4 and S5 (shown in Fig.~\ref{slitpos}) within this region, aligned along the axis from the cometary head towards the diffuse tail. We integrated the flux density within each aperture and fitted a single Gaussian profile to each spectrum. The results are shown in Fig.~\ref{slit}. It is evident that the velocity is increasingly blue shifted as we move from NW to SE. The shift in velocity is $\sim$8~km/s as we traverse from S1 to S5 and the LSR velocities of ionized gas regions within the apertures are listed in Table~\ref{tb2}. 
The presence of a velocity gradient across the cometary axis implies a large scale motion of gas in the \hii~region. Similar ionized gas motions across few other cometary \hii~regions have been reported previously. For instance, \citet{1994ApJ...429..268G} found velocity gradients of 4.5, 8 and 12~km/s towards three cometary \hii~regions, G13.87+0.28, G32.80+0.19 and G61.48+0.09, respectively. Using the near-infrared Br$\gamma$ line mapping, \citet{1999MNRAS.305..701L} detected a large velocity gradient ($\sim$20~km/s) within the cometary \hii~region G29.96-0.02. These represent compact and ultracompact \hii~regions (size $\leq$1~pc) powered by massive stars whose spectral types are earlier than O8. The peak electron densities are of the order of 10$^3-$10$^4$~cm$^{-3}$. The estimated molecular hydrogen column densities towards the cometary-like \hii~regions G32.80+0.19 and G61.48+0.09 lie within the range 5$\times$10$^{21}-$3$\times$10$^{22}$~cm$^{-2}$ \citep{1995ApJ...453..727G}. These values are similar to the mean column density of 2.5$\times$10$^{22}$~cm$^{-2}$ estimated towards G351.69--1.15 (Paper I). 

%%%%%%%%%%%%%%%%%%%% Fig8 %%%%%%%%%%%%%%%%%%%%%%%%%%%%%%%%%%%%%%%%%%%%%%%%%

%%%%%%%%%%%%%%%%%%%% %%%%%%%%%%%%%%%%%%%%%%%%%%%%%%%%%%%%%%%%%%%%%%%%%

%%%%%%%%%%%%%%%%%%%% Fig9 %%%%%%%%%%%%%%%%%%%%%%%%%%%%%%%%%%%%%%%%%%%%%%%%%
%%%%%%%%%%%%%%%%%%%% %%%%%%%%%%%%%%%%%%%%%%%%%%%%%%%%%%%%%%%%%%%%%%%%%

\par In order to understand the mechanism(s) responsible for the ionized gas motion, it is instructive to examine the local density structures as well as the kinematics of the ambient molecular cloud. For this, we use transitions of several molecular species and the results are presented in the next section. 

\subsection{Photodissociation region (PDR)}

\subsubsection{G351.69--1.15}
The RRLs of carbon are believed to emanate from the PDRs associated with the \hii~region \citep{1977A&A....57..341P}. From the fit of the carbon line towards G351.69--1.15 with a Gaussian profile, we observe that the velocity of the carbon line is $-12.6$~km/s while its width is relatively narrow ($\Delta$V=5.4~km/s).  The narrow width of the carbon line hints at the emission originating from a relatively cooler medium, i.e., from the PDR. The central velocity of the carbon line is red shifted with respect to that of hydrogen by 0.6~km/s. Comparing the velocity of the carbon line with the MALT90 molecular line data (discussed later), we see that the carbon RRL is blue shifted by 4.3~km/s with respect to ambient cloud velocity. Keeping in view the fact that the carbon RRL emission is weak and detected close to the peak of hydrogen RRL, it is possible that the PDR medium is set into motion by the expansion of the \hii~region and we are likely tracing the carbon RRL emission close to the \hii~region. The origin of carbon RRL could either be due to the spontaneous emission in the \cii~region and/or due to stimulated emission by the strong background radiation field of the \hii~region \citep{1998ApJ...504..375K}. In the case of G351.69--1.15, we detect C172$\alpha$ RRL near the continuum peak which could possibly indicate the predominance of stimulated emission. We have not detected carbon RRL towards other regions of the continuum emission. In order to trace carbon RRLs towards the diffuse envelope and to study the velocity structure of PDR, we need sensitive, high resolution observations with larger integration times. 
%%%%%%%%%%%%%%%%%%%%%%%%%%%%%%%%%%% Table  %%%%%%%%%%%%%%%%%%%%%%%%%%%%%%%%%%%%
\begin{table}
\footnotesize
\caption{Central velocity towards different aperture positions}
\begin{center}

\hspace*{-0.5cm}
\begin{tabular}{l c } \hline \hline
 &  \\
Aperture      & V$_{LSR}$ (km/s)\\
\hline\\
S1 & $-8.6\pm1.2$ \\
S2& $-10.9\pm0.6$ \\
S3 & $-12.5\pm0.7$ \\
S4& $-13.9\pm0.6$ \\
S5 & $-16.5\pm0.9$ \\
\hline 

\end{tabular}
\label{tb2}
\end{center}
\end{table}

%%%%%%%%%%%%%%%%%%%%%%%%%%%%%%%%%%%  %%%%%%%%%%%%%%%%%%%%%%%%%%%%%%%%%%%%
\subsubsection{G351.63--1.25}
\begin{figure}
\hspace*{-0.4cm}
\centering
\includegraphics[scale=0.3]{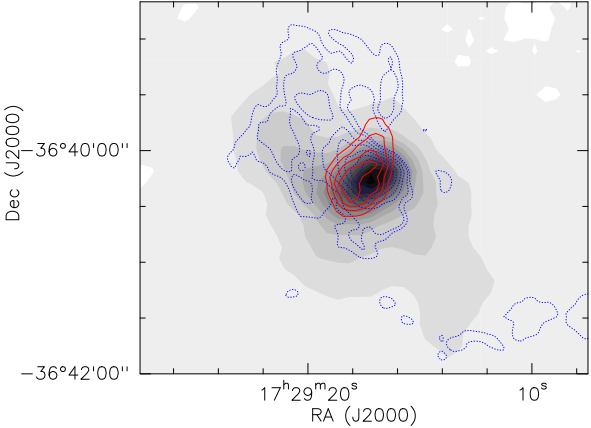}
\caption{870~$\mu$m cold dust emission towards G351.63--1.25 overlaid with integrated intensity contours of carbon RRL (red) and 1280~MHz radio continuum contours (dotted blue). The contour levels of the carbon RRL emission are from 7~Jy/Beam*m/s to 40~Jy/Beam*m/s in steps of 8~Jy/Beam*m/s. The contour levels of the 1280~MHz emission are same as that of Fig.~\ref{continuum}(b).}

\label{17258c}
\end{figure}

The integrated intensity map of carbon RRL towards G351.63--1.25$\textit{A}$ is shown in Fig.~\ref{17258c}. The peak of the carbon RRL is shifted with respect to the H172$\alpha$ peak by $\sim 7\arcsec$. Carbon line emission is believed to arise in the PDR and also the in the PIM located outside the HII region where hydrogen can be partially ionized giving rise to the narrow hydrogen line. The temperature of this region cannot be larger than 1145K. In addition, we observe that most of the emission overlaps with the radio continuum emission region. This suggests that the carbon RRL could be from the PDR in front of the \hii~region, along our line-of-sight. We also see a small extension of the emission towards northwest-southeast direction, where there is no hydrogen RRL or continuum emission. If the emission arises in the homogeneous and isotropic partially ionized layer in front of the \hii~region, the the carbon RRL emission should follow the continuum emission. Detection of C172$\alpha$ emission towards outer regions where there is weak or no radio continuum emission suggests that these originate within regions whose physical properties such as $n_e$, $T_e$ and emission measure (EM) are different from that of the PIM surrounding the \hii~region \citep[e.g.,][]{1998ApJ...504..375K}. From the molecular line data, the velocity of molecular cloud is estimated as $\sim -12$~km/s \citep{{1977PASAu...3..152B},{1978MNRAS.183..711G}}. The carbon line also has a velocity of $-12.4$~km/s, indicating that the PDR is  stationary with respect to the molecular cloud. Alternately, it is possible that we are tracing the PDR emission from the surface layers rather than regions close to the \hii~region that could involve the motion of ionized gas. 
%%%%%%%%%%%%%%%%%%%% Fig10 %%%%%%%%%%%%%%%%%%%%%%%%%%%%%%%%%%%%%%%%%%%%%%%%%

%%%%%%%%%%%%%%%%%%%%%%%%%%%%%%%%%%%%%%%%%%%%%%%%%%%%%%%%%%%%%%%%%%%%%

\section{MOLECULAR LINE EMISSION FROM G351.69--1.15}

\subsection{Velocity profile of the cloud}
\begin{figure*}
\vspace*{-0.3cm}
\centering
\includegraphics[scale=0.29]{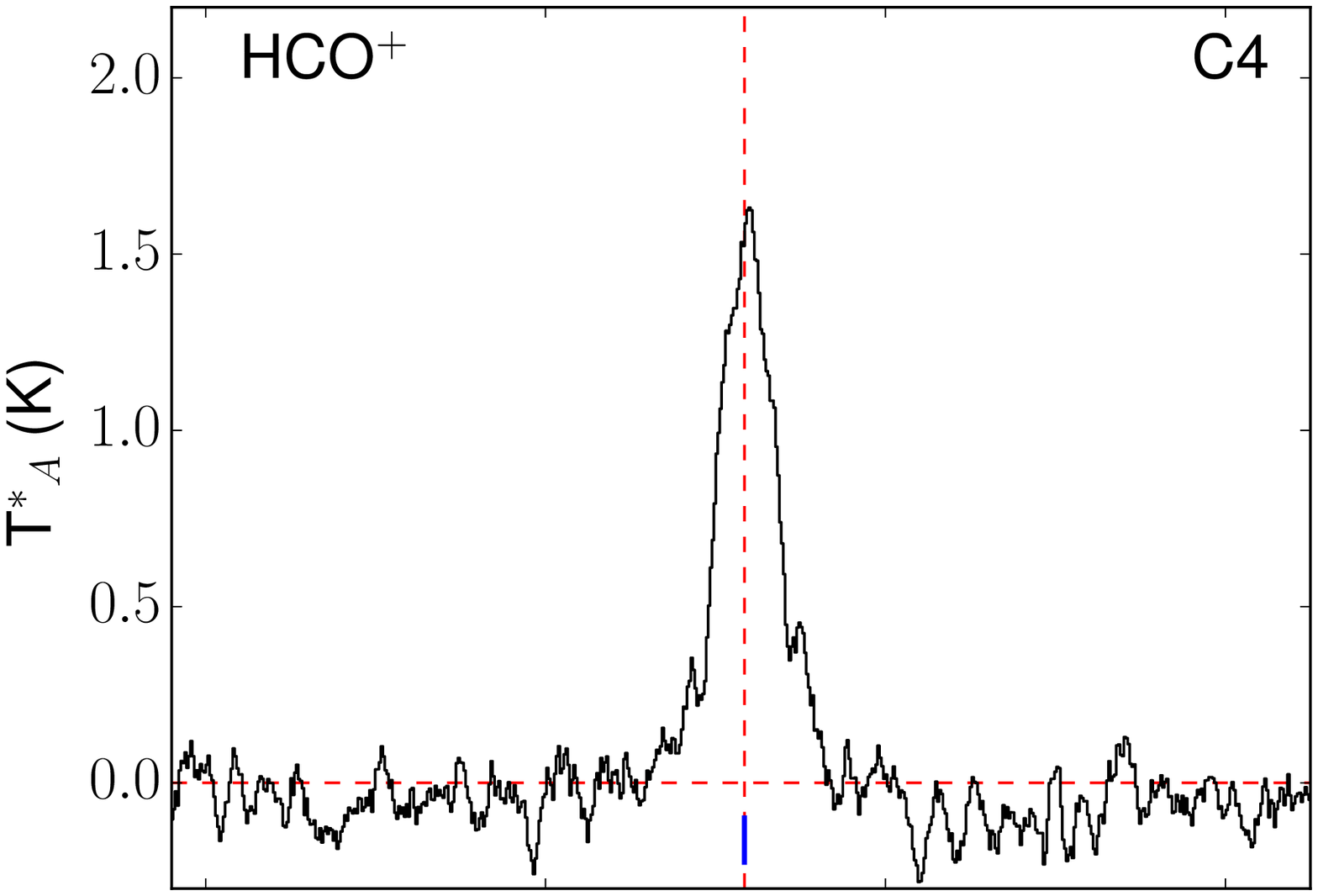} \quad \includegraphics[scale=0.29]{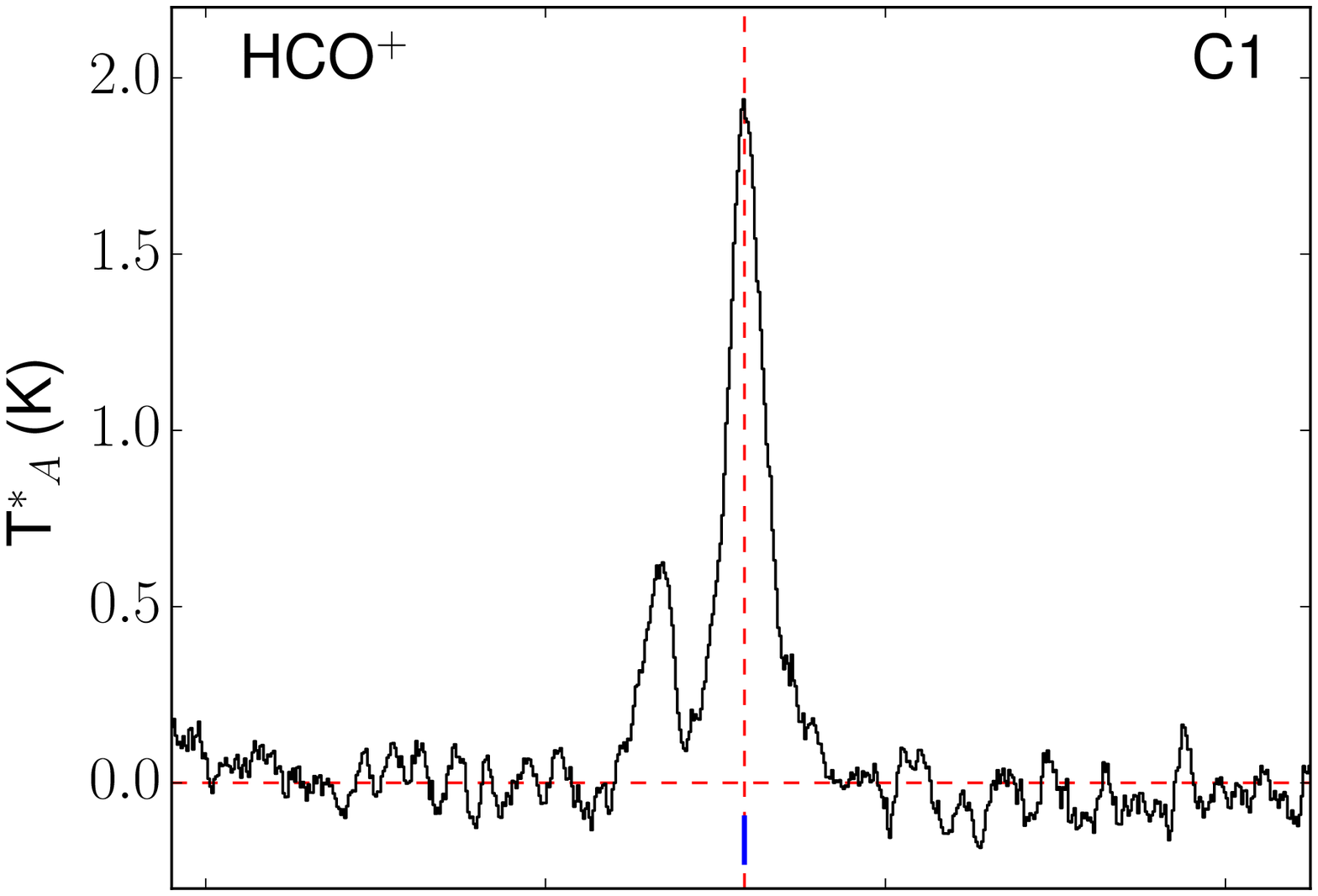} \quad \includegraphics[scale=0.29]{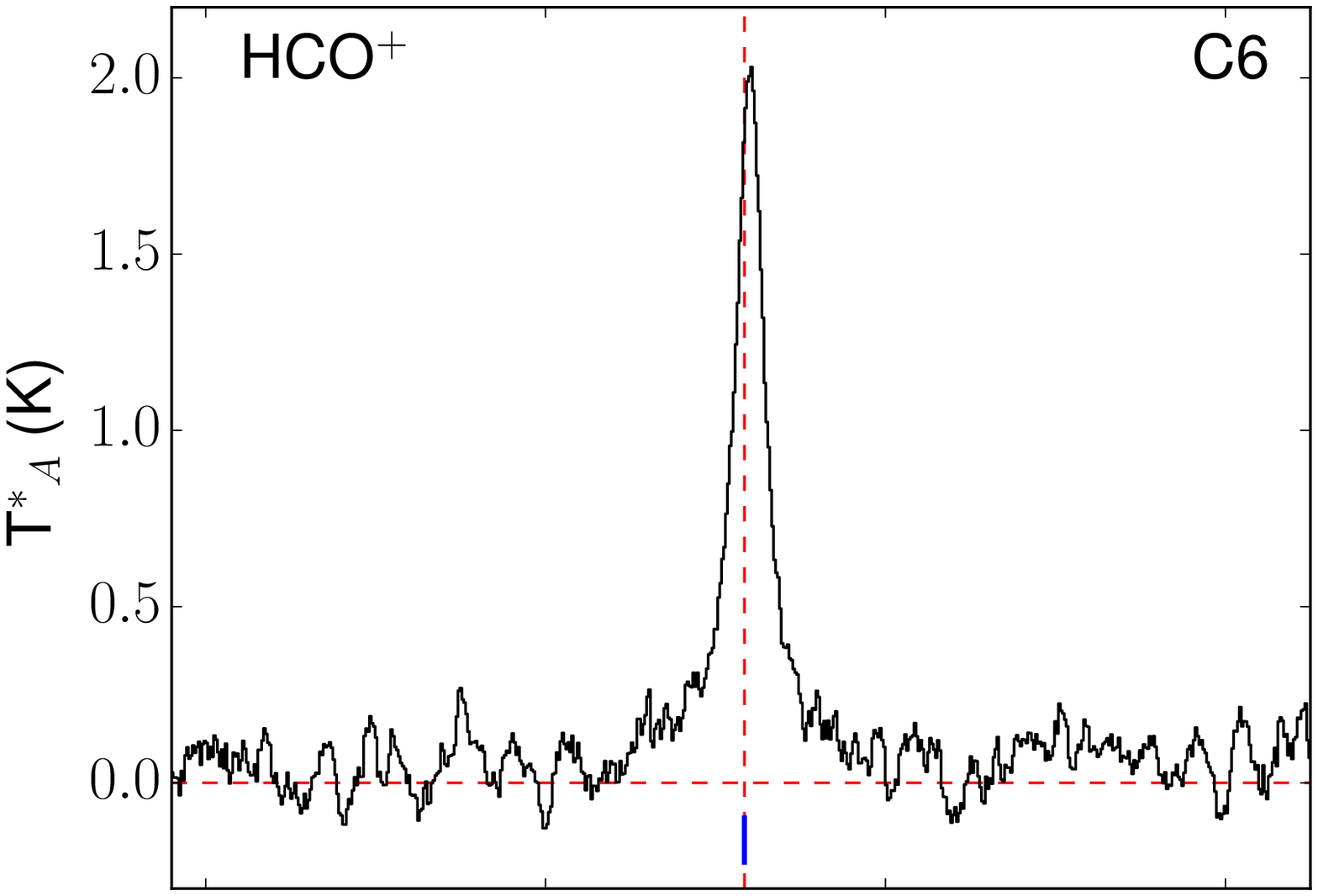} \quad  \includegraphics[scale=0.29]{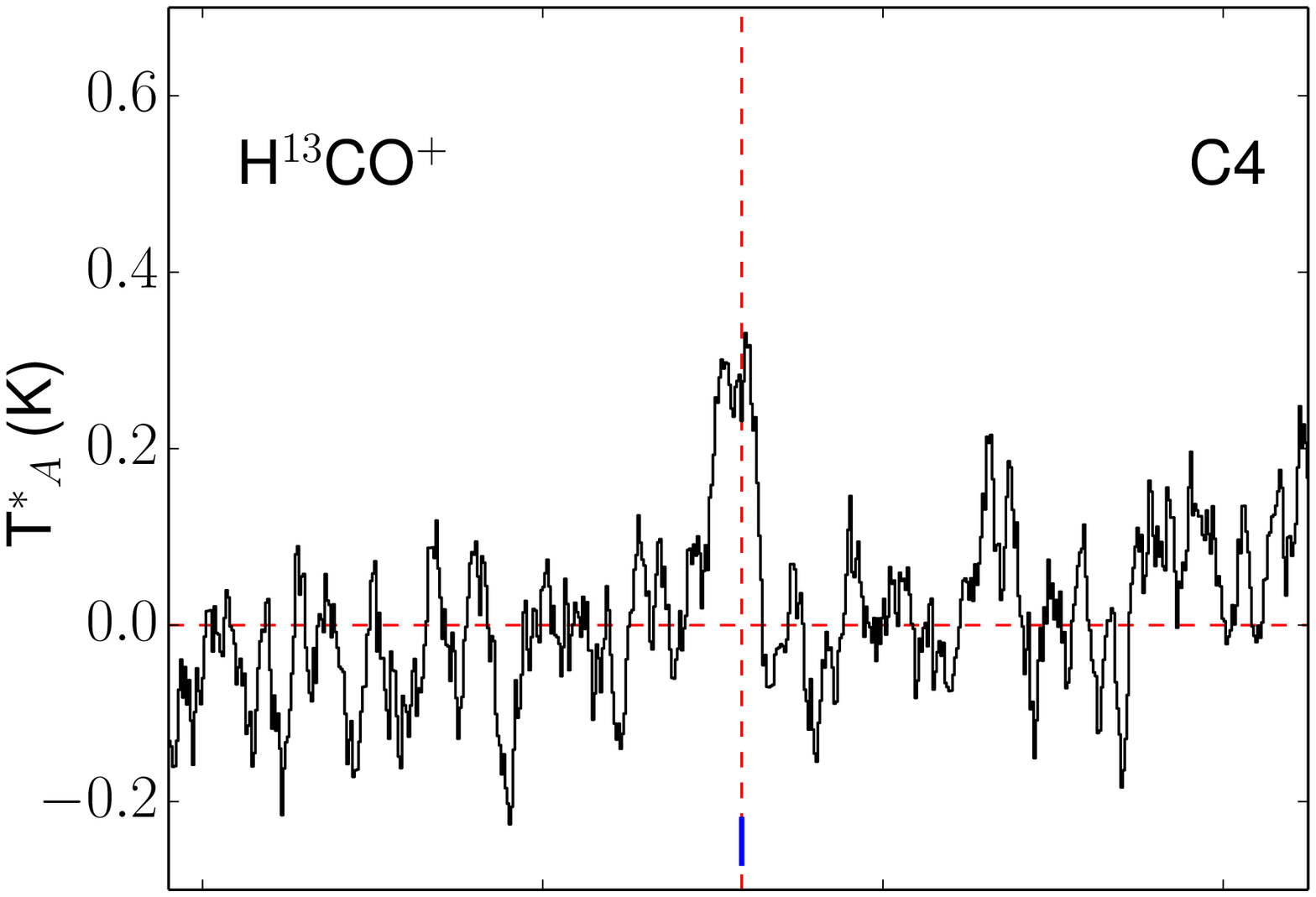} \quad \includegraphics[scale=0.29]{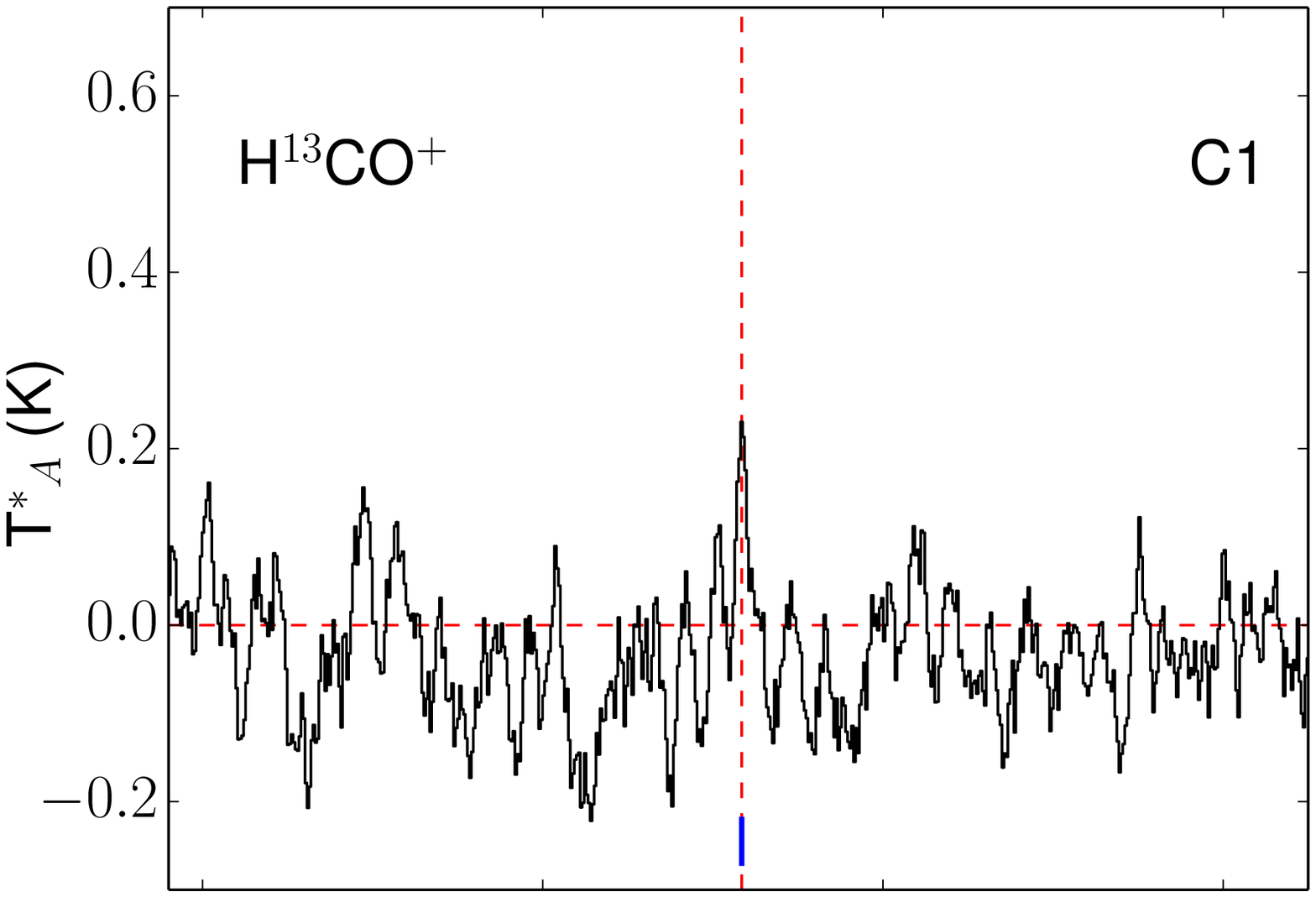}  \quad \includegraphics[scale=0.29]{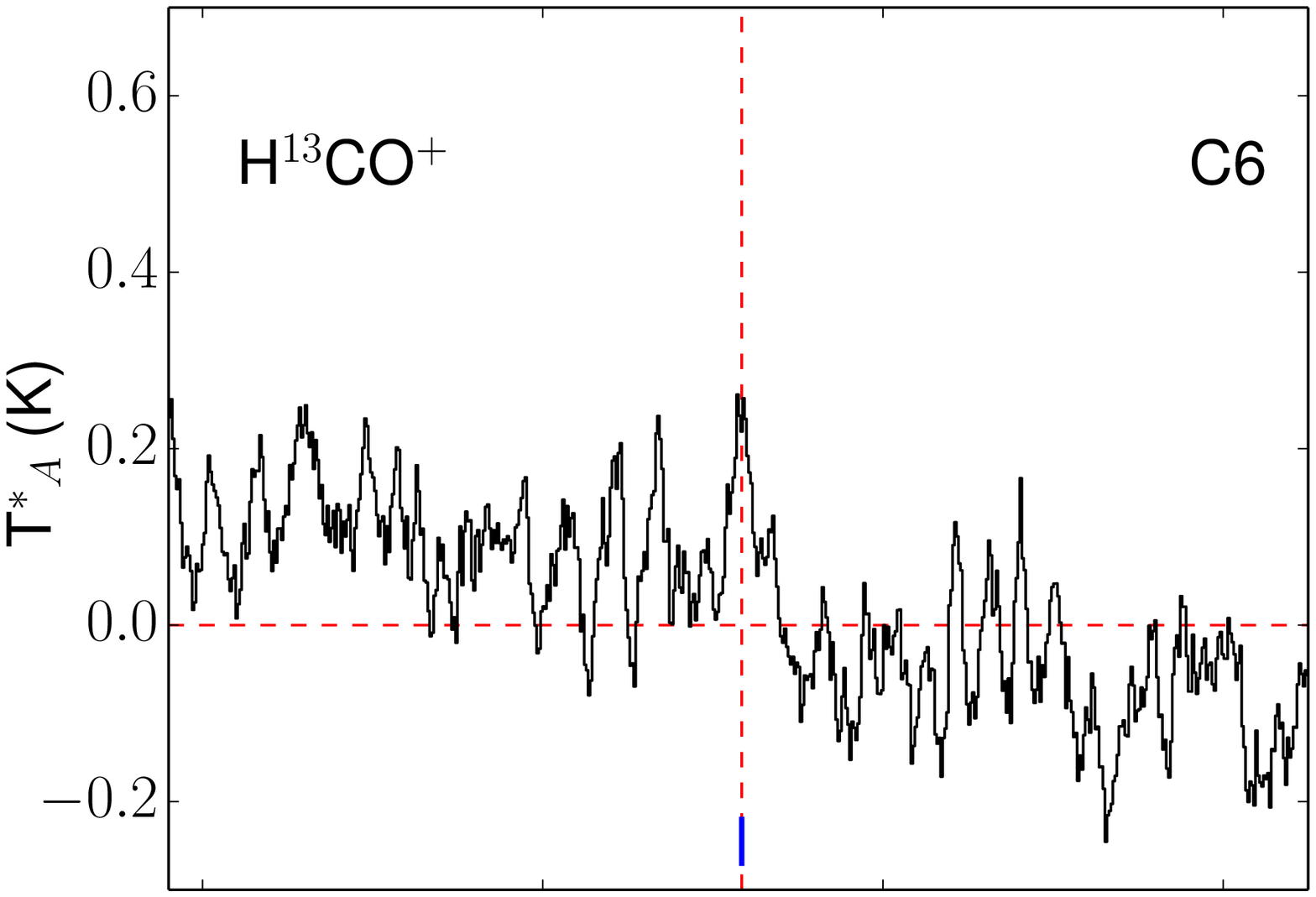} \quad \includegraphics[scale=0.29]{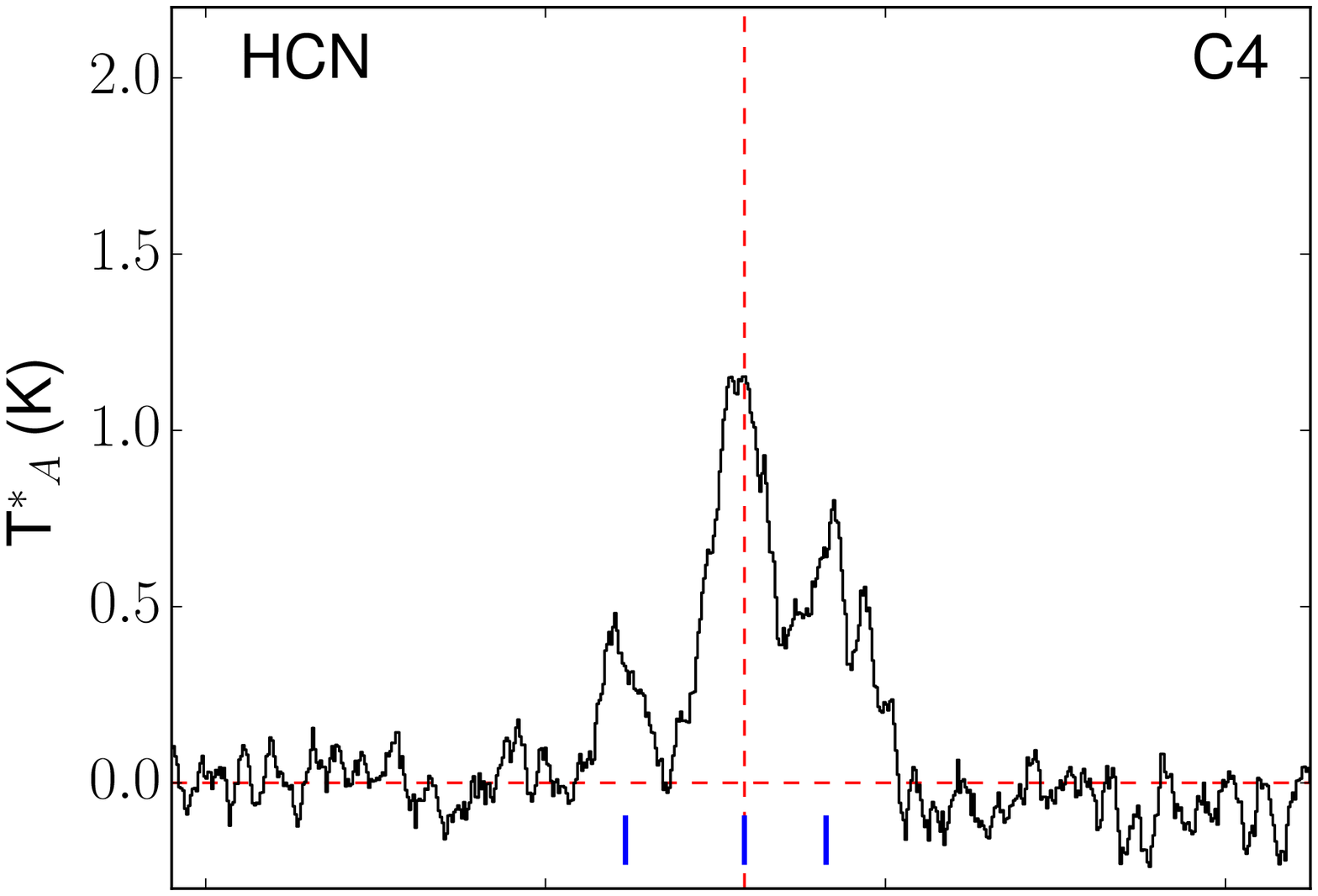} \quad \includegraphics[scale=0.29]{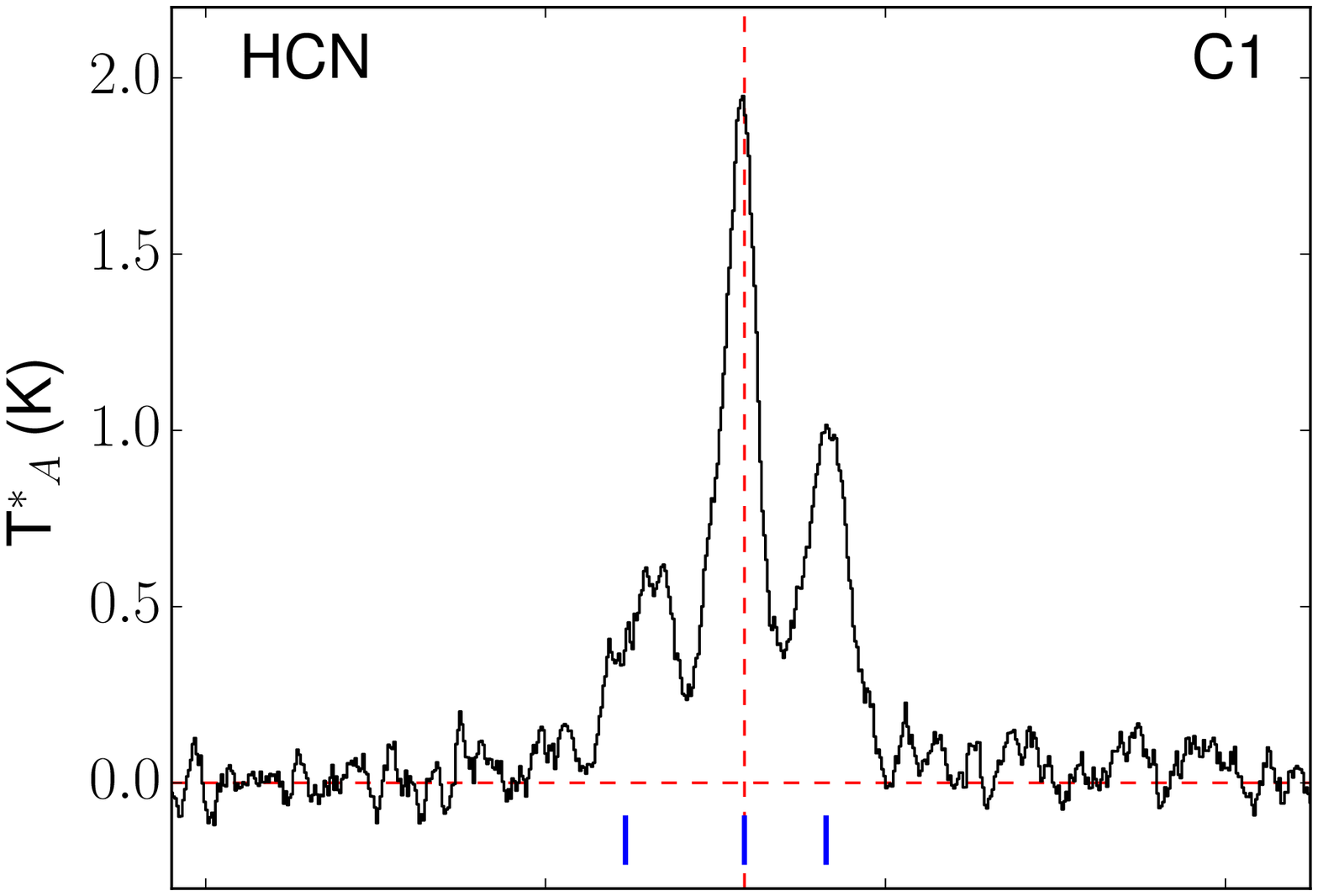} \quad \includegraphics[scale=0.29]{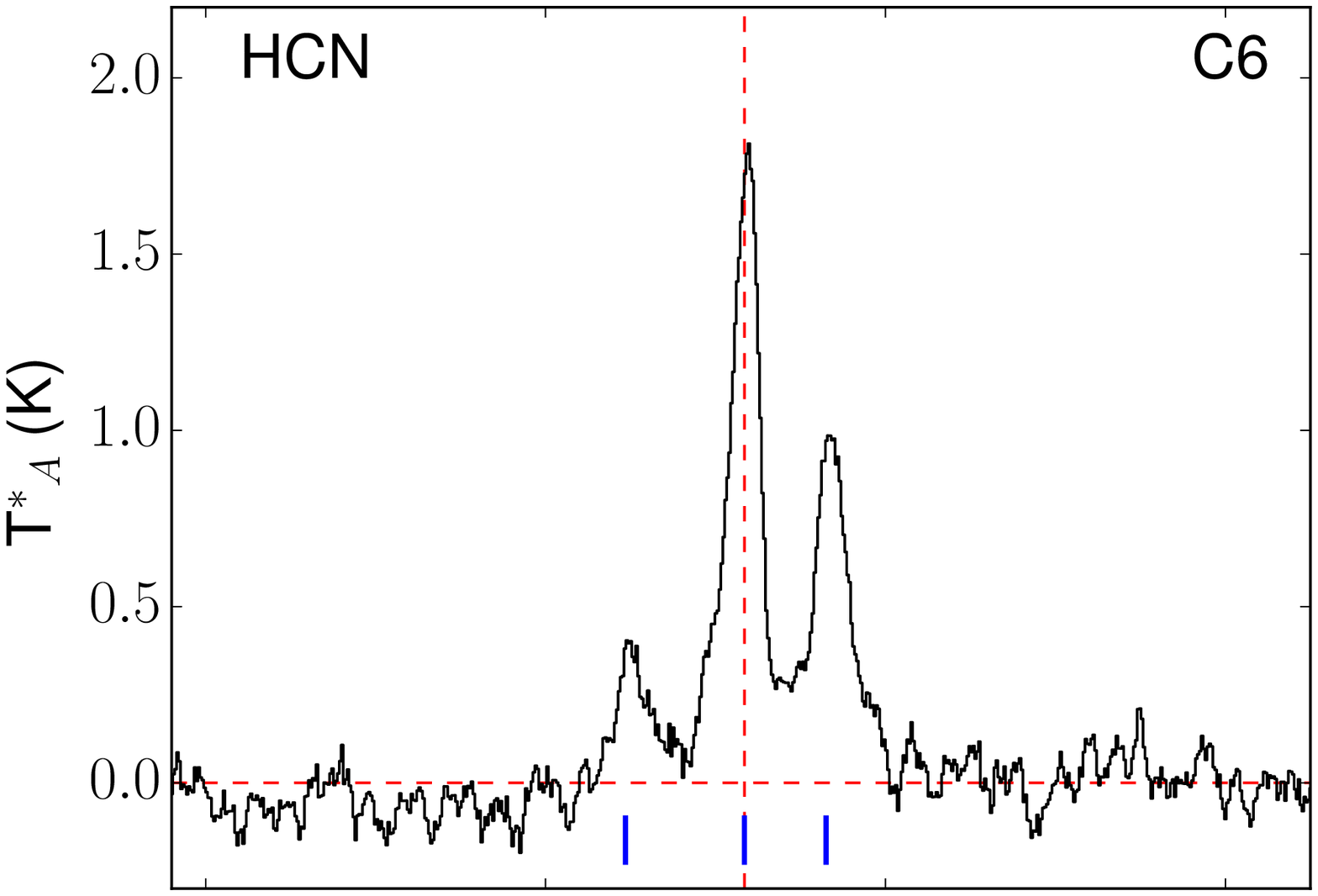} \quad \includegraphics[scale=0.29]{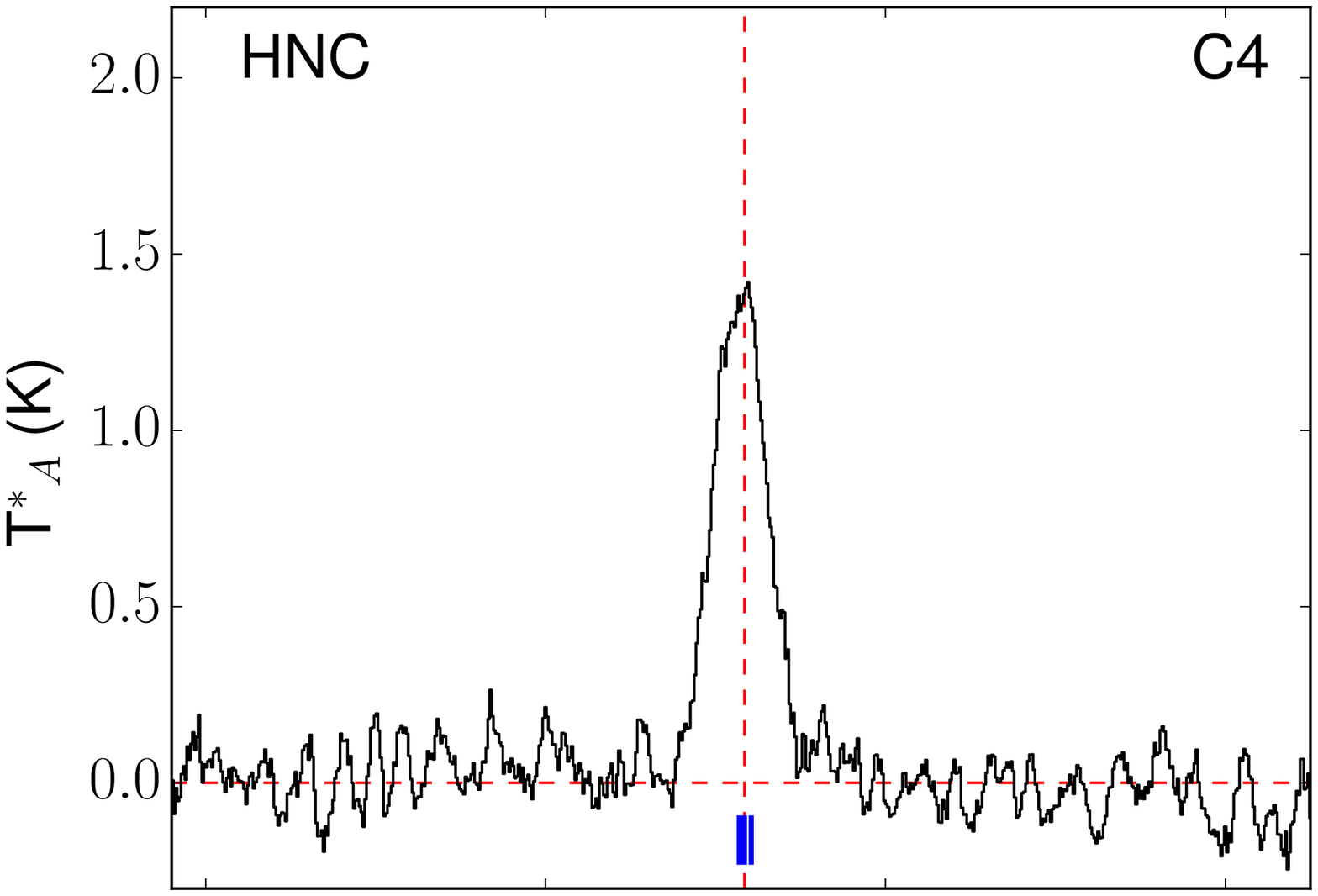} \quad \includegraphics[scale=0.29]{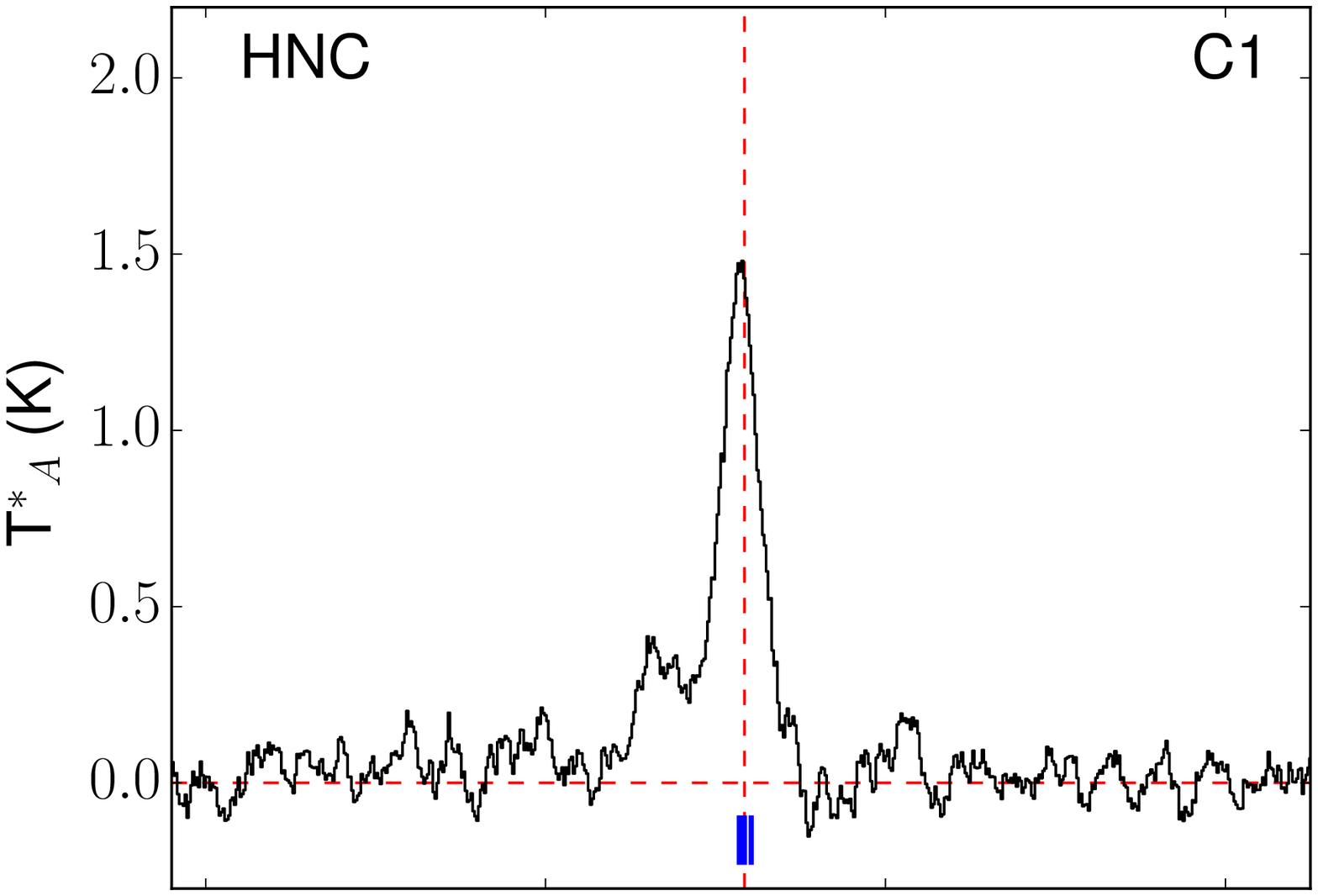} \quad \includegraphics[scale=0.29]{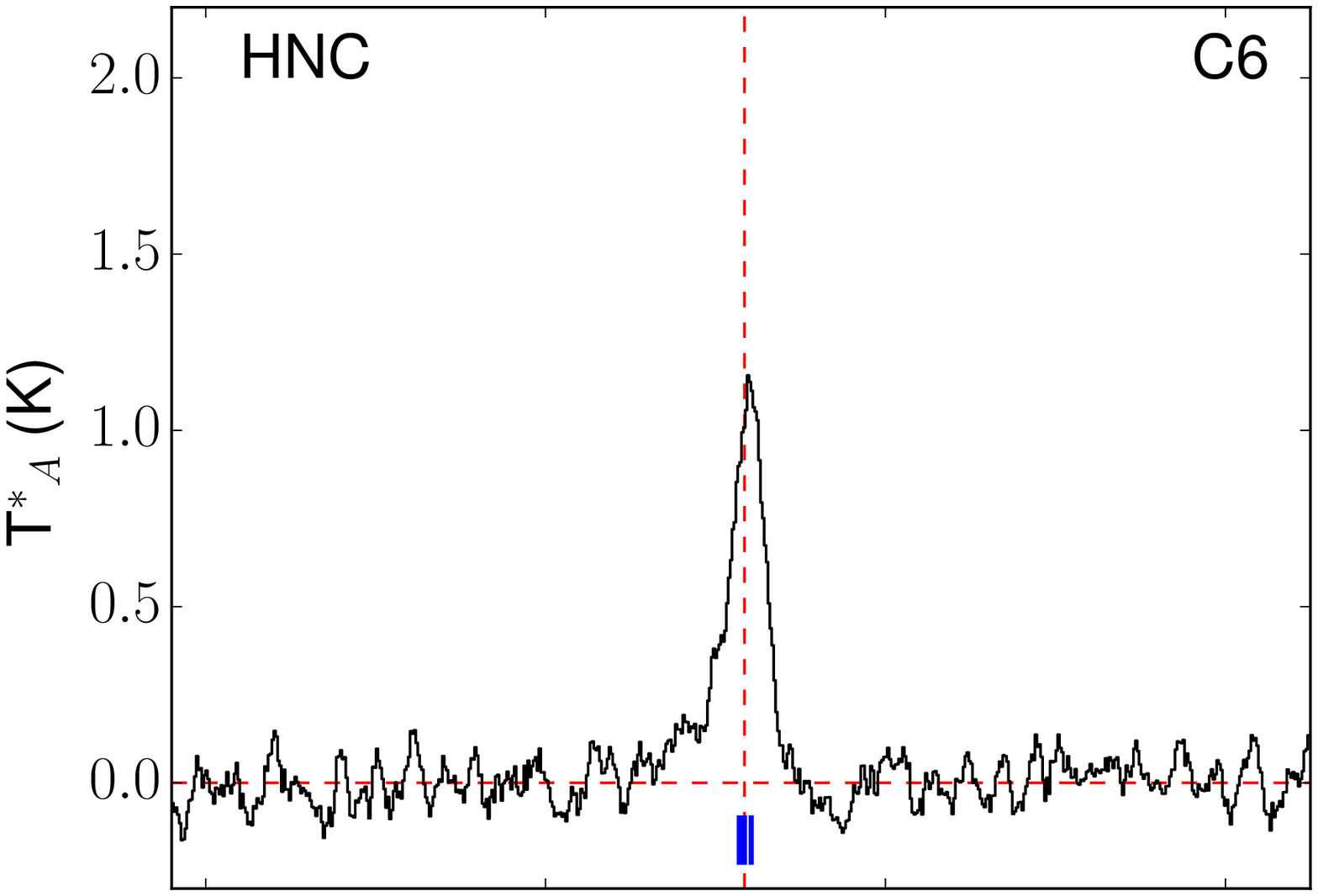} \quad \includegraphics[scale=0.29]{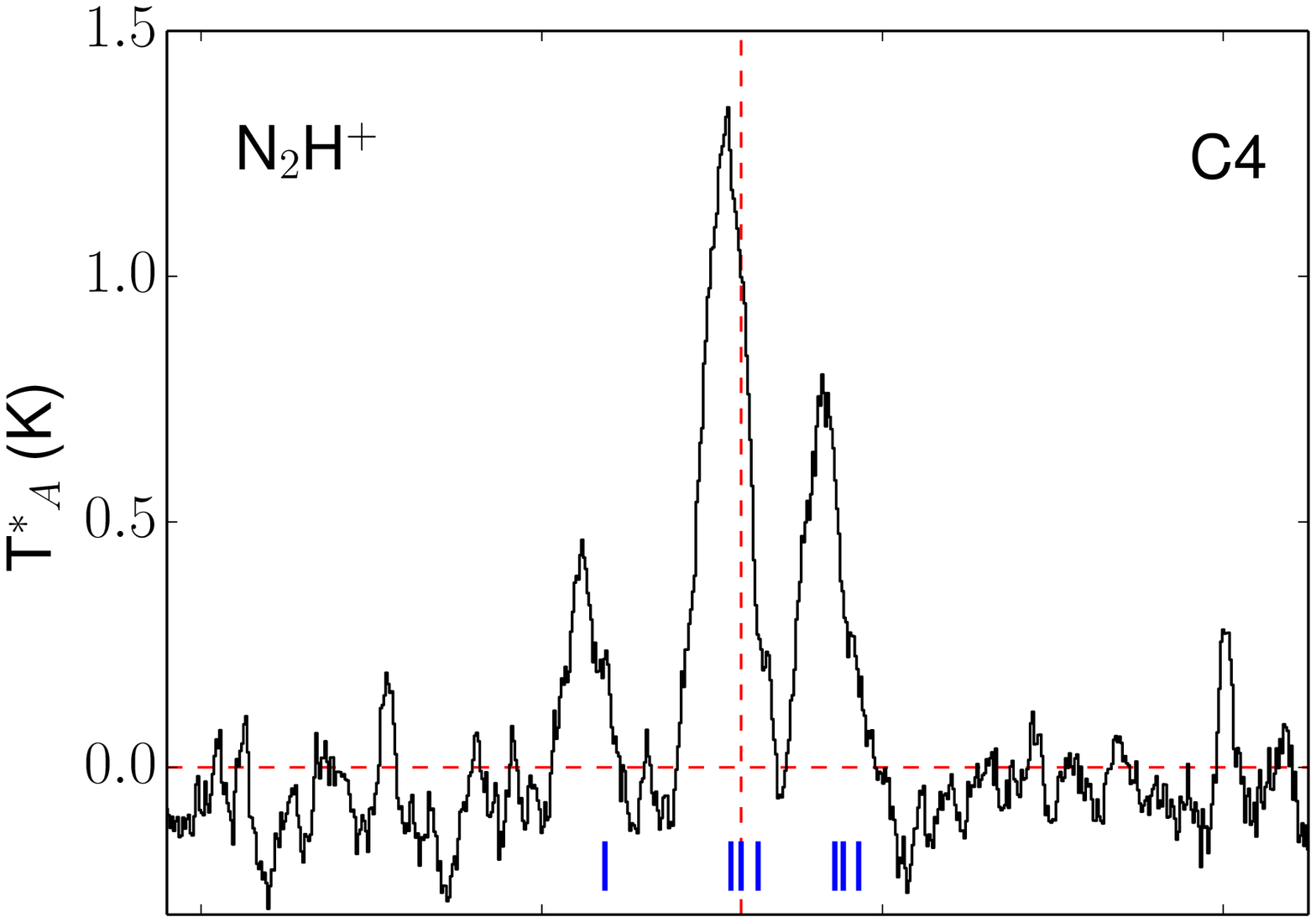} \quad \includegraphics[scale=0.29]{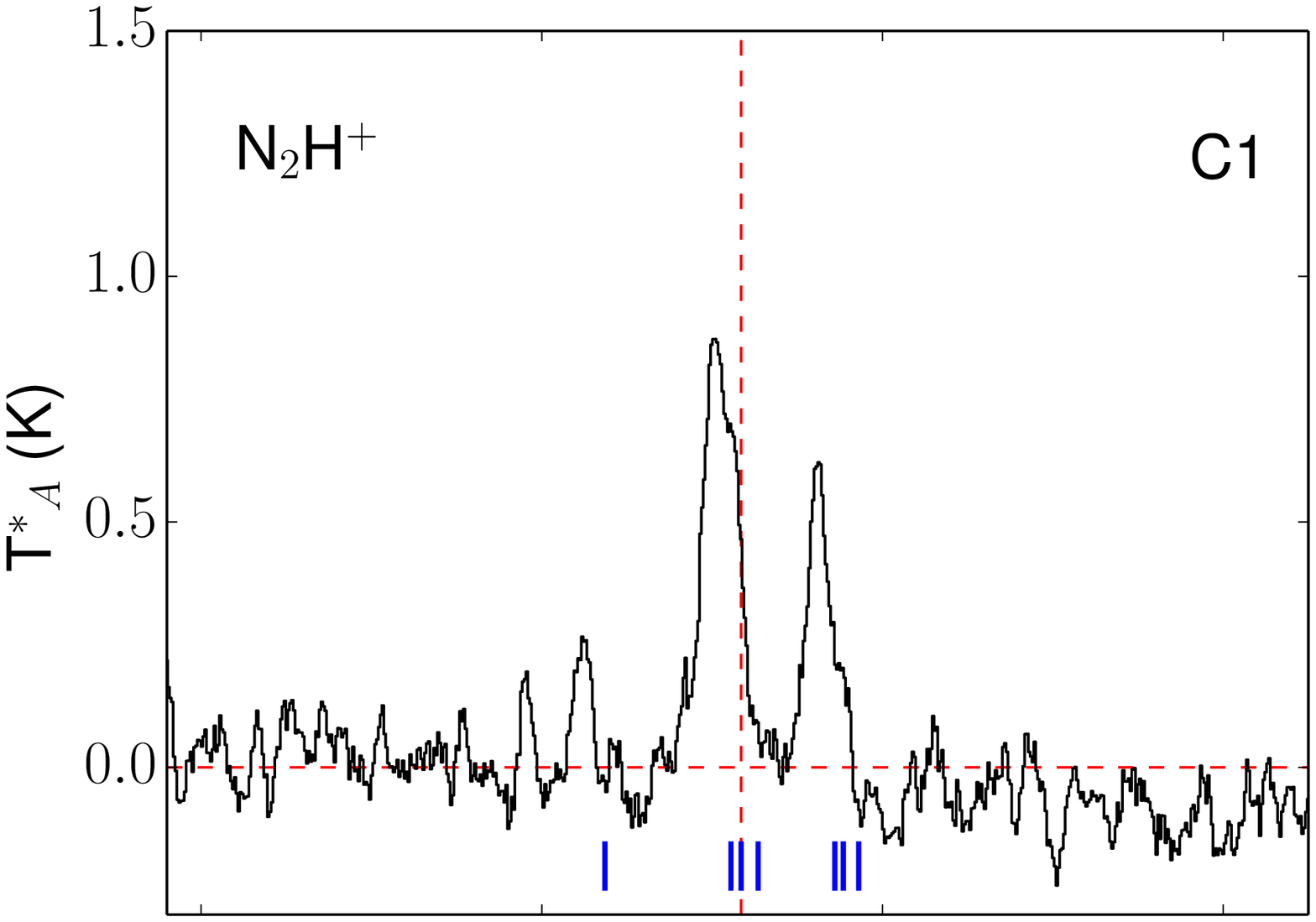} \quad \includegraphics[scale=0.29]{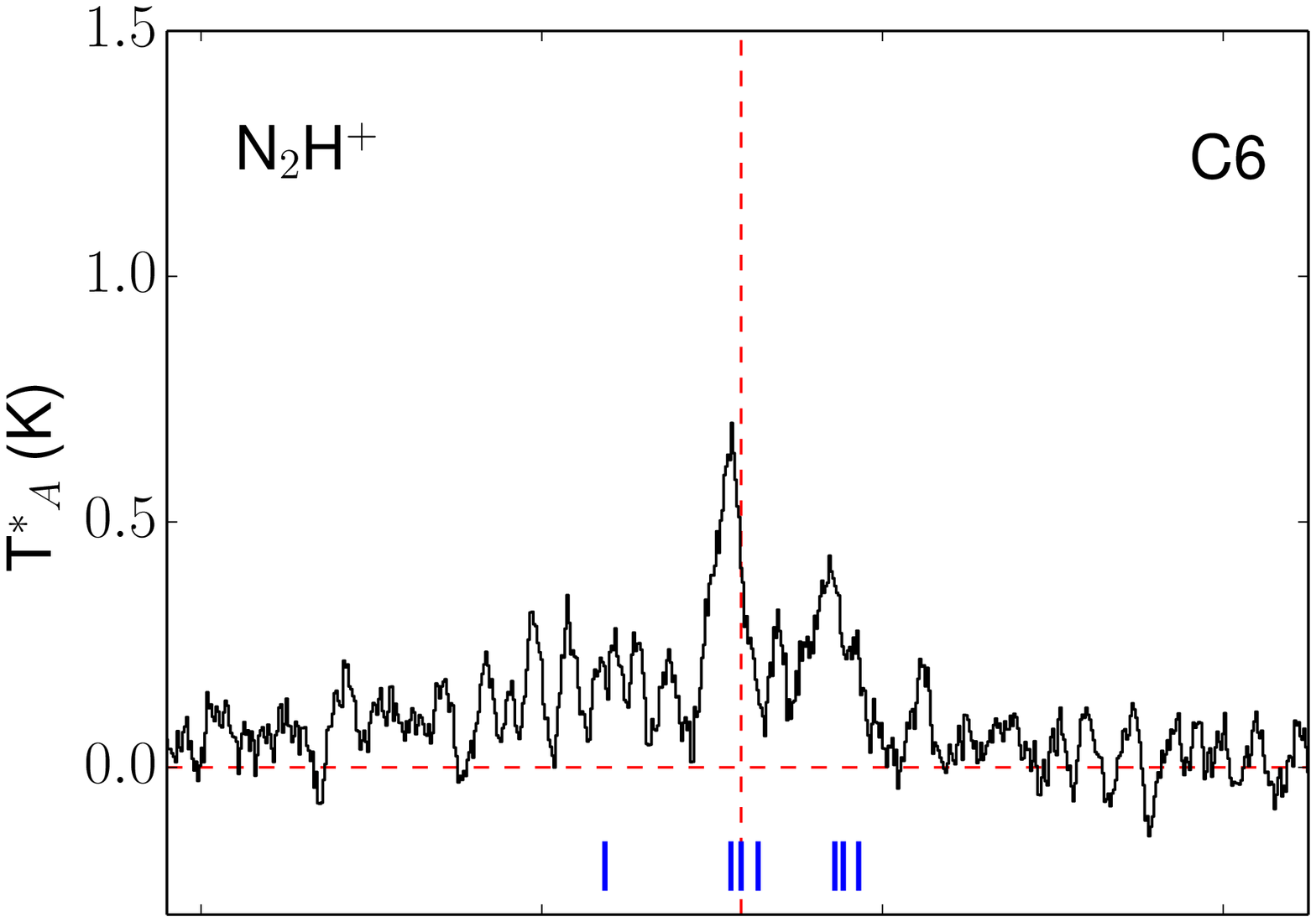} \quad \includegraphics[scale=0.29]{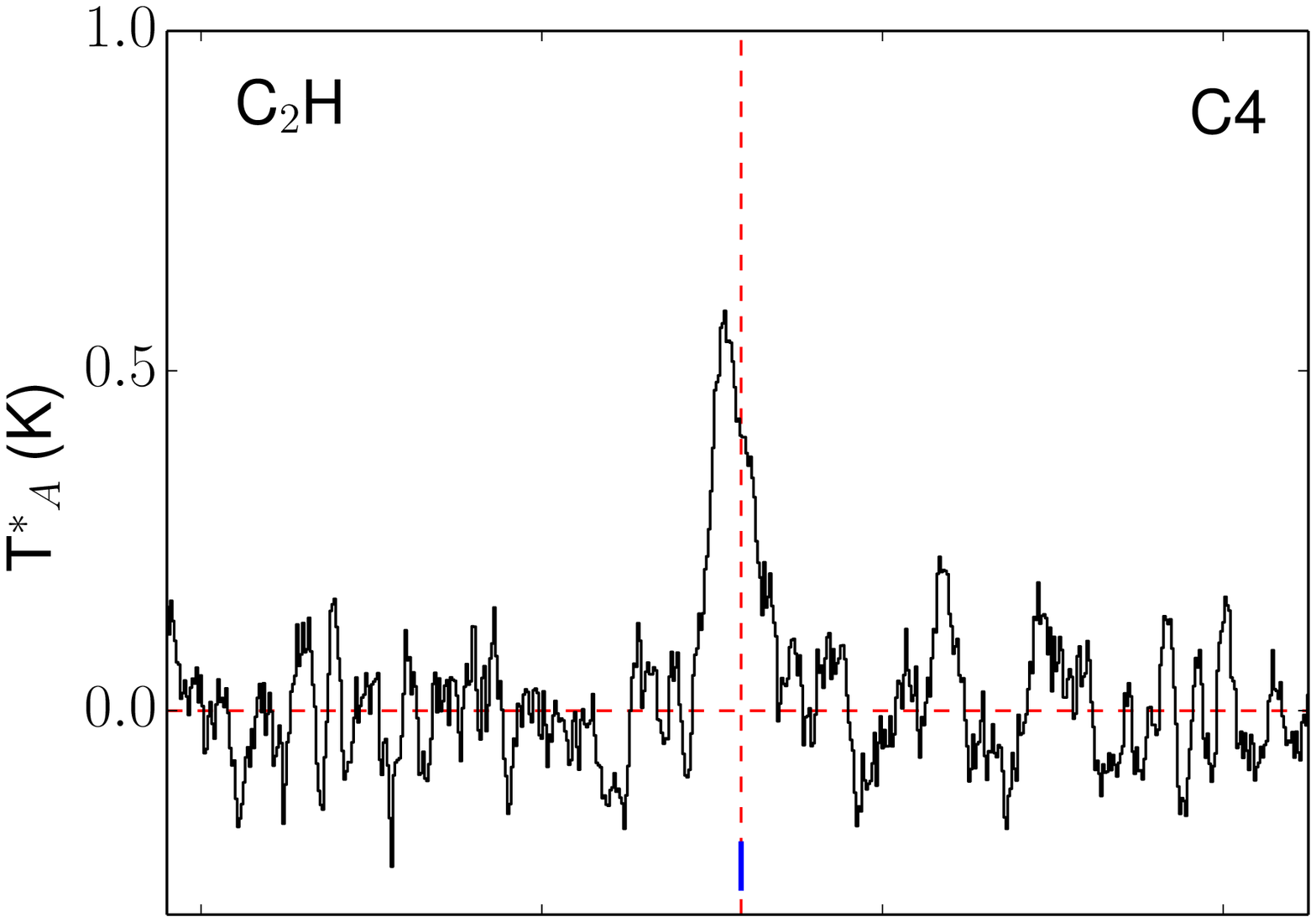} \quad \includegraphics[scale=0.29]{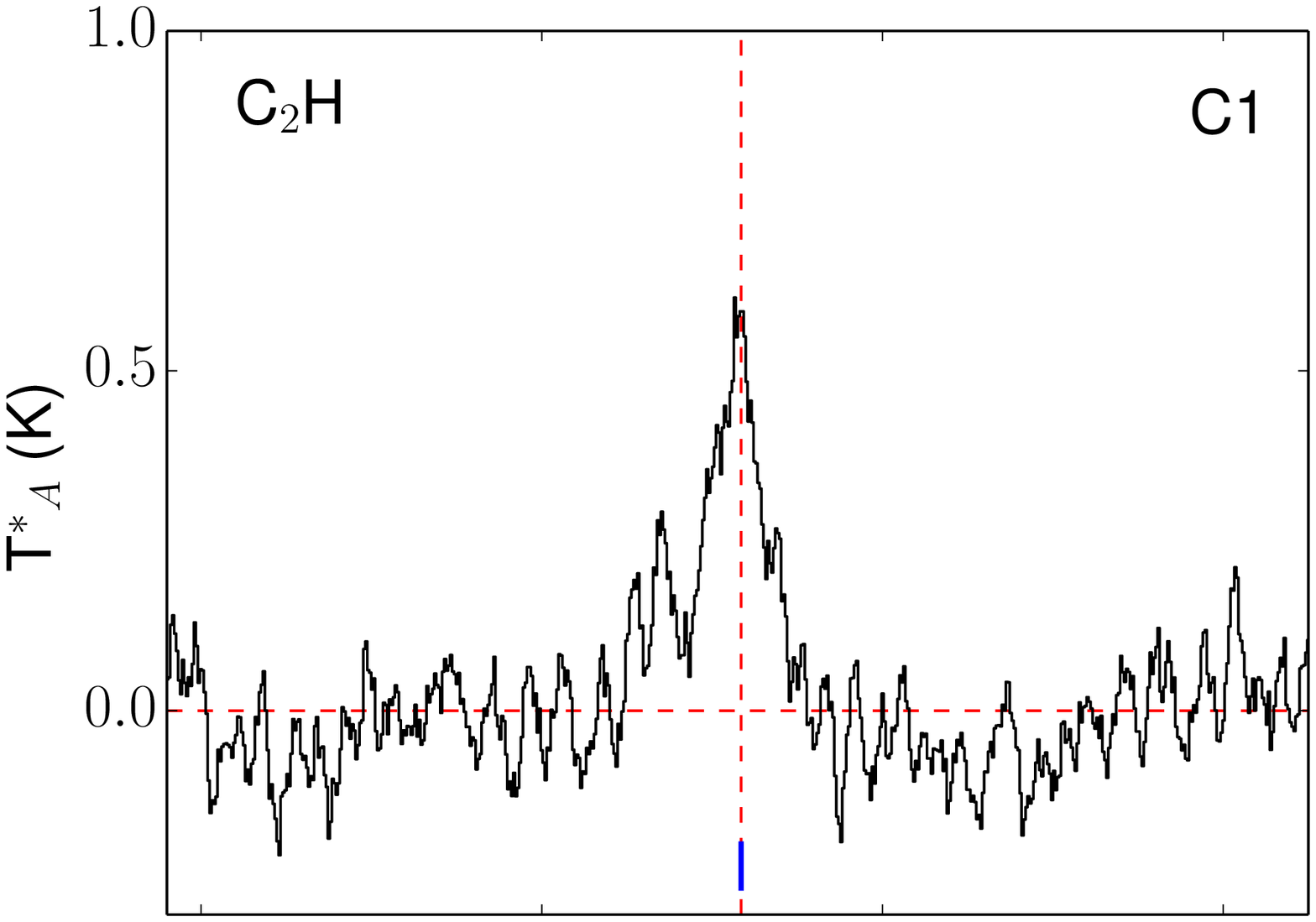} \quad \includegraphics[scale=0.29]{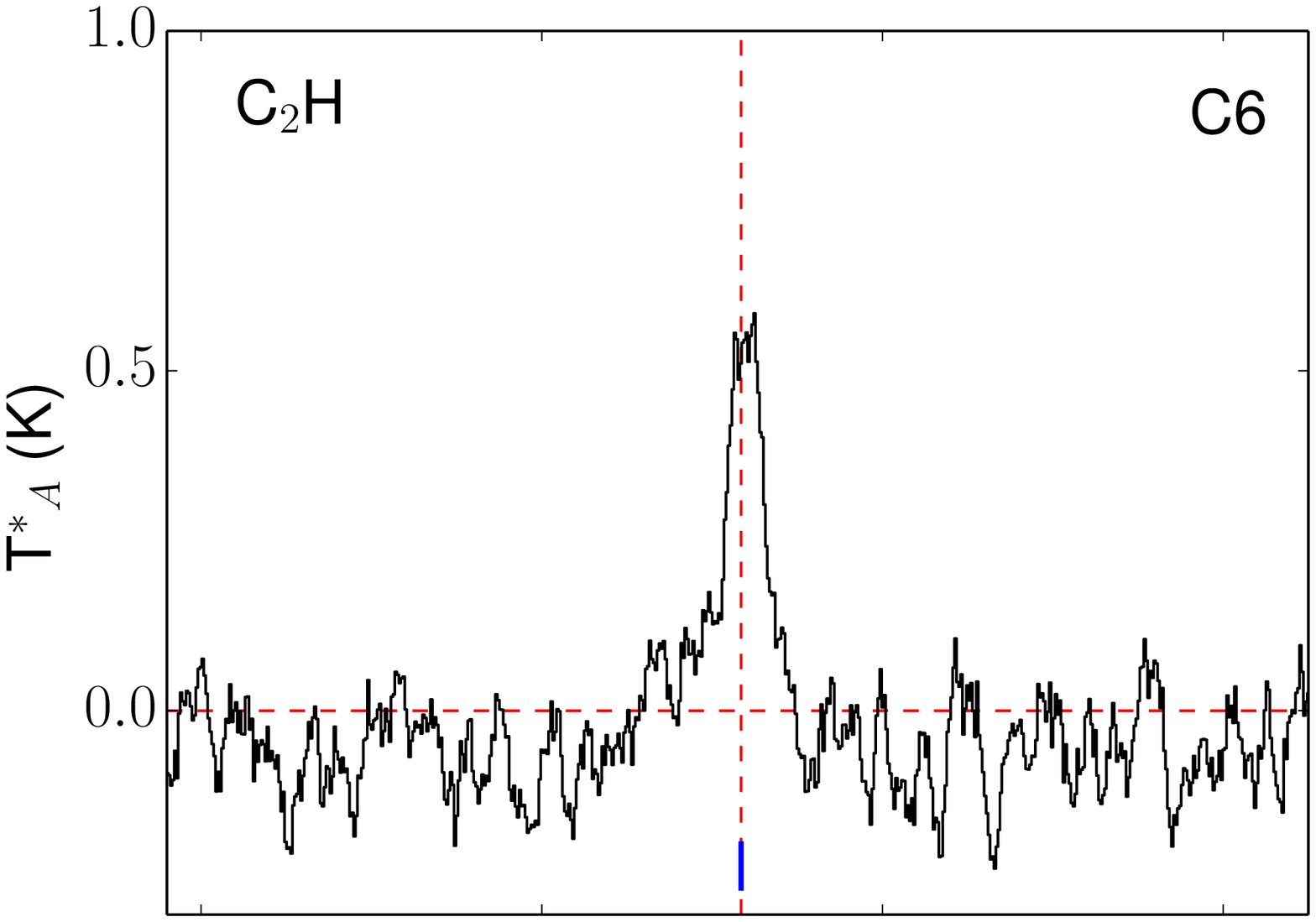} 

\vspace*{-0.1cm}

\caption{Velocity profiles of the molecular lines towards three positions, which are peaks of clump C4, C1 and C6 (Fig.\ref{malt90mom}). The spectra are boxcar smoothed by 7 channels that corresponds to a velocity smoothing of 0.8~km/s. From top to bottom, the lines are HCO$^+$, H$^{13}$CO$^+$, HCN, HNC, N$_2$H$^+$ and C$_2$H. The x-axis is the LSR velocity in units of km/s and y-axis is the antenna temperature corrected for atmospheric attenuation in units of K. The dashed lines represent the central velocity of H$^{13}$CO$^+$ line ($-8.3$~km/s; vertical) and zero intensity (horizontal). The hyperfine components are marked with blue vertical lines.}
\label{malt90spec}
\end{figure*}

The properties of the molecular cloud associated with G351.69--1.15 are investigated using the molecular line data from MALT90 survey. Six molecular species were detected towards this region: HCO$^+$, H$^{13}$CO$^+$, HCN, HNC, N$_2$H$^+$ and C$_2$H.  In order to probe the kinematics of the cloud, we have considered the spectra of these molecules towards three different locations, that correspond to the peaks of three cold dust clumps: C4, C1 and C6. These clumps are shown in Fig.~\ref{350}(a). Our justification for considering C1 and C6 is that they lie close to the region covered by the rectangular apertures for measuring the velocities of ionized gas (Section 4.3). In addition, we consider the clump C4 as it is devoid of ionized gas emission. While the former two clumps are in active/evolved star formation stage, the latter is in an intermediate stage (Paper I). The molecular spectra towards C4, C1 and C6 (clump positions as we move from north to south respectively) are presented in Fig.~\ref{malt90spec}.

%%%%%%%%%%%%%%%%%%%% Fig11 %%%%%%%%%%%%%%%%%%%%%%%%%%%%%%%%%%%%%%%%%%%%%%%%%

%%%%%%%%%%%%%%%%%%%%  %%%%%%%%%%%%%%%%%%%%%%%%%%%%%%%%%%%%%%%%%%%%%%%%%

\par Among the six molecular species, five are detected at the locations of all the three clumps. H$^{13}$CO$^+$ is detected solely towards the clump C4. We could estimate the LSR velocity using either HCO$^+$ or H$^{13}$CO$^+$ as their spectra involve a single transition, i.e. without any hyperfine components. However, we prefer to use the latter as it is likely to be optically thin.  The LSR velocity is $\sim -8.3$~km/s estimated from the H$^{13}$CO$^+$ spectrum towards its peak emission (see Fig.~\ref{h13copfit}). The hyperfine components of HCN and N$_2$H$^+$ are clearly resolved and detected towards the three clumps. 

\subsection{Wing emission and additional velocity components in the HCO$^+$ spectrum}
We carefully examined the HCO$^+$ spectrum towards the peak positions of three clumps: C1, C4 and C6, as the spectrum towards each location reveals distinct features. Towards C4, we observe additional components in close proximity but on either side of the brightest component.  C1 shows an additional well resolved velocity component while C6 shows line emission that is characterized by broad wings. To the spectrum towards C1, we fitted two Gaussian profiles for the two components. In the case of C6, we fit a narrow as well as a broad component as a single Gaussian profile provides an inadequate fit evident from the large residuals to the spectral line. The results of the fits to the HCO$^+$ spectra of these clumps are  shown in Fig.~\ref{hcopmulti} and presented in Table~\ref{tb3} . In this section, we present the results of our analysis of these features.

\subsubsection{C4}

The profile of HCO$^+$ towards C4 shows three components as is evident from Fig.~\ref{hcopmulti}(a) and the central brightest component is broader than that observed towards the other clumps. We fit three Gaussian profiles to the spectrum  and the best fit parameters are listed in Table~\ref{tb3}. This broadening is also apparent in the HNC spectrum towards C4. The velocity of the central component is $-$8.1~km/s whereas the velocities of blue and red components are $-11.5$ and $-4.7$~km/s. Both these components are separated by $3.4$~km/s from the central peak. Many star forming regions display broad wings in the HCO$^+$ profiles that have been interpreted to be the result of molecular outflows \citep[e.g.][]{{1997ApJ...484..256G},{2010ApJ...712..674Z}}. The broadening of a spectral line could arise due to turbulence close to the protostellar object that could be the outcome of outflows, fragmentation, or unresolved velocity gradients \citep{2014A&A...565A.101T}. In addition, hydrodynamic simulations of a collapsing cloud in the absence of outflows  could also give rise to optically thick lines (e.g. HCO$^+$) that become broader with time \citep{2013ApJ...771...24S}. \citet{2002ApJ...573..215G} suggest that line broadening of molecules could also originate in flows created by photoevapourating clumps. In Paper I, we identified an extended green object (EGO-1) close to the peak emission of the clump C4 that is likely to trace a massive protostellar object. In addition, we detected a H$_2$ knot near EGO-1. We speculated that this could be the outcome of an outflow. Keeping this in view, we regard the broad features along with the secondary peaks as spectral signatures sampling the proposed outflow close to the protostellar object.

\subsubsection{C1}

Towards the clump C1, apart from the primary component there is an additional, blue shifted component of HCO$^+$. This component is also conspicuous among the spectra of HNC and C$_2$H, towards C1.  The velocity of the primary component is $-8.2$~km/s with a width of $\Delta$V$\sim2.7$~km/s. The velocity of the secondary component is $-13.3$~km/s and its width is 2.1~km/s. This line is relatively weak with an intensity that is nearly one-third of the primary feature as seen in Fig.~\ref{hcopmulti}(b). In many cases, an asymmetry in the line intensity is observed that is characteristic of infall or outflow motion \citep[e.g.][]{1994ApJ...431..767W,1997ApJ...484..256G,2008ApJ...688L..87V}. In our case, we do not observe any noticeable asymmetry in the primary line intensity. Rather, we observe a well separated blue component that 
could be a potential cloud situated along the same line-of-sight.  
Such multiple velocity components that are separated by few km/s have been interpreted as overlapping cores characteristic of high mass star forming regions \citep{2014A&A...565A.101T}.

%%%%%%%%%%%%%%%%%%%%%%%%%%%%%%%%%%%%%%%%%%%%%%%Table 3%%%%%%%%%%%%%%%%%%%%%%%%%%%%%%%%%%%%%%%%%%%%%%%%%%%%%%%%%%%%%%

\begin{table}
\footnotesize
\caption{HCO$^+$ line parameters}
\begin{center}

\hspace*{-0.5cm}
\begin{tabular}{l r r r} \hline \hline
  & & &\\
Source            & V (km/s) &   $\Delta$V$^a$ (km/s) &$\Delta$v$^b$(km/s)\\
\hline
\multirow{3}{*}{C4} & -4.72$\pm$0.08& 0.89$\pm$0.18 &3.36\\
& -8.08$\pm$0.03& 3.62$\pm$0.07 &0\\
& -11.45$\pm$0.06& 0.46$\pm$0.16 &$-3.37$\\
\hline
\multirow{2}{*}{C1}& -8.14$\pm$0.02 &2.73$\pm$0.05 &0\\
& -13.3$\pm$0.05& 2.12$\pm$0.12 &$-5.22$\\
\hline

\multirow{2}{*}{C6}&-7.92$\pm$0.02& 1.83$\pm$0.05 &0.32\\
& -8.25$\pm$0.19 & 9.61$\pm$0.54 &0\\
\hline
\end{tabular}\\
\label{tb3}
\scriptsize{$^a$ : Line width, $^b$ : separation in velocity from the brightest component}
\end{center}
\end{table}

%%%%%%%%%%%%%%%%%%%%%%%%%%%%%%%%%%%%%%%%%%%%%%%%%%%%%%%%%%%%%%%%%%%%%%%%%%%%%%%%%%%%%%%%%%%%%%%%%%%%%%%%%%%%%%%%

%%%%%%%%%%%%%%%%%%%% Fig12 %%%%%%%%%%%%%%%%%%%%%%%%%%%%%%%%%%%%%%%%%%%%%%%%%

\begin{figure}
\hspace*{-0.4cm}
\centering
\includegraphics[scale=0.35]{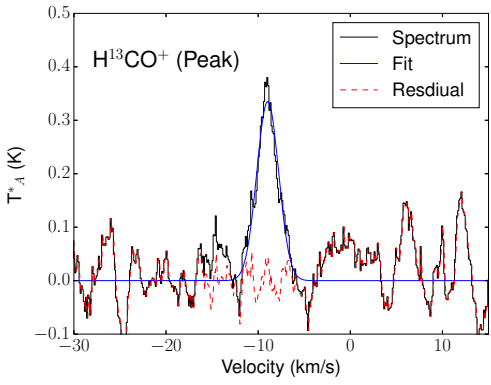} 
\caption{Spectrum of H$^{13}$CO$^+$ line towards the location of its peak ($\alpha_{J2000} =17^h$ 29$^m$02.65$^s$, $\delta_{J2000}=-36^\circ$~32$\arcmin$~57.6$\arcsec$) fitted with single component Gaussian function.}
\label{h13copfit}
\end{figure}

%%%%%%%%%%%%%%%%%%%% %%%%%%%%%%%%%%%%%%%%%%%%%%%%%%%%%%%%%%%%%%%%%%%%%

\subsubsection{C6}
\begin{figure*}
\hspace*{-0.5cm}
\centering
\includegraphics[scale=0.28]{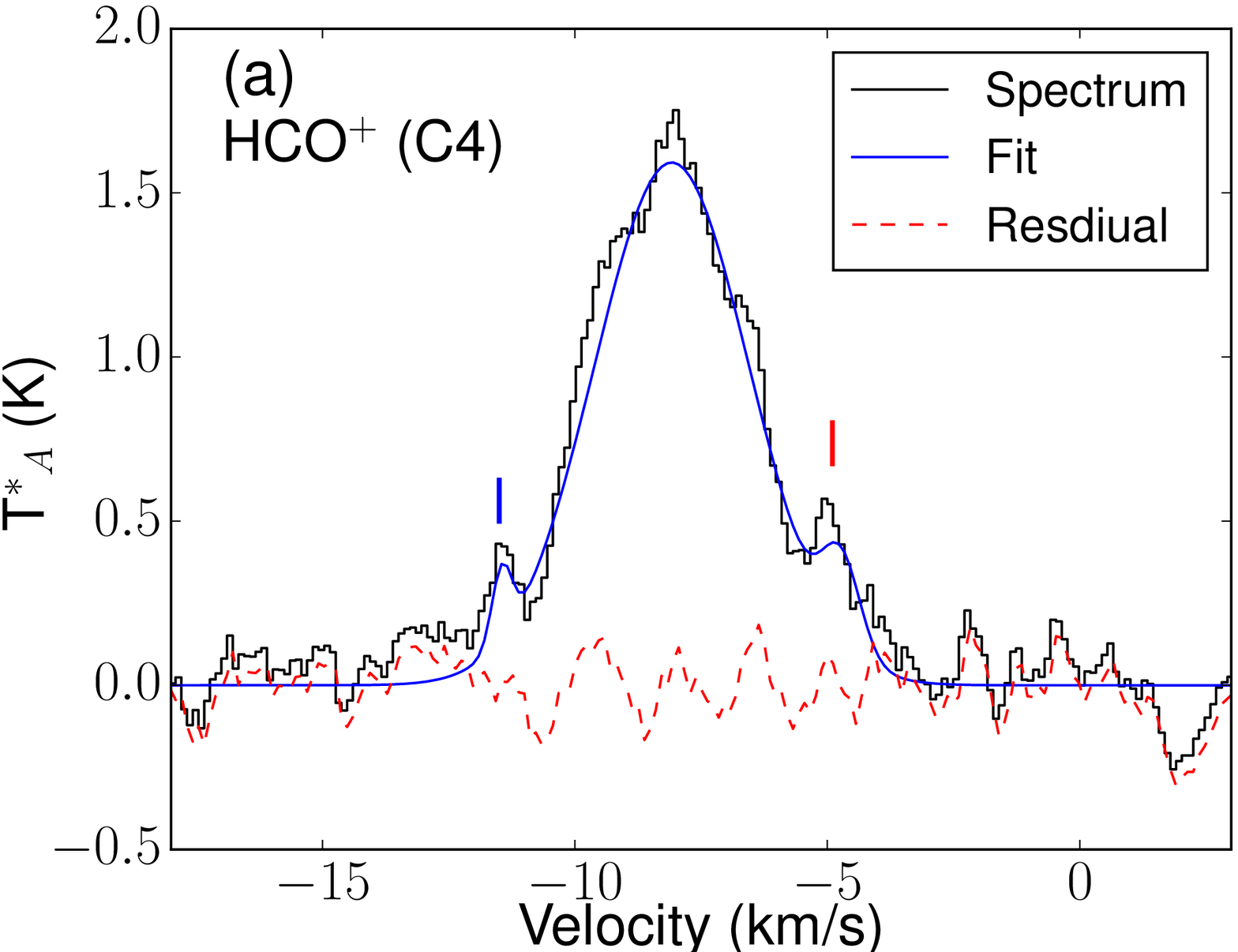} \quad \includegraphics[scale=0.28]{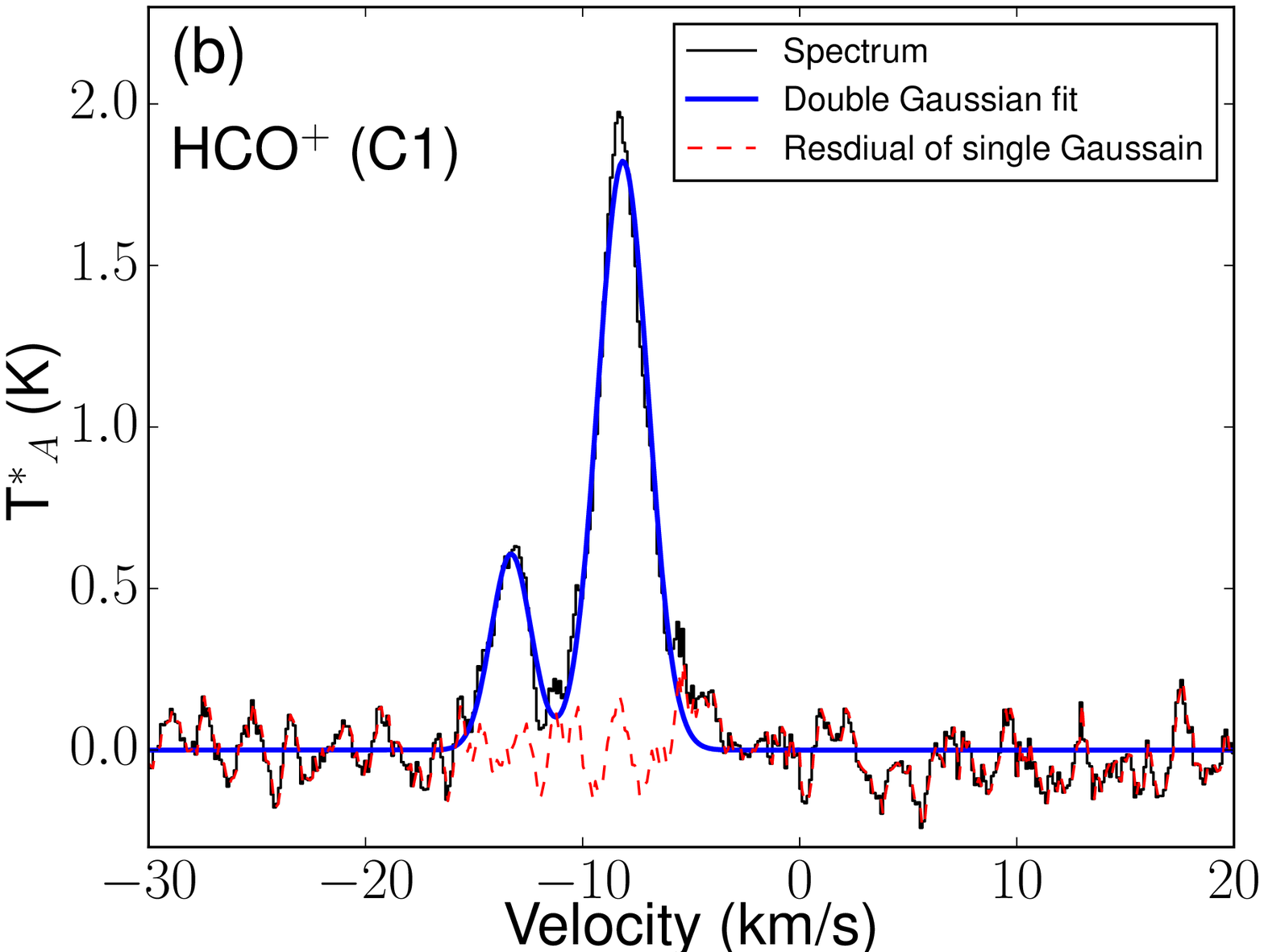}  \quad \includegraphics[scale=0.28]{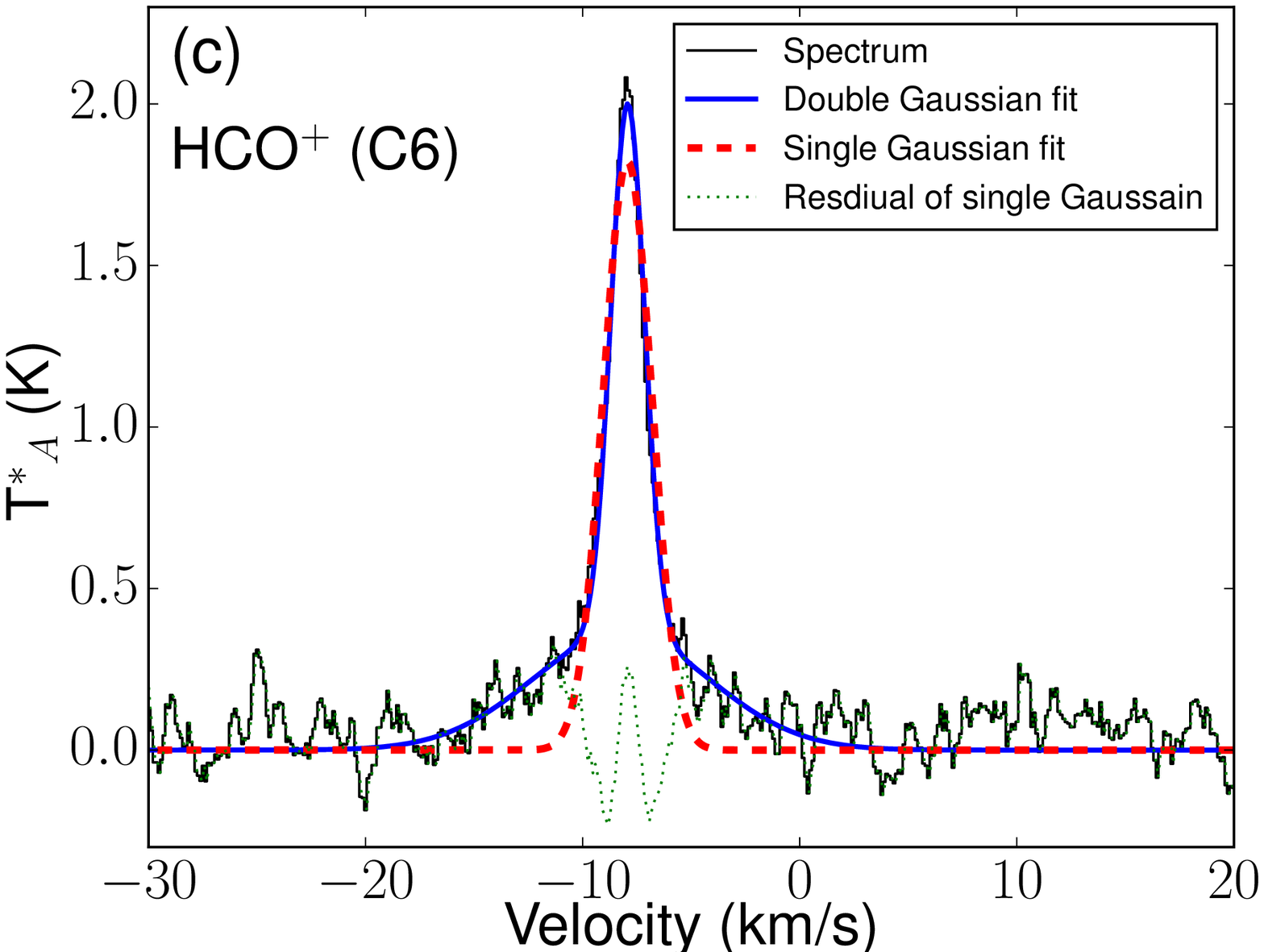}
\caption{(a) The HCO$^+$ spectrum towards peak of C4 fitted with 3 Gaussians. The red and blue peaks on either side of the central line are marked with vertical lines. (b) The HCO$^+$ spectrum towards the location C1. Spectrum is fitted with two Gaussians. (c) HCO$^+$ spectrum towards C6. Fit to the spectrum with a single Gaussian is shown with red dashed line whereas the two component Gaussian fit is represented by blue line. The residual of the single Gaussian fit (green dotted line) is also shown in the spectrum.}
\label{hcopmulti}
\end{figure*}

The HCO$^+$ spectrum towards the clump C6 displays excess emission in the line wings that cannot be fitted with a single Gaussian. In the single Gaussian fit, the residuals show a distinct pattern with positive values at the line center, and increasingly negative values towards the line wings and positive values thereafter. This pattern manifests the failure of a single Gaussian fit and has been recognized by \citet{1990ApJ...352..522D} who explain this feature as the result of an additional broad, low-intensity component. Fig.\ref{hcopmulti}(c) shows the single and double Gaussian fits to the observed profile. From the plot, it is apparent that the spectrum is better fit with the double Gaussian profile. The central velocities of the fitted components are $-$8.2 and $-$7.9~km/s respectively, and the corresponding line widths are 9.5 and 1.8~km/s, respectively. 

\par A study of molecular line profiles towards towards nearby star forming regions by \citet{1990ApJ...359..344F} showed that nearly all the profiles exhibit excess line emission towards the line wings. They argued that the excess emission is not associated with the protostellar activity, and is the direct observational signature of turbulence within the gas clouds instead. It is possible that the excess line emission towards C6 emanates due to turbulence in the molecular cloud caused by the expansion of ionized gas towards a lower density region.

\subsection{Column Densities of HCO$^+$ and H$^{13}$CO$^+$ }

We obtain an estimate of the column densities of the isotopologues HCO$^+$ and H$^{13}$CO$^+$ towards G351.69--1.15 using simplistic assumptions. Under conditions of local thermodynamic equilibrium, these are: (i) H$^{13}$CO$^+$ line is optically thin, (ii) HCO$^+$ line is optically thick, and (iii) the excitation temperatures of HCO$^+$ and H$^{13}$CO$^+$ species are the same. As the H$^{13}$CO$^+$ line has a poor SNR across the region, we consider the line intensities towards the peak of H$^{13}$CO$^+$ emission that is in the vicinity of clump C2 shown in Fig.~\ref{h13copfit}.  We proceed under the assumption that the H$^{13}$CO$^+$ profile is optically thin. We would like to remark that the column densities obtained this way would act as lower limits, in case the H$^{13}$CO$^+$ line is optically thick. The expression for the excitation temperature of an optically thick HCO$^+$ line is given below \citep{2013RAA....13...28Y}.

\begin{equation}
T_{\textrm{ex}} = \frac{h\nu_0}{k}\left[\textrm{ln}\left(1+\frac{h\nu_0/k}{T_{\textrm{max}}(\textrm{HCO}^+)+J_\nu(T_{\textrm{bg}})}\right)\right]^{-1}
\end{equation}

\begin{equation}
J_\nu(T) = \frac{h\nu_0}{k}\frac{1}{(e^{h\nu_0/k\,T}-1)}
\end{equation}

\noindent Here $\nu_0$ is the rest frequency of the HCO$^+$ line that is taken as 89188.526~MHz in the present case, $T_{\textrm{bg}}$ is the temperature of the background radiation taken as 2.73~K \citep{{2006MNRAS.367..553P},{2015MNRAS.451.2507Y}} while $h$ and $k$ represent Planck constant and Boltzmann constant, respectively.  $T_{\textrm{max}}(\textrm{HCO}^+)$ is the brighteness temperature of the maximum line intensity and is estimated using $T_{\textrm{max}}(\textrm{HCO}^+)$=T$^*_A$/$\eta_{MB}$, where T$^*_A$ is the antenna temperature corrected for atmospheric attenuation and  $\eta_{MB}$ is the main beam efficiency. We have adopted $\eta_{MB}$ as 0.49 \citep{2005PASA...22...62L} and estimated  $T_{\textrm{max}}(\textrm{HCO}^+)$ as 4.7~K. Using the above expressions and assuming a beam filling factor of unity, we estimated the excitation temperature $T_{\textrm{ex}}\sim7.7$~K. The excitation temperature is used to obtain the optical depth of the H$^{13}$CO$^+$ line in the following way \citep{2013RAA....13...28Y}.

\begin{equation}
\tau = -\textrm{ln}\left[1-\frac{T_{\textrm{max}}(\textrm{H}^{13}\textrm{CO}^+)}{[J_\nu(T_{\textrm{ex}})-J_\nu(T_{\textrm{bg}})]}\right]
\end{equation}
%%%%%%%%%%%%%%%%%%%% Fig13 %%%%%%%%%%%%%%%%%%%%%%%%%%%%%%%%%%%%%%%%%%%%%%%%%

%%%%%%%%%%%%%%%%%%%% %%%%%%%%%%%%%%%%%%%%%%%%%%%%%%%%%%%%%%%%%%%%%%%%%
\noindent The optical depth of the H$^{13}$CO$^+$ line is determined to be $\tau\sim0.18$. The total column density of the H$^{13}$CO$^+$ line, $N(\textrm{H}^{13}\textrm{CO}^+)$, is calculated using Eqn. (2) of \citet{2014A&A...562A...3M}. The H$^{13}$CO$^+$ column density is $\sim 3.2\times$10$^{13}$cm$^{-2}$. The values of various parameters used for the estimation of column density are given in Table~\ref{h13cop}. Assuming the abundance ratio [HCO$^+$]/[H$^{13}$CO$^+$] $\sim50$ \citep[e.g.][]{2015MNRAS.451.2507Y}, the HCO$^+$ column density is estimated as 1.6$\times$10$^{15}$cm$^{-2}$. We compare these values with those towards other molecular clouds. \citet{2006MNRAS.367..553P} estimated the H$^{13}$CO$^+$ and HCO$^+$ column density towards 79 massive star forming regions and their derived H$^{13}$CO$^+$ column densities lie in the range 0.5 - 14.3$\times$10$^{13}$~cm$^{-2}$ while the HCO$^+$ column densities lie in the range 2.5 - 71.4$\times$10$^{14}$~cm$^{-2}$. Our values fall well within this range. It is to be noted that although we have assumed a beam filling factor $\sim1$, if the gas is clumpy within the beam area, the true filling factor will be less than 1. In that case, the beam-averaged column density will be a lower limit to the source-averaged value.

\par We also estimated the fractional abundances of HCO$^+$ and H$^{13}$CO$^+$ molecules by dividing the molecular column density by the H$_2$ column density, $\chi$(HCO$^+$)=$N$(HCO$^+$)/$N$($\rm{H}_2$). For this, we have considered $N(\rm{H}_2)\sim1.7\times10^{22}$~cm$^{-2}$ towards clump C2 (Paper I). Based on this, the calculated fractional abundances are 9.4$\times$10$^{-8}$ and 1.9$\times$10$^{-9}$ for HCO$^+$ and H$^{13}$CO$^+$ molecules, respectively. A comparison of the HCO$^+$ abundance with that of other star forming regions is presented in detail in the following subsection. 

%%%%%%%%%%%%%%%%%%%%%%%%%%%%%%%Table 4%%%%%%%%%%%%%%%%%%%%%%%%%%%%%%%%%%%%%%%%%%%%%%%%%%%%%%%%%%%%%%%%%%%%%%%%%%
\begin{table}
\footnotesize
\caption{Line parameters for H$^{13}$CO$^+$}
\begin{center}

\hspace*{-0.5cm}
\begin{tabular}{l c } \hline \hline
 &  \\
Parameter      & Value\\
\hline\\
Frequency$^*$ (MHz) & 86754.33 \\
$E_u/k^*$ (K)& 4.16 \\
$S\mu^{2*}$ (D$^2$)& 15.2 \\
Rotational constant B$^*$ (MHz)&43377\\
$T_{\textrm{ex}}$ (K)& 7.7 \\
$\tau$ & 0.12 \\
$N(\textrm{H}^{13}\textrm{CO}^+)$ (cm$^{-2}$)& 2.3$\times$10$^{13}$\\
\hline 
\multicolumn{2}{l}{$^*$\footnotesize{Values adopted from SPLATALOGUE spectral line catalog available}}\\
\multicolumn{2}{l}{\footnotesize{at http://www.cv.nrao.edu/php/splat/}}

\end{tabular}
\label{h13cop}
\end{center}
\end{table}
%%%%%%%%%%%%%%%%%%%%%%%%%%%%%%%%%%%%%%%%%%%%%%%%%%%%%%%%%%%%%%%%%%%%%%%%%%%%%%%%%%%%%%%%%%%%%%%%%%%%%%%%%

%%%%%%%%%%%%%%%%%%%% Fig14 %%%%%%%%%%%%%%%%%%%%%%%%%%%%%%%%%%%%%%%%%%%%%%%%%

\begin{figure*}

\centering
\hspace*{-0.5cm}
\includegraphics[height=45.5mm,width=.36\textwidth]{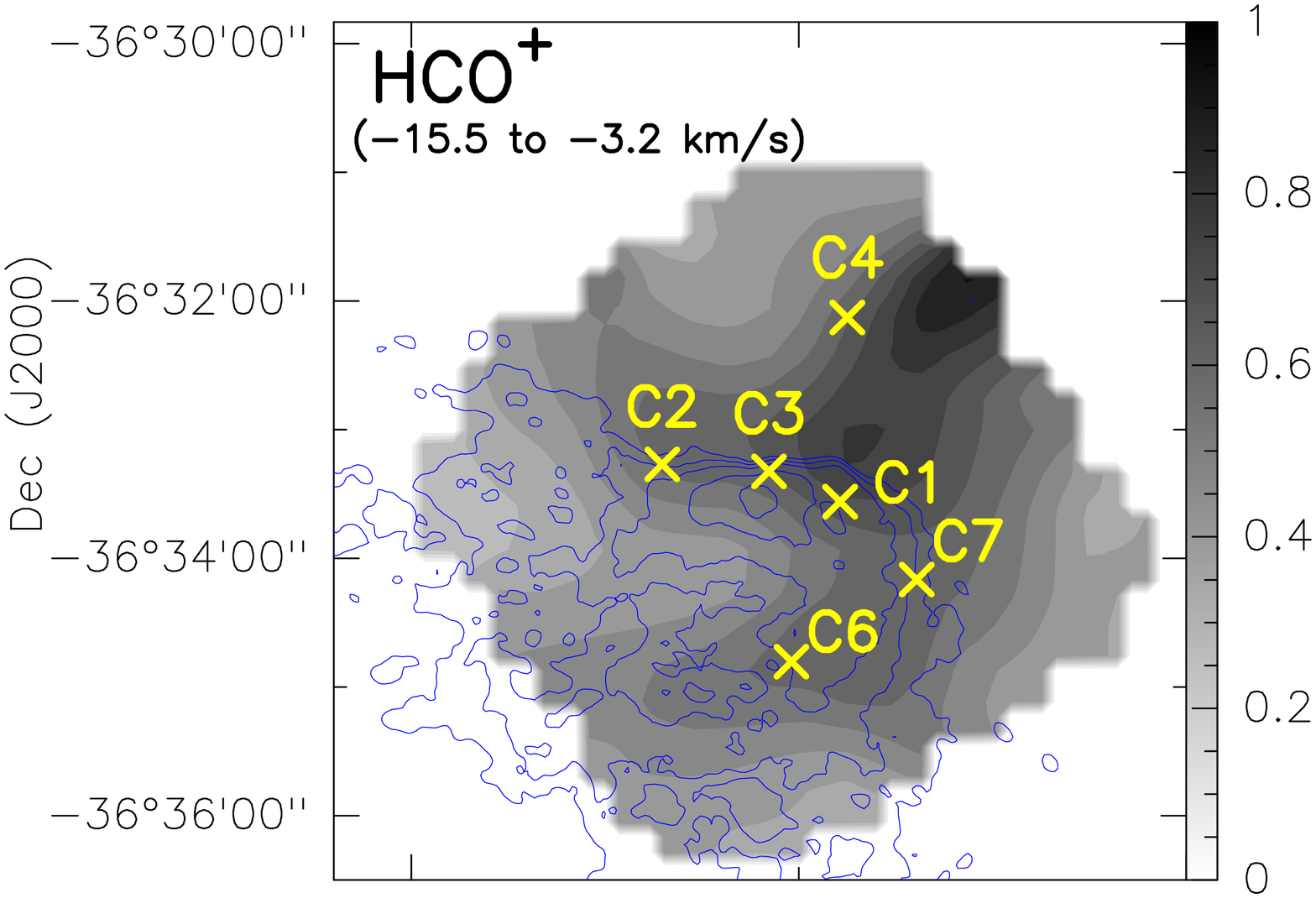}  \hspace{0.3em} \quad \includegraphics[width=.28\textwidth]{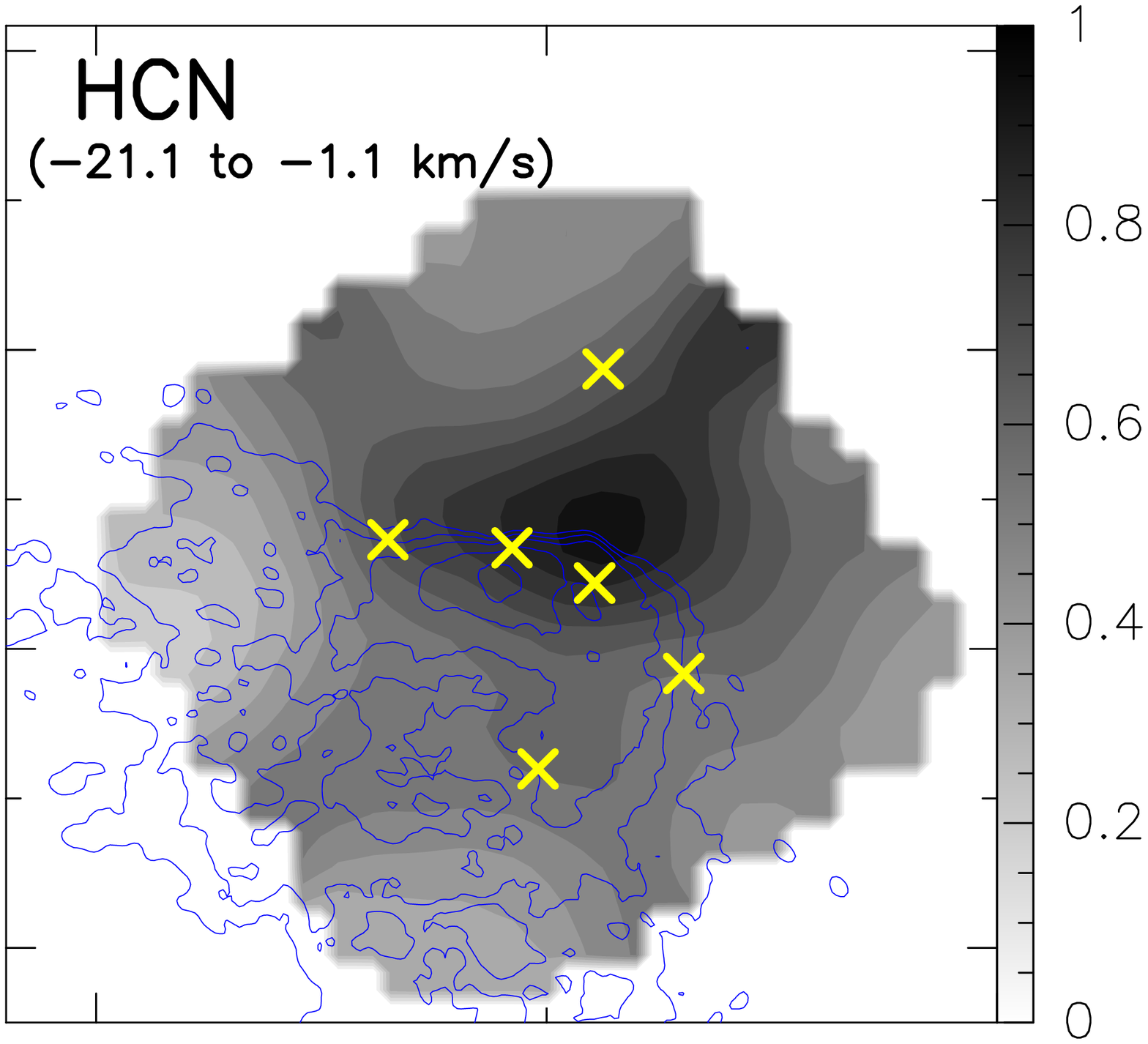} \hspace{0.3em} \quad \includegraphics[width=.28\textwidth]{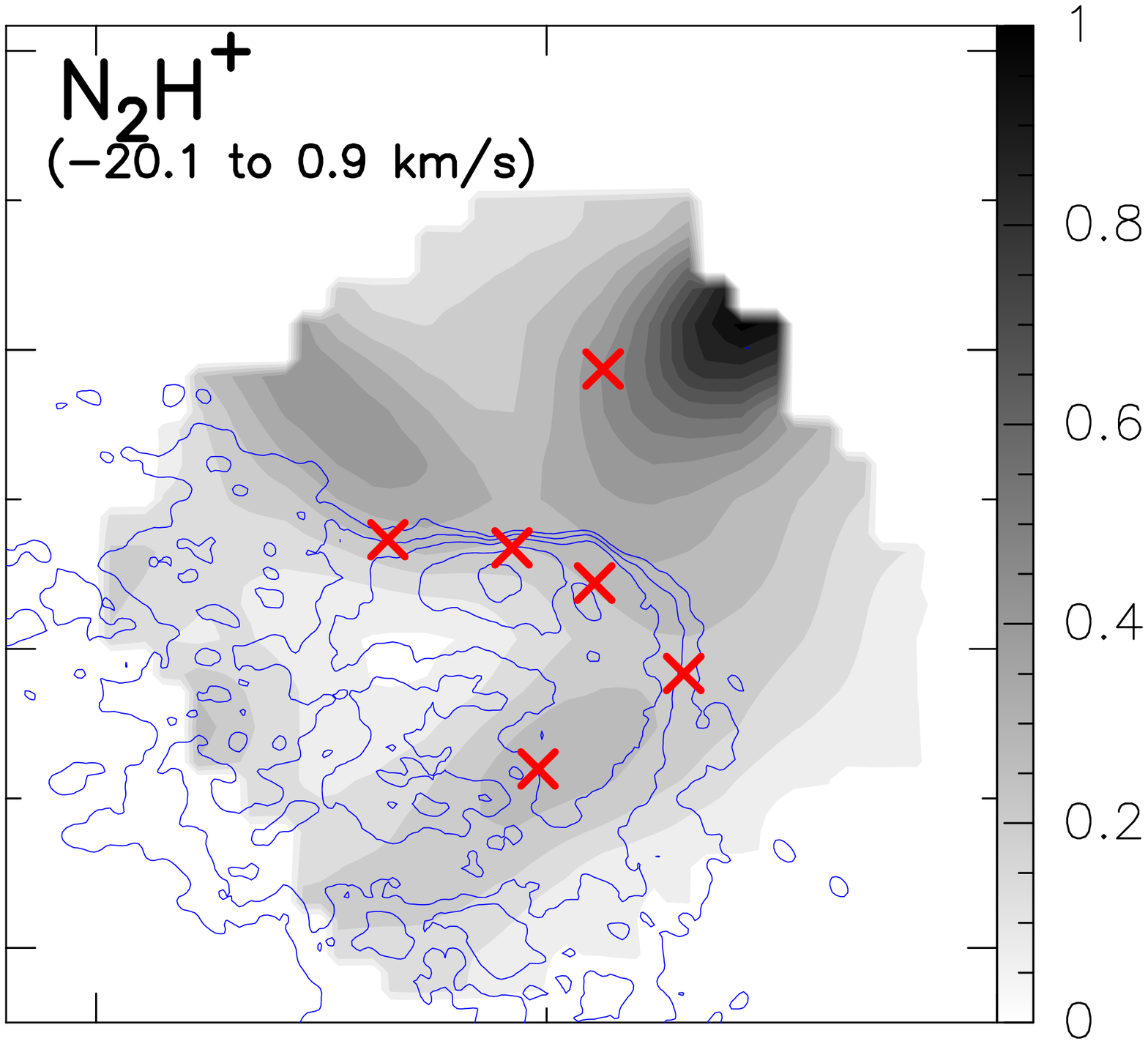} \quad 

\vspace*{0.3cm}
\hspace*{-0.5cm}
\includegraphics[height=52mm,width=.37\textwidth]{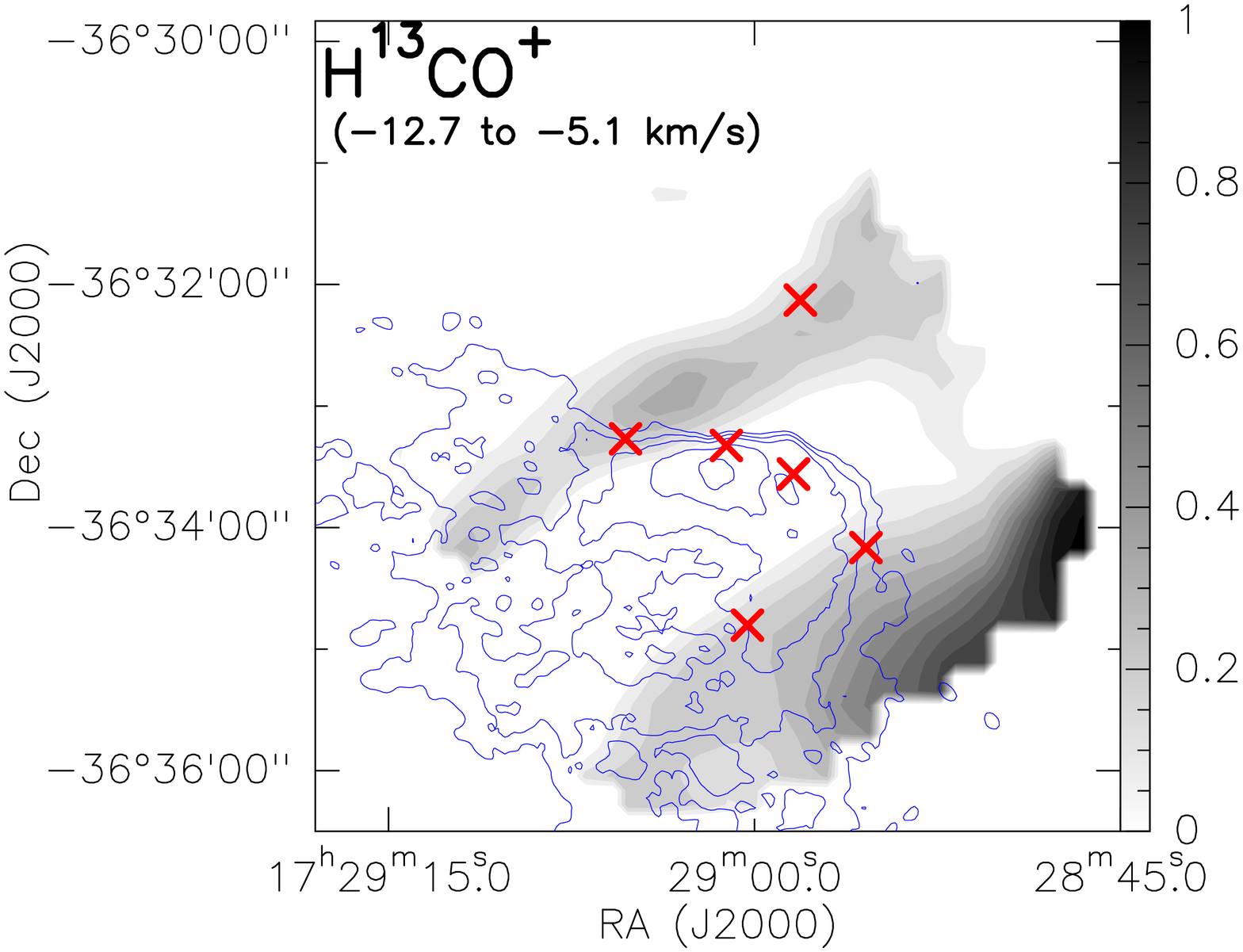}  \hspace{-0.9em} \quad \includegraphics[width=.30\textwidth]{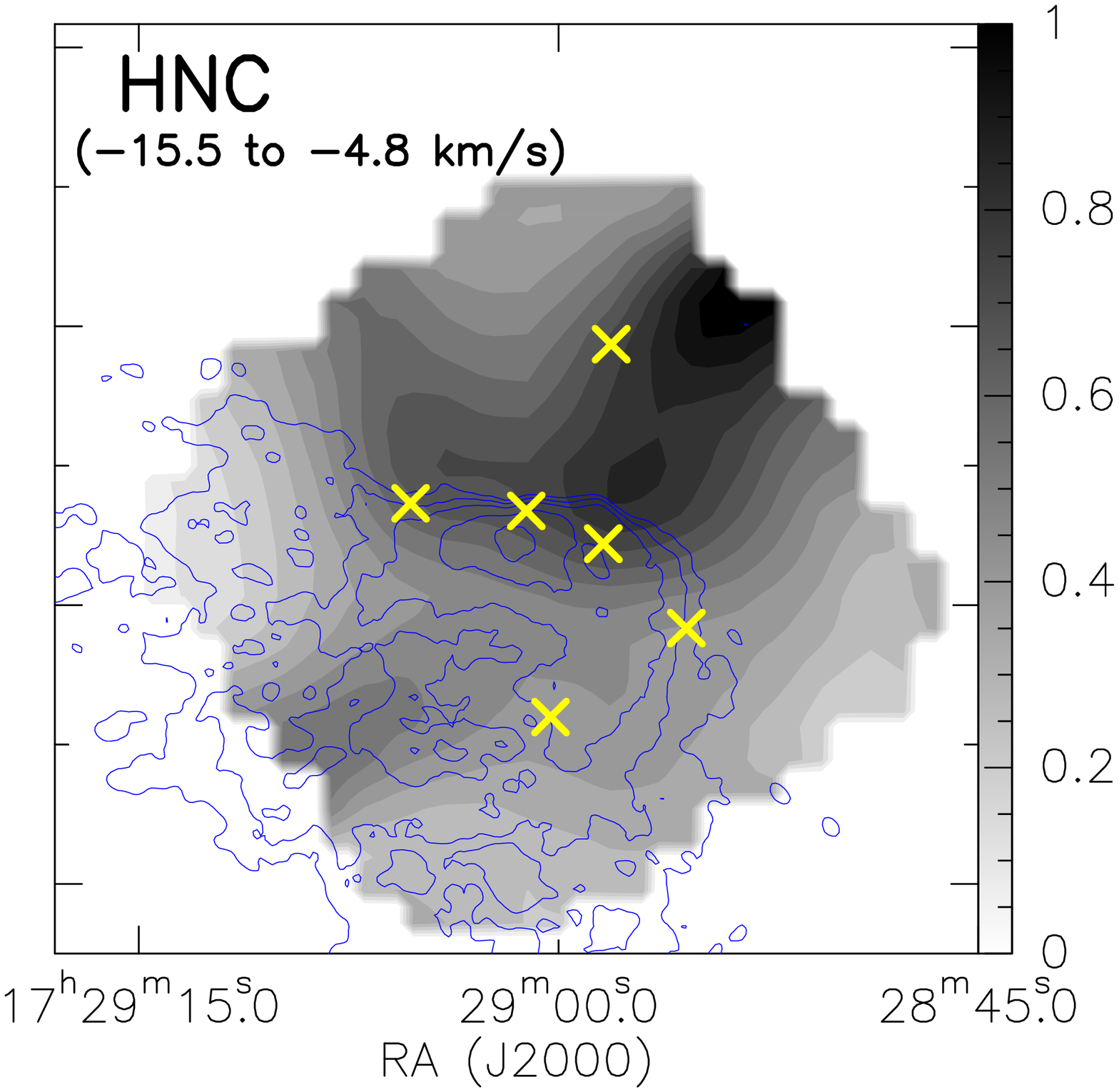} \hspace{-0.9em} \quad \includegraphics[width=.30\textwidth]{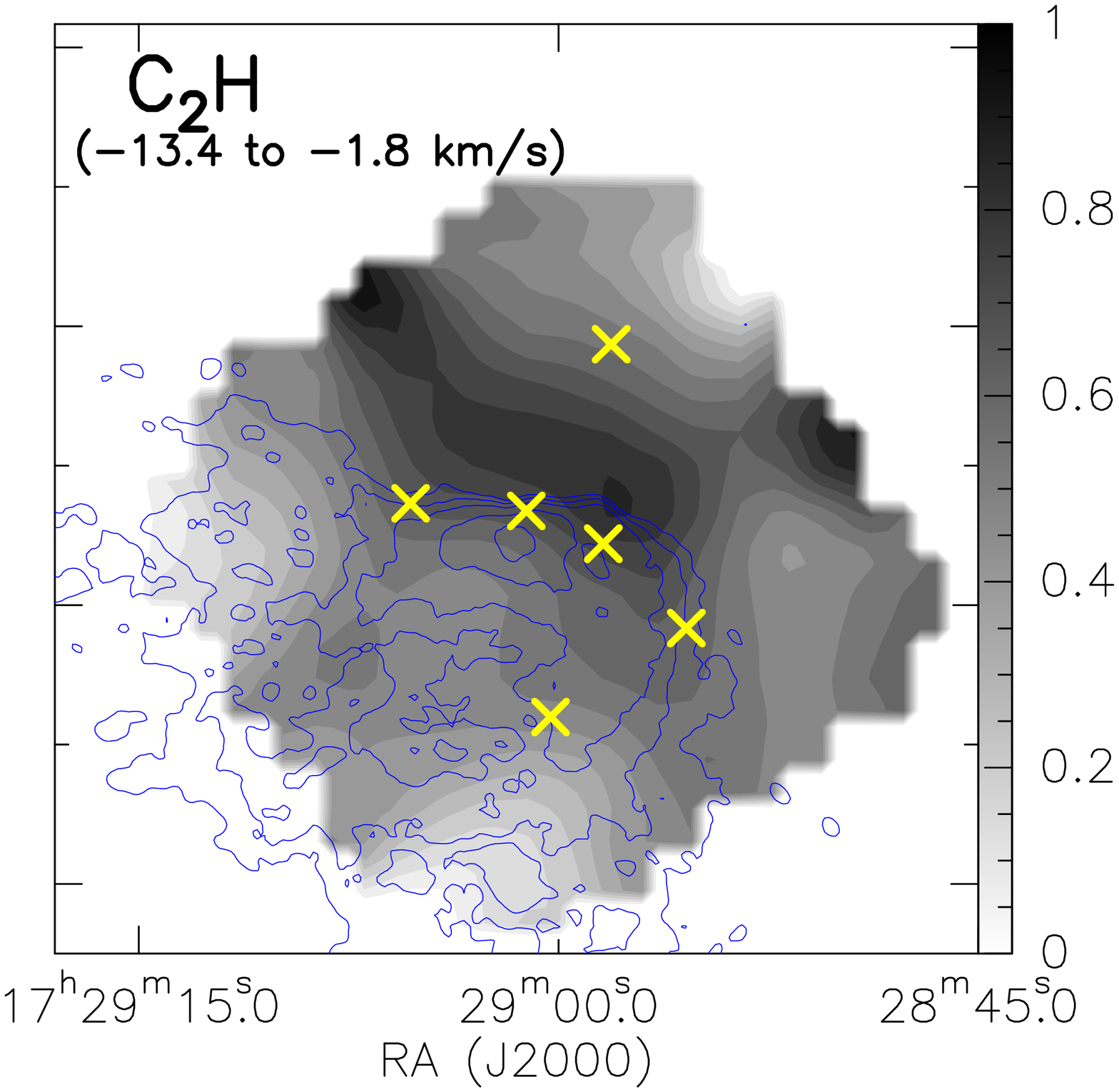}
\caption{Integrated intensity maps of MALT90 molecular species detected towards G351.69--1.15 (HCO$^+$, HCN, N$_2$H$^+$, H$^{13}$CO$^+$, HNC and C$_2$H). The levels are represented in terms of percentage of peak where the peak values for HCO$^+$, HCN, N$_2$H$^+$, H$^{13}$CO$^+$, HNC and C$_2$H molecules are 9.6, 13.7, 13.9, 3.1, 12.2 and 4.1~K~km/s respectively. The images are overlaid with 1280~MHz radio continuum contours. Also marked are the peak positions of all the clumps that are in the field of view of MALT90 data. The range of velocities used for constructing moment map of individual species are also given in the panels.}
\label{malt90mom}
\end{figure*}

%%%%%%%%%%%%%%%%%%%% %%%%%%%%%%%%%%%%%%%%%%%%%%%%%%%%%%%%%%%%%%%%%%%%%
\subsection{Molecular line emission as a chemical tracer of dense gas}

In this section we examine the spatial distribution of the molecular emission and attempt to understand the processes or conditions that the molecules trace. The integrated intensity maps of the molecular species detected towards G351.69--1.15 are presented in Fig.~\ref{malt90mom}. From the zeroth moment map, we can see that the peak emission of the molecular lines is located to the north or north-west of the \hii~region, above the cometary head. The properties and emission morphology of the individual species are discussed below.

\subsubsection{HCO$^+$ and H$^{13}$CO$^+$ (Formylium)}
\mbox{ }\\
 In G351.69--1.15, we observe extended emission in HCO$^+$ towards north-west of the radio emission. The emission shows two peaks and one of these lie close to the radio peak (angular separation ~50$\arcsec$). This peak lies near the brightest cold dust clump C1. The second peak lies in the vicinity of clump C4. We also detect diffuse emission that is extended towards south-west. The HCO$^+$ ion is primarily formed by the gas-phase ion-neutral interaction H$_3^+$~+~CO~$\rightarrow$~HCO$^+$~+~H$_2$ \citep{1973ApJ...185..505H} and gets destroyed by the dissociation recombination reaction HCO$^+$~+~e$^-$~$\rightarrow$~CO~+~H \citep{2007SerAJ.175....1N}. This line is believed to be a good infall and outflow tracer \citep[e.g.,][]{{2001A&A...376..271C},{2004MNRAS.351.1054R},{2005A&A...442..949F}}. The HCO$^+$ abundance is found to increase in regions where shocks are generated, such as protostellar objects associated with outflows \citep{2014A&A...562A...3M}. \citet{2012ApJ...756...60S} and \citet{2013ApJ...777..157H} studied a sample of molecular clumps and concluded that HCO$^+$ abundance also increases as a function of evolution. In Section 5.3, we have estimated the HCO$^+$ abundance $\chi$(HCO$^+$) as 9.4$\times$10$^{-8}$ towards clump C2, that is classified as an active/evolved clump (Paper I). This is comparable to the median value of 5.7$\times$10$^{-8}$ determined by \citep{2012ApJ...756...60S} for a sample of evolved clumps in 37 infrared dark clouds. The median values of $\chi$(HCO$^+$) for intermediate and quiescent clouds are lower by a factor of 3 than our estimate implying that C2 is at a later evolutionary stage as proposed in Paper I.
 
\par The formation mechanism of H$^{13}$CO$^+$ is similar to that of HCO$^+$ where there is $^{13}$CO instead of $^{12}$CO. It can also be formed through the isotope transfer process HCO$^+$~+~$^{13}$CO~$\rightarrow$~H$^{13}$CO$^+$~+~$^{12}$CO \citep{2014A&A...562A...3M}. It is a high density gas tracer and is generally assumed to be optically thin. H$^{13}$CO$^+$ is mainly destroyed by the fast dissociative recombination with electrons H$^{13}$CO$^+$~+~e$^-$~$\rightarrow$~$^{13}$CO~+~H \citet{2009A&A...498..771G}. We detected a weak H$^{13}$CO$^+$ emission towards G351.69--1.15.  The maximum is seen towards the west of the clumps C7 and C6 and could be a result of edge effects. We detect a peak near clump C2. A local peak is also observed towards clump C4. In the H $^{13}$CO$^+$ spectrum towards C4, we see a double peaked profile that could be due to two components or due to a self-absorbed profile. This could also arise from noise effects. There is no emission towards the radio continuum peak unlike the HCO$^+$ counterpart.

\par A comparative study on the fractional abundance of HCO$^+$ and H$^{13}$CO$^+$ towards shielded cores and UV irradiated cloud edges in the Horsehead nebula shows that the abundances towards PDR regions are a factor of 4 lower than that in the shielded cores \citep{2009A&A...498..771G}. Towards G351.62--1.15, the apparent lack of H$^{13}$CO$^+$ near the cometary head and interior of the \hii~region could be attributed to the destruction of this species by UV radiation and high density electrons ($n_e\sim$10$^2$ cm$^{-3}$). The HCO$^+$ molecule on the other hand, is relatively strong (by a factor of $\sim$50), hence it is still detected within the \hii~region.

\subsubsection{HCN (Hydrogen cyanide) and HNC (Hydrogen isocyanide)}
 HCN and HNC have hyperfine structures with three hyperfine components due to the nuclear quadrupole moment of $^{14}$N.  The HCN molecule and its metastable geometrical isomer HNC are mainly formed through the dissociative recombination reaction HCNH$^+$~+~e$^-$~$\rightarrow$~HCN~+~H or HNC~+~H \citep{1978ApJ...222..508H}. 
The distribution of HCN emission is similar to HCO$^+$ but we discern a single peak that is near clump C1. Most of the emission is concentrated towards north of the \hii~region above the clumps C1, C2 and C3. We detect weak emission towards the radio peak with the emission displaying a southward extension. The overall morphology of HNC emission is similar to that of HCN. In the integrated intensity map, we also detect a second peak towards C4 similar to the case of HCO$^+$.  
HCN is abundant in gas phase and found extensively in the cold central regions of star forming cores \citep{2012MNRAS.420.1367L}. HNC is enhanced in very cold gas and both these molecules are tracers of infall and outflows in star forming regions \citep[e.g.,][]{{2004ApJ...612..342A},{2007ApJ...664..928S},{2013prpl.conf1B081S}}. HCN and HNC are thought to be correlated and previous studies indicate that the ratio [HNC]/[HCN] is close to or slightly higher than unity (0.6--4.5) towards various molecular clouds \citep{{1998ApJ...503..717H},{2000ApJ...542..870D}}. 

\subsubsection{N$_2$H$^+$ (Diazenylium)} 
 N$_2$H$^+$ has been regarded as a dense gas tracer in early stages of star formation scenario.  It has 15 hyperfine transitions out of which 7 have different frequencies. Three of the hyperfine components are resolved in high mass star forming regions owing to their turbulent line widths \citep{2015MNRAS.451.2507Y}. It is mainly formed through the gas phase interaction H$_3^+$~+~N$_2$~$\rightarrow$~N$_2$H$^+$~+~H$_2$. N$_2$H$^+$ is more resistant to freeze-out on grains compared to carbon bearing species \citep{2007ARA&A..45..339B} and is mainly destroyed in electron recombination process N$_2$H$^+$~+~e$^-$~$\rightarrow$~N$_2$~+~H or NH~+~N \citep{2014A&A...562A...3M}. The morphology of N$_2$H$^+$ emission towards G351.69--1.15  is mostly distributed north of the cometary \hii~region although it differs in the pattern of peak emission with the other species discussed. We see a a strong peak near C4 similar to what we observe towards HCO$^+$ and HNC molecules. However, the peak towards C1 observed in the HCO$^+$, HCN and HNC lines is not perceived. In addition, we see another peak near clump C2 like the H$^{13}$CO$^+$. As N$_2$H$^+$ and H$^{13}$CO$^+$ molecules trace the optically thin, high density gas, the seconday peak near C2 implies the presence of high density material in the proximity of C2. We also identify a weak, local enhancement in N$_2$H$^+$ emission near the clump C6.
\let\cleardoublepage\clearpage
\subsubsection{C$_2$H (Ethynyl)}

The C$_2$H emission towards our region of interest displays a different morphology from the other species. However, a common attribute  is the detection of peak emission close to C1. We also detect two additional peaks to the north-west and north-east of C1, that are not correlated with the dust clumps or emission from the other species. C$_2$H is believed to form through the photodissociation of acetylene molecule C$_2$H$_2$~+~$h\nu$~$\rightarrow$~C$_2$H~+~H \citep{1993A&A...276..473F} and is considered to be a good tracer of PDRs \citep{2012A&A...543A..27G}. However, \citet{2008ApJ...675L..33B} suggested that C$_2$H could trace dense gas in the early stages of star formation but is converted into other molecules in the hot-core phase.
%%%%%%%%%%%%%%%%%%%% Fig15 %%%%%%%%%%%%%%%%%%%%%%%%%%%%%%%%%%%%%%%%%%%%%%%%%
\begin{figure}
%\hspace{-2.3cm}

%\centering
\includegraphics[trim=4cm 5cm 1cm 5cm,width=0.6\textwidth]{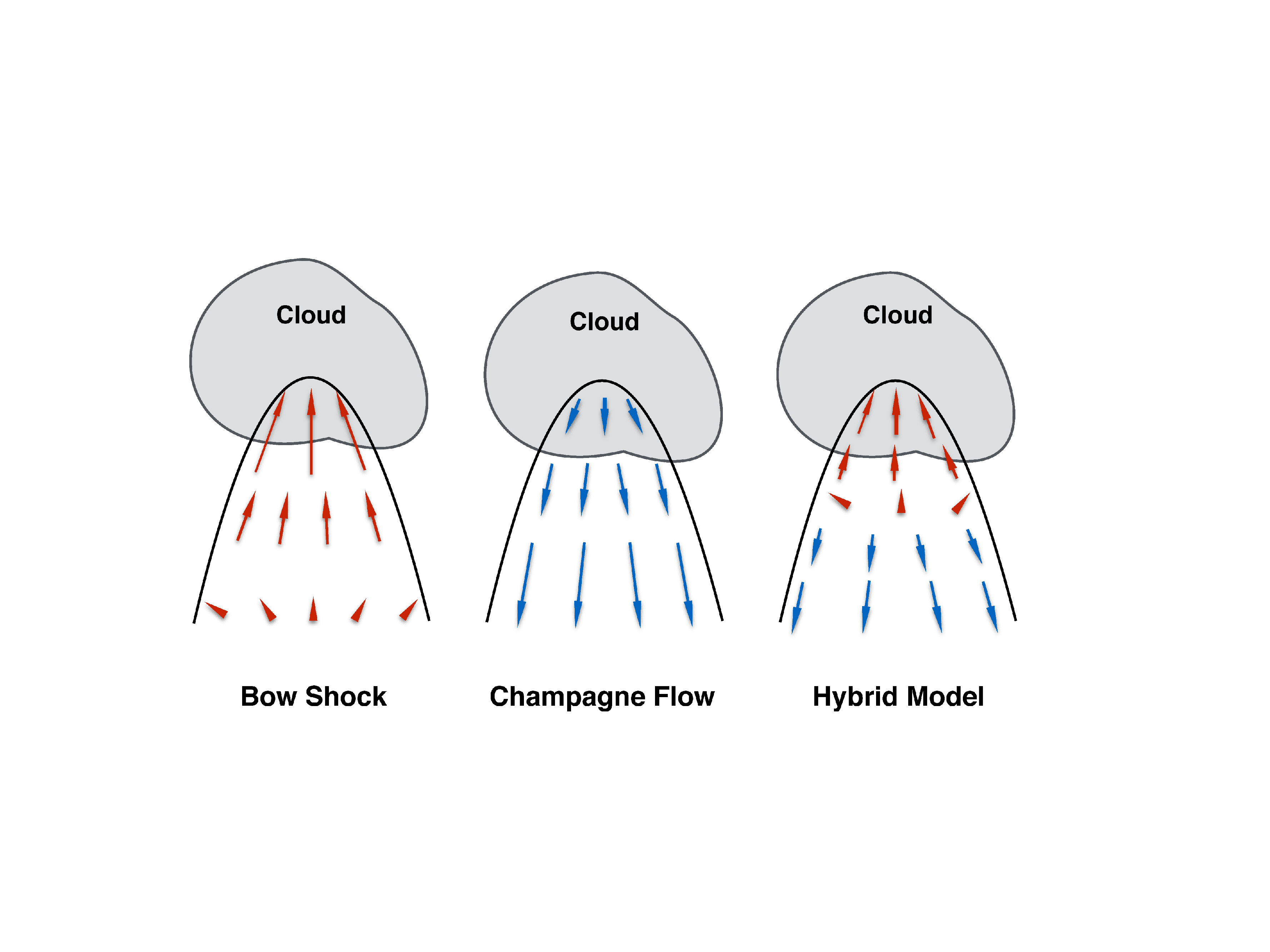}
\caption{Schematic representations of velocity fields in cometary \hii~regions for three models : (left) bow shock, (middle) champagne flow, (right) hybrid model \citep[adapted from][]{2003ApJ...596..344C}. The velocities are with respect to that of ambient molecular cloud velocity.}
\label{models}
\end{figure}

%%%%%%%%%%%%%%%%%%%% %%%%%%%%%%%%%%%%%%%%%%%%%%%%%%%%%%%%%%%%%%%%%%%%%
 \citet{2013ApJ...773..123S} mapped the C$_2$H emission towards G028.23--00.19 and found that the emission predominantly comes from the cold, central part of the molecular cloud rather than the PDR dominated outer layers. 
In the G351.69--1.15 region, as the peak near C1 is observed with high density gas tracers such as HCO$^+$, HCN and HNC, we assert that the C$_2$H emission close to C1 could possibly be originating from the molecular cloud and not from the PDR.  Also, most of the emission is distributed to the north of the cometary head and reduces considerably as we move further north. We interpret that this emission close to the ionization front  could have contributions from both, the dense molecular gas as well as the PDR. \\

\par In a nutshell, the six different molecular species identified in this region are distributed mostly towards the north and north-west of the radio peak consistent with the cold dust emission detected towards this region. Most of these molecules are high density tracers and the emission is clumpy in nature. This further confirms the presence of local density inhomogeneities within this region. The high density cloud material appears to constrict the flow of ionized gas beyond the cometary head. The ionized plasma is more extended in the regions of low molecular gas density i.e., towards the south-east. We compare the kinematics and distribution of molecular gas  with that of ionized gas in the next section.

\section{KINEMATIC MODELS OF THE \hii~REGIONS}
\subsection{Kinematics of G351.69--1.15: Bow shock or Champagne flow?}

In this section, we analyze the velocity structure of G351.69--1.15 using the RRL as well as molecular line data. We correlate the kinematics with the existing models to gain an insight into the origin of the cometary morphology. The velocity structures of  ionized and molecular gas across the cometary axis are disparate in the prominent models.
The champagne flow model \citep{{1979A&A....71...59T},{1983A&A...127..313Y}} explains the cometary morphology as an outcome of the asymmetrical expansion of the \hii~region within a dense cloud, towards a region of lower gas density. In this model, as the ionizing star is stationary, the velocity of cometary head is expected to be the same as that of the molecular cloud whereas a gradual increase in velocity is expected towards the cometary tail. The ionized gas can attain large velocities of the order of $\sim$30~km/s due to acceleration by a strong pressure as if through the nozzle \citep{1979ApJ...233...85B}.

\par In the bow shock model \citep{{1990ApJ...353..570V},{1991ApJ...369..395M}}, the cometary morphology is explained as bow shocks of ionizing stars with strong stellar winds moving supersonically through the molecular cloud. In this model, as the star is moving at large velocities with respect to the molecular cloud, the velocity of the cometary head is expected to be high. As we move towards the tail region, the ionized gas velocity is expected to be similar to the velocity of the molecular cloud. Thus, the velocity structure in this case is the reverse of what is expected for the champagne flow model. We have presented the schematic of velocity fields in these models in Fig.~\ref{models} \citep{2003ApJ...596..344C}. 

%%%%%%%%%%%%%%%%%%%% Fig16 %%%%%%%%%%%%%%%%%%%%%%%%%%%%%%%%%%%%%%%%%%%%%%%%%

\begin{figure}
\vspace{7cm}
\hspace{-0.1cm}
\centering
\includegraphics[scale=0.35]{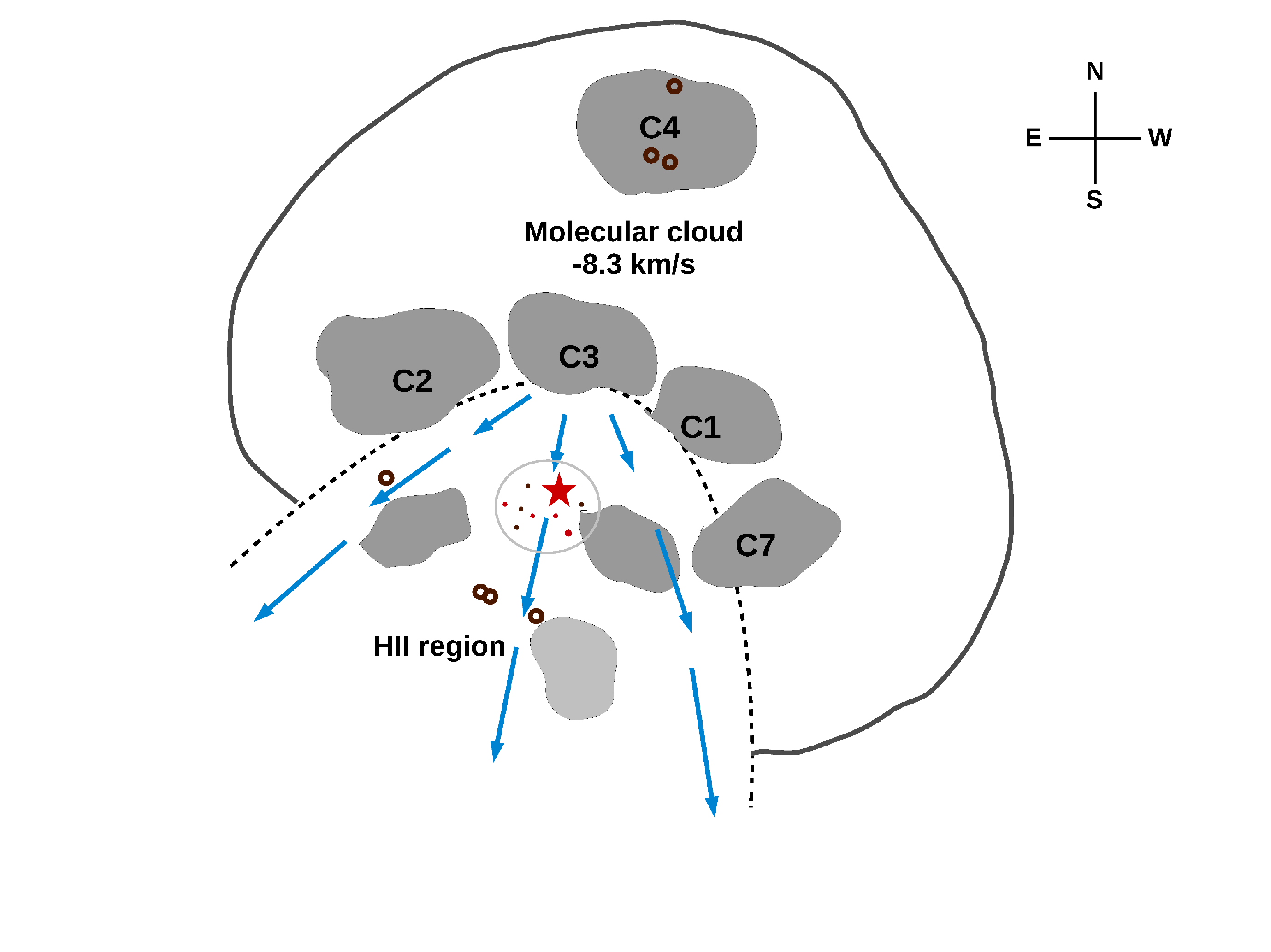}
\vspace{-0.9cm}
\caption{The champagne flow model for G351.69--1.15. The velocity field of RRL line with respect to molecular cloud is represented with blue arrows within the cometary \hii~region (dotted line). The molecular cloud with a central velocity of $-8.3$~km/s is shown (solid line) along with the locations of high density clumps (grey). Open circles mark the positions of detected H$_2$ knots (see Paper I). The dots represent the near-infrared cluster along with an asterisk denoting IRS-1, the brightest star in the cluster. The image is not to scale.}
\label{17256model}
\end{figure}

%%%%%%%%%%%%%%%%%%%%% %%%%%%%%%%%%%%%%%%%%%%%%%%%%%%%%%%%%%%%%%%%%%%%%%

\par Apart from the pure bow shock and champagne flow models, there are \textquotedouble{hybrid} models that incorporate the effect of stellar winds into existing models \citep{{1994ApJ...432..648G},{2006ApJS..165..283A}}. \citet{2006ApJS..165..283A} present the radiation-hydrodynamic simulations for various conditions such as pure bow shocks, pure champagne flows with steep and shallow density gradients, champagne flows with stellar winds, champagne flows with stellar winds and stellar motions etc. These models predict different velocity structures as compared to bow shock and champagne flow models. \citet{2003ApJ...596..344C} investigated the gas kinematics towards the cometary \hii~regions in DR21 region and they found large velocities at the cometary heads suggestive of bow shocks. They also reported an increase in velocity towards the ionized tail of the southern \hii~region, that is consistent with the champagne flow model. They concluded that the gas kinematics is indicative of a \textquotedouble{hybrid} model where gas flow is bow shock-like in the cometary head whereas it is champagne flow-like in the cometary tail. This hybrid model scenario was further confirmed with sensitive H66$\alpha$ observations by \citet{2014A&A...563A..39I}. The representative hybrid velocity structure is also displayed in Fig.~\ref{models}.

\par The MALT90 molecular line data shows that the velocity of the HCO$^+$ molecular line is $-8.3$~km/s. This is consistent with velocities of C$^{17}$O~(1--0) and C$^{17}$O~(2--1) lines \citep[$-8.1$ and $-8.5$~km/s respectively;][]{2005A&A...432..921F}. Our  analysis of RRL velocities using apertures (Section 4.3) shows that there exists a velocity gradient across the cometary axis of G351.69--1.15. 
The molecular line data on the other hand shows a nearly constant velocity across the entire cloud. The velocity of the molecular medium is consistent with the velocity of ionized gas near the cometary head ($-8.6$~km/s) from aperture S1. As we move towards the tail i.e. towards apertures S2, S3, etc, there is an increase in the velocity of ionized gas. Near aperture S5, the velocity difference between ionized and molecular gas is as large as $\sim$8~km/s (from the aperture analysis). This favors the champagne flow model. This corroborates with our results from Paper I. While the models predict velocity gradients as large as 30~km/s between the cometary head and tail in champagne flows \citep{1979ApJ...233...85B}, other observational studies have reported velocity gradients of the order of $\sim$10~km/s \citep[e.g.][]{1994ApJ...429..268G} towards cometary \hii~regions.

\par We would like to note that although we have attributed the cometary morphology to a simple champagne flow model, we cannot neglect the possibility of a stellar wind emanating from the ionizing source. In our earlier work, we detected near-infrared H$_2$ emission near the cometary head, believed to be signatures of shocked gas (see Paper I). This could be due to the stellar wind interacting with the ambient medium. However, the possibility of shocked gas due to the expansion of the \hii~region cannot be ruled out. 
As the velocity information from RRL close to the cometary head suffers from poor SNR, it is difficult to distinguish between these two possibilities and {further investigation requires sensitive, high angular resolution information}. 

\par A schematic that represents our understanding for the formation of the cometary \hii~region towards G351.69--1.15 is shown in Fig.~\ref{17256model}. This scenario is consistent with what was speculated in Paper I. Here, we briefly explain this using a figure. The ionizing star IRS1, shown in Fig.~\ref{350}(a) and depicted as an asterisk is believed to be responsible for the \hii~region (Paper I). The expansion of the \hii~region is hindered by the presence of dense clumps such as C2, C3, C1 and C7 towards the northern region. The velocity of the molecular and ionized gas are similar here. Towards the south, the density of molecular gas is low, that is evident from the dust continuum as well as the molecular line emission maps. The infrared cluster that lies in this region is also shown in the schematic.
As the ionized gas expands out, it expands to regions of lower density, towards the south-east where there are fewer clumps than that of the head region. The freely flowing ionized gas accelerates here leading to relative velocities as high as 12~km/s (evident from the first moment map). We have also marked on the figure, the approximate location of few H$_2$ knots detected in this region. Thus, the density gradients drive the formation of the cometary head of the \hii~region towards G351.69--1.15. 

%%%%%%%%%%%%%%%%%%%% Fig17 %%%%%%%%%%%%%%%%%%%%%%%%%%%%%%%%%%%%%%%%%%%%%%%%%
\begin{figure}
\hspace{-0.5cm}
\centering
\includegraphics[scale=0.4]{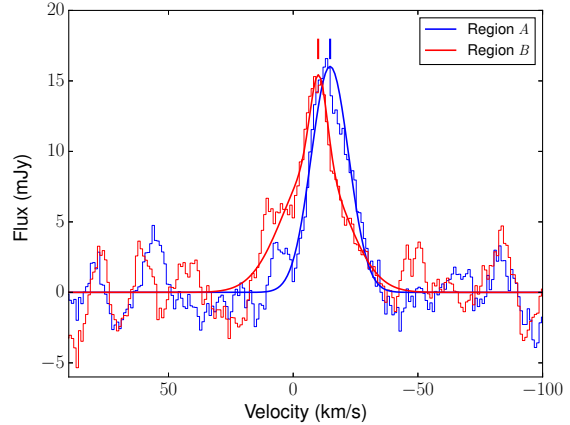} 
\caption{Overlay of the H172$\alpha$ RRL spectrum towards G351.63--1.25 $\textit{A}$ and $\textit{B}$ (see Fig.~\ref{17258spec}). To compare the spectra of the two regions, we have scaled the spectrum of region $\textit{B}$ by a factor of 3.2. The central velocities of both the regions are marked with vertical lines. The velocities of the two regions differ by 4.8~km/s. }
\label{17258double}
\end{figure}

%%%%%%%%%%%%%%%%%%%%  %%%%%%%%%%%%%%%%%%%%%%%%%%%%%%%%%%%%%%%%%%%%%%%%%
%%%%%%%%%%%%%%%%%%%% Fig18 %%%%%%%%%%%%%%%%%%%%%%%%%%%%%%%%%%%%%%%%%%%%%%%%%
\begin{figure*}
\hspace{-1.8cm}
\centering
\includegraphics[scale=0.34]{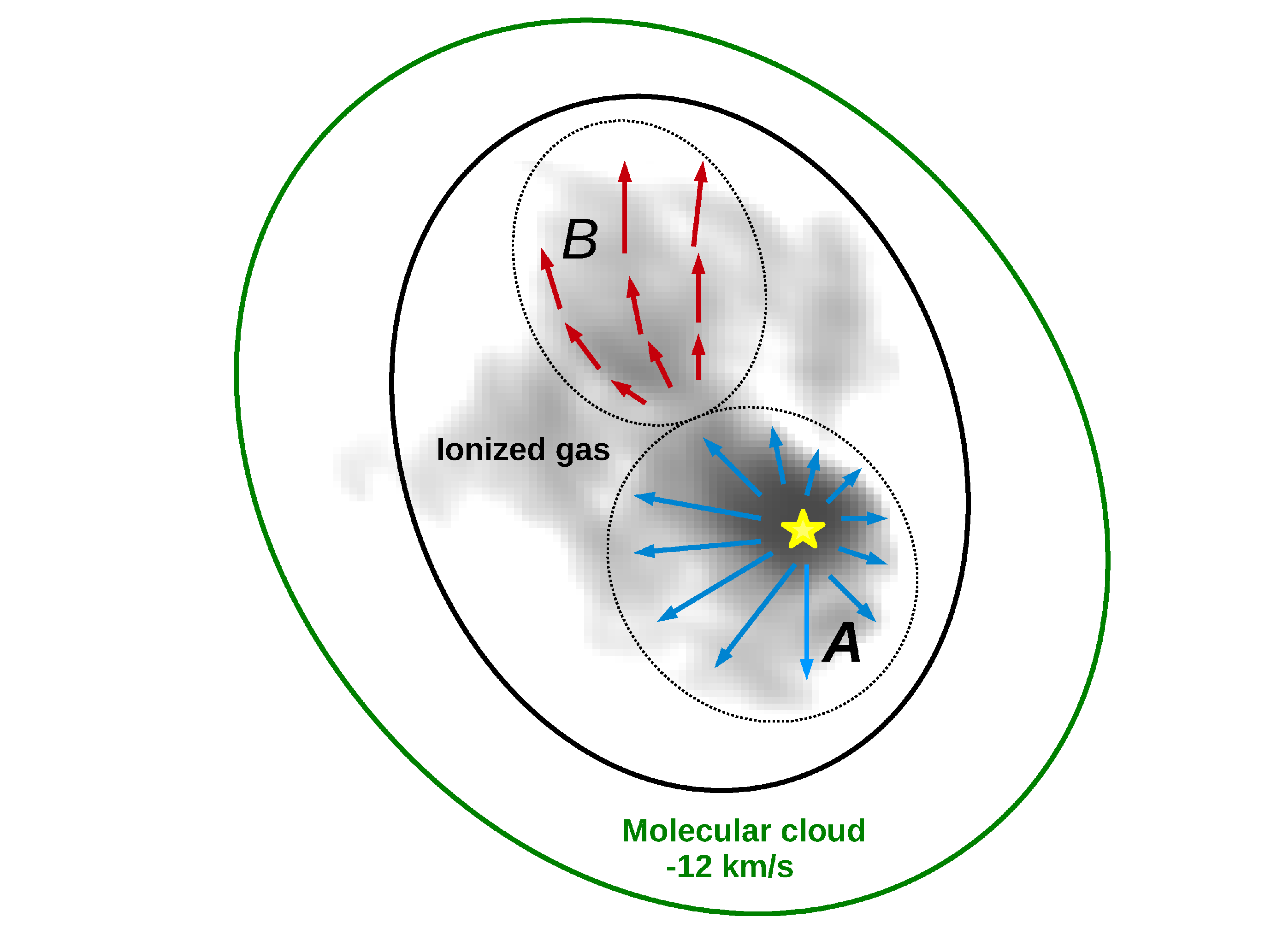} \quad \hspace{-1.5cm} \includegraphics[scale=0.38]{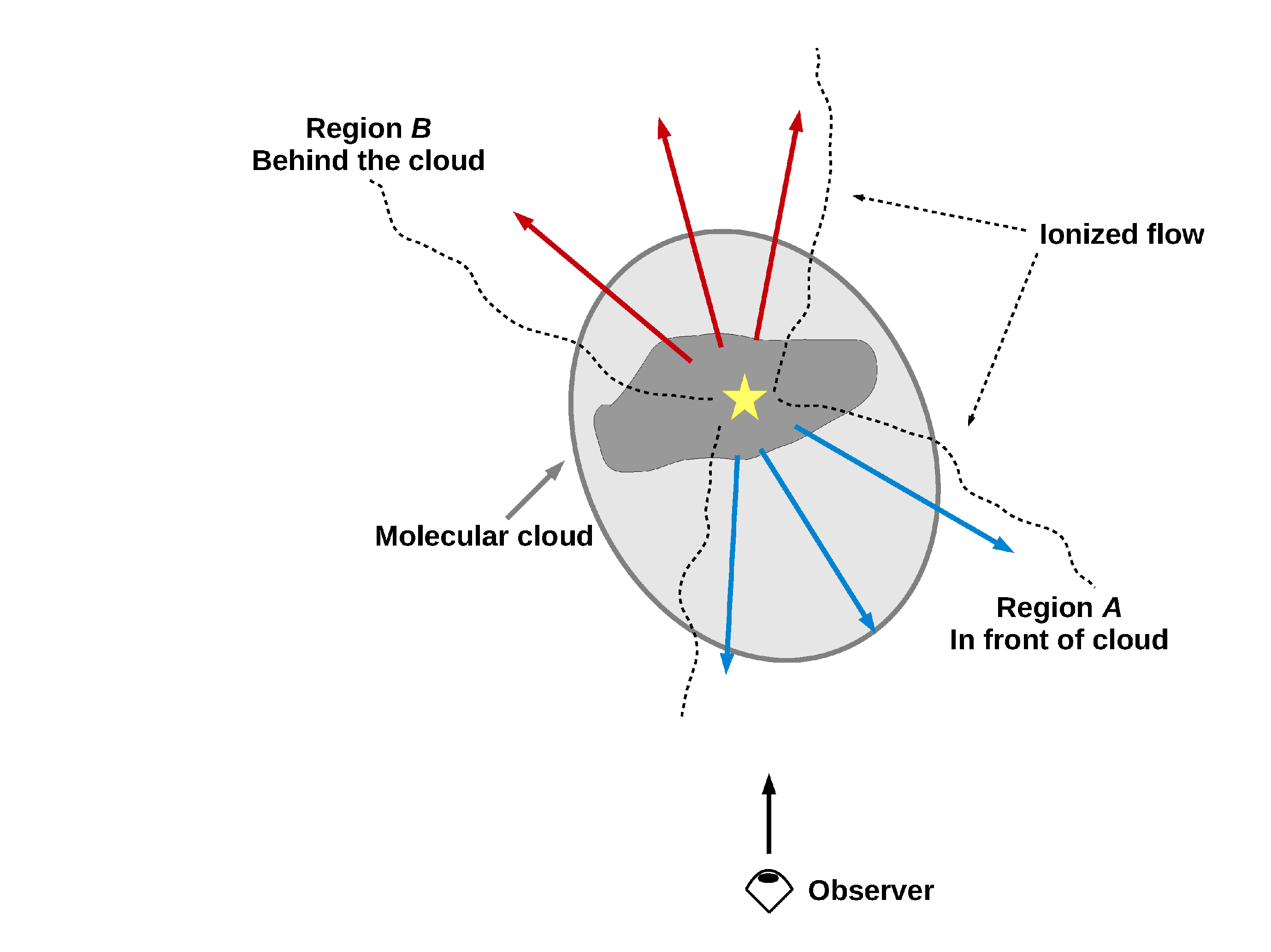}
\caption{(Left) The pictorial representation of RRL velocity field observed towards G351.63--1.25. The molecular cloud is represented with a green ellipse and ionized gas encircled with a black ellipse. The dashed ellipses encircle regions $A$ and $B$, which are overlaid on 1280 MHz greyscale image. The blue arrows represent velocities that are blue shifted with respect to the molecular cloud. Similarly, red arrows denote the red shifted velocities. The location of radio peak is marked as an asterisk. (Right) Schematic diagram showing the orientation of G351.63--1.25 with respect to the observer. The bipolar \hii~region is aligned at an angle to the observer's line-of-sight so that the expansion of \hii~region is not exactly directed to the observer. The images are not to scale.}
\label{17258th}
\end{figure*}

%%%%%%%%%%%%%%%%%%%%  %%%%%%%%%%%%%%%%%%%%%%%%%%%%%%%%%%%%%%%%%%%%%%%%%
\subsection{Bipolar nature of G351.63--1.25}
In order to understand the kinematics of G351.63--1.25, a comparison of the velocities of both, ionized and molecular gas, is vital. In the absence of MALT90, we use the molecular line information from earlier single pointing observations. The molecular line velocity is $-12.0$~km/s for CO and $-11.7$~km/s for CS observations, towards G351.63--1.25 \citep{{1977PASAu...3..152B},{1982PASAu...4..434W}}. However, being single dish and single pointing observations, its is difficult to segregate the emission from \textit{A} and \textit{B} regions. In our case, we clearly see a distinction in the RRL velocity components towards these two regions and the difference between the LSR velocities integrated over the regions is 4.8~km/s (Fig.~\ref{17258double}). Towards \textit{A}, the velocity close to the peak emission is $-11$~km/s from the first moment map. Towards the extended envelope, the velocity appears to increase upto $-16$~km/s. Assuming that the molecular line velocity towards \textit{A} is $\sim-12$~km/s, then the ionized gas velocity would appear blue shifted towards the outer regions. This could be interpreted as a champagne flow due to density gradients in the region. Towards region \textit{B}, the velocity structure is different compared to region \textit{A}. The overall velocity structure is red shifted compared to that of \textit{A}. In addition, we see a gradient in the north-south direction, i.e. as we proceed from the southern end of region B towards the northern side, the velocity gets red shifted by $\sim5$~km/s. A pictorial representation of the velocity field is shown in Fig.~\ref{17258th} (Left). 

\par In order to comprehend the velocity field better and compare it with the morphology of the molecular cloud, we have used the dust continuum map at 870~$\mu$m from ATLASGAL. Fig.~\ref{350}(b) presents the overlay of 1280~MHz radio continuum contours over the  cold dust emission at 870~$\mu$m. The extent of dust emission is larger than the ionized gas emission although we see ionized gas emission pockets to the north-west and south-west where there is little cold dust emission.  In addition, we observe that the morphology of dust emission towards the central region near the peak is flattened in a direction that is approximately perpendicular to the ionized gas distribution. Our group's earlier work on this region attributed this morphology to the champagne flow model, where the bipolar nature of  
ionized gas towards G351.63--1.25 is constrained to move towards the low density regions, perpendicular to the direction of flattening (Paper~II). 

\par Based on the above, we arrive at the following schematic of the \hii~region G351.63--1.25, shown in Fig.~\ref{17258th} (Right). 
We assume that regions \textit{A} and \textit{B} belong to the same molecular cloud and the kinematics can be understood in terms of (i) density distribution, and (ii) projection effects. We are of the opinion that the exciting star lies within the region \textit{A}, at the location of the ionized gas peak emission that is close to the dust continuum peak. This is shown as an asterisk in Fig.~\ref{17258th}. The \hii~region is believed to be inclined with respect to the observer's line-of-sight, with region \textit{B} located farther away. As \textit{B} is located at the farther end of the molecular cloud, the ionized gas from \textit{B} is likely to expand away from the observer. This would give rise to the red-shifted components in the velocity field. The region \textit{A}, on the other hand, is relatively blue-shifted with respect to the cloud and hence it is possibly located near the edge of the cloud closer to the observer. This would explain the relative blue-shifted emission towards the observer. In this scenario, due to the inclination with respect to the observer's line-of-sight, the observer sees emission that would be increasingly blue-shifted towards one side with respect to the other side. This could explain the larger blue-shift in emission from the east and south-east as compared to the western edge. This would also be consistent with the larger red-shift observed towards the north-western edge of region \textit{B}. Thus, the kinematics of the bipolar \hii~region G351.63--1.25 has origins in the champagne flow of ionized gas out of a flat molecular cloud and the direction of the density gradient is responsible for the \hii~region silhouette. However, this proposed qualitative model although consistent with observations, is not exclusive and we cannot rule out the presence of local velocity gradients due to the clumpy medium. High resolution molecular line observations would aid immensely in confirming the scenario.

\section{CONCLUSIONS}
Based on the analysis of RRLs and molecular line data towards the \hii~regions G351.69--1.15 and G351.63--1.25, we arrive at the following conclusions.

\subsection{G351.69--1.15}

\begin{enumerate}
\item{We detect 172$\alpha$ RRLs of hydrogen and carbon in the cometary \hii~region G351.69--1.15. The hydrogen RRL traces the ionized gas and is similar in morphology to that of the continuum emission whereas the carbon RRL originates from the PDR in the vicinity.}

\item{The effects of pressure and dynamical broadening are observed in the line profile of hydrogen RRL. The effect of dynamical broadening is significant compared to that of the pressure broadening.}

\item{A velocity gradient across the cometary axis of the \hii~region has been detected. RRL velocity at the cometary tail is blueshifted by $\sim$12~km/s compared to that towards the cometary head.}

\item{Six molecular species (HCO$^+$, H$^{13}$CO$^+$, HCN, HNC, N$_2$H$^+$ and C$_2$H) detected in the region tracing the properties such as density, chemical robustness, PDR etc. The line profiles reveal a multitude of information such as outflow signatures and presence of turbulence.}

\item{The RRL velocity at the cometary head is nearly the same as that of the molecular cloud. Towards the tail, the ionized gas is blue shifted with respect to ambient cloud suggesting champagne flow as the mechanism responsible for cometary morphology.}  
\end{enumerate}

\subsection{G351.63--1.25}

\begin{enumerate}
\item{We have identified two dust clumps in 870~$\mu$m cold dust emission. The total cloud mass is estimated to be 2860~M$_\odot$.}

\item{Hydrogen and carbon RRLs are detected towards G351.63--1.25$\textit{A}$ region and an additional RRL of an unknown species, possibly the sulfur RRL is also detected towards this region. Only hydrogen RRL is observed towards region $\textit{B}$.}

\item{Among the line broadening effects observed towards G351.63--1.25, thermal broadening is the major contributor while the effects of dynamical and pressure broadening are relatively small.}

\item{The velocities of the two regions G35.16--1.3$A$ and $B$ are relatively blue shifted and red shifted with respect to the molecular cloud velocity. The velocity difference between the two regions is 4.8~km/s.} 
 
\item{The bipolar morphology of this \hii~region is explained as a champagne flow of the ionized gas out of a flat molecular cloud.} 
\end{enumerate}

\bigskip
\noindent \textbf{ACKNOWLEDGEMENT}\\

\par We thank the referee for a critical reading of the manuscript and highly appreciate the comments and suggestions, which significantly improved the quality of the paper. We thank the staff of GMRT, who made the radio observations possible. GMRT is run by the National Centre for Radio Astrophysics of the Tata Institute of Fundamental Research. The ATLASGAL project is a collaboration between the Max-Planck-Gesellschaft, the European Southern Observatory (ESO) and the Universidad de Chile. It includes projects E-181.C-0885, E-078.F-9040(A), M-079.C-9501(A), M-081.C-9501(A) plus Chilean data. This publication made use of data products from $Herschel$ (ESA space observatory). We thank Sebastien Bardeau for his help in molecular line analysis using the GILDAS CLASS software. 
  
\bibliography{reference}

\appendix
\onecolumn
\noindent\textbf{APPENDIX}
\section{HERSCHEL HI-GAL SURVEY}

\begin{figure*}
\hspace*{-0.4cm}
\centering
\includegraphics[scale=0.25,angle=90]{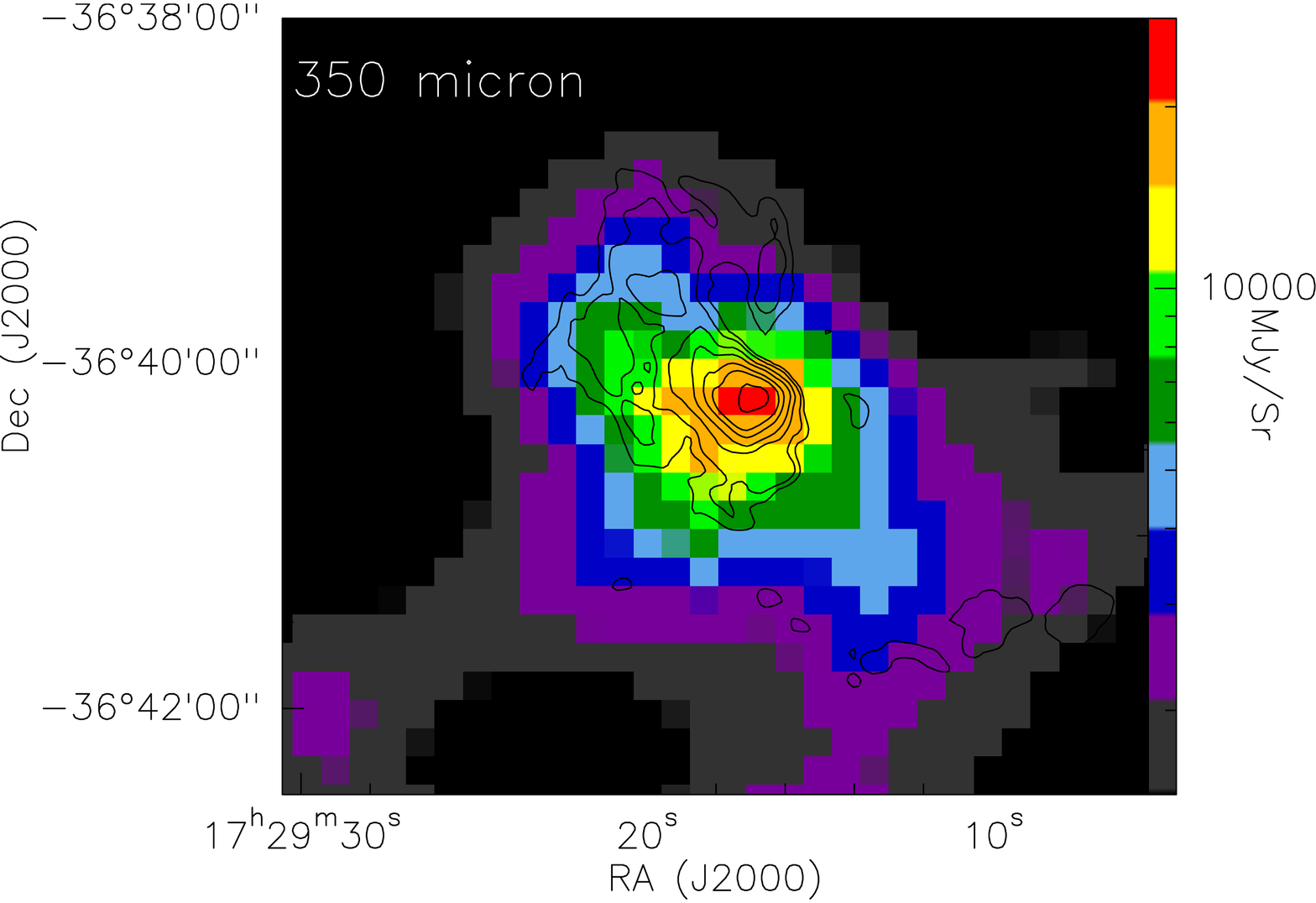} \quad \includegraphics[scale=0.25,angle=90]{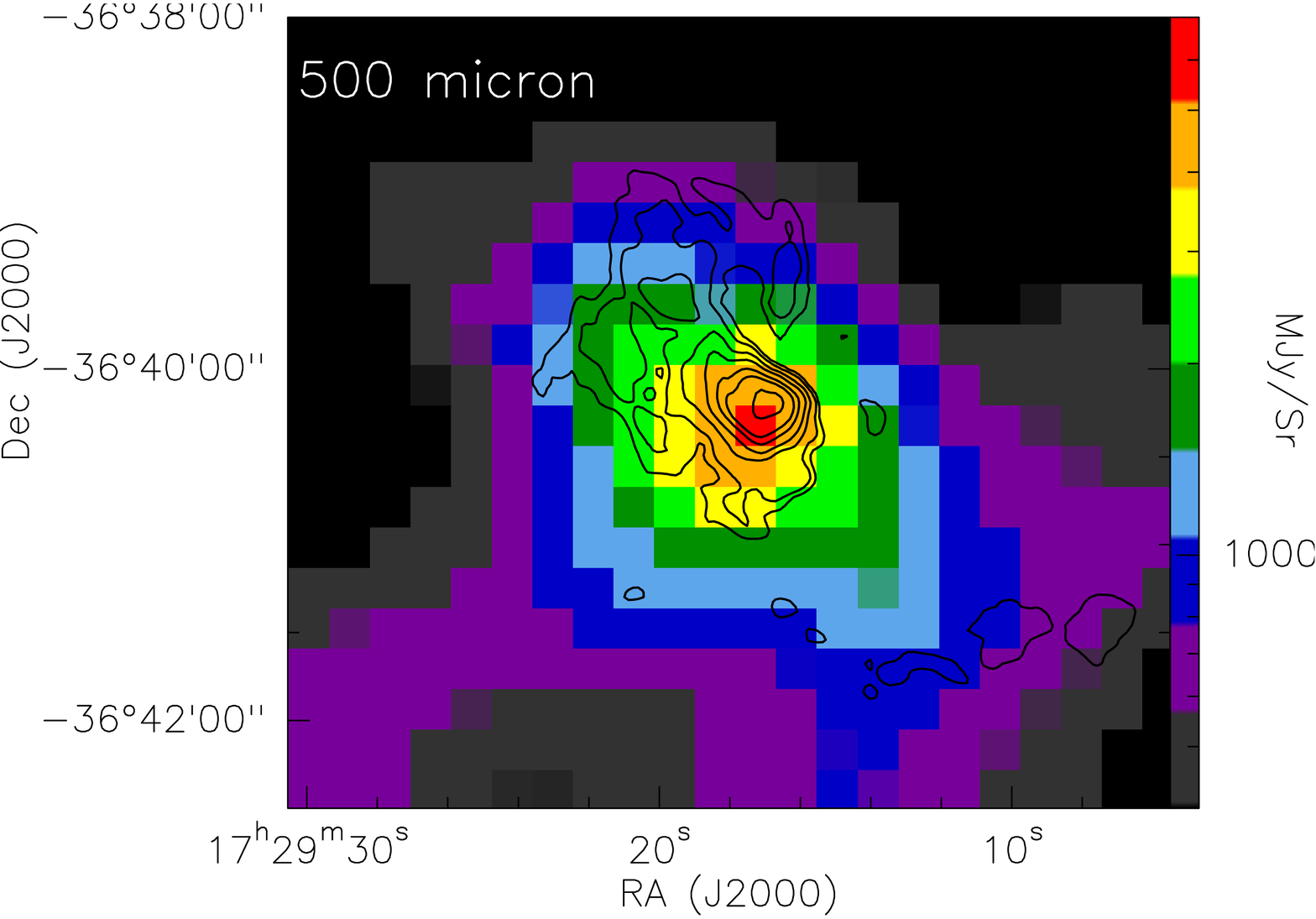} \caption{ (a) Herschel 350~$\mu$m dust continuum map of the \hii~region G351.63--1.25 overlaid with 1280~MHz radio contours. (b) Herschel 500~$\mu$m cold dust map overlaid with 1280~MHz radio contours.}
\label{herschel}
\end{figure*}

The \textit{Herschel Space Observatory} is a 3.5-meter telescope capable of observing in the far-infrared and submm spectral range 55-671~$\mu$m \citep{2010A&A...518L...1P}. The images obtained from the Herschel Archive are part of the Herschel Hi-GAL survey \citep{2010PASP..122..314M} carried out using the Photodetector Array Camera and Spectrometer \citep[PACS,][]{2010A&A...518L...2P} and the Spectral and Photometric Imaging Receiver \citep[SPIRE,][]{2010A&A...518L...3G}. The Hi-GAL observations were carried out in $\lq$parallel mode' covering 70, 160 $\mu$m (PACS) as well as 250, 350 and 500~$\mu$m (SPIRE). Towards G351.63--1.25, we find that there is no coverage by PACS. Among the SPIRE images available, the 250~$\mu$m image is saturated. We, therefore, use the sensitive 350~$\mu$m and 500~$\mu$m Hi-GAL images to understand the cold dust distribution. The pixel sizes in the images are 10$\arcsec$ and 14$\arcsec$ at 350 and 500~$\mu$m, respectively. The resolution is  24.9$\arcsec$ at 350~$\mu$m and 36$\arcsec$ at 500~$\mu$m. The 350 and 500~$\mu$m images of the G351.63--1.25 region are shown in Fig.~\ref{herschel}. The emission morphology is similar to that observed at 870~$\mu$m as seen in Fig.~\ref{350}(b) except the non-detection of clump C2. This is due to the lower resolution of the 350 and 500~$\mu$m maps.

%%%%%%%%%%%%%%%%%%%% Fig19 %%%%%%%%%%%%%%%%%%%%%%%%%%%%%%%%%%%%%%%%%%%%%%%%%

%%%%%%%%%%%%%%%%%%%%%%%%%%%%%%%%%%%%%%%%%%%%%%%%%%%%%%%%%%%%%%%%%%%%%

\end{document}